\documentclass{article} 
\usepackage{iclr2027_conference,times}

\usepackage{hyperref}
\hypersetup{hidelinks}
\usepackage{url}
\usepackage{booktabs}
\usepackage{array}
\usepackage{amsmath}
\usepackage{amsfonts}
\usepackage{amssymb}
\usepackage{graphicx}
\usepackage{needspace}
\usepackage{float}
\usepackage{siunitx}
\usepackage[most]{tcolorbox}
\usepackage{colortbl}
\usepackage{placeins}
\usepackage{etoc}
\usepackage{placeins}
\AtBeginDocument{\etocdepthtag.toc{main}}

\definecolor{broadblue}{HTML}{0072B2}
\definecolor{rootorange}{HTML}{A85800}
\definecolor{authorpurple}{HTML}{9C3F74}
\definecolor{probeband}{HTML}{EEF2F4}
\definecolor{goodgreen}{HTML}{1B7A3D}
\definecolor{badred}{HTML}{B3261E}
\definecolor{weakamber}{HTML}{B8860B}

\graphicspath{{figures/}}

\newtcolorbox{paperbox}[1][]{
  enhanced,
  breakable,
  colback=black!2,
  colframe=black!25,
  boxrule=0.5pt,
  arc=2pt,
  left=6pt,
  right=6pt,
  top=6pt,
  bottom=6pt,
  title=#1,
  fonttitle=\bfseries
}

\title{MUSES: Retrieving the Literature That Shapes a Researcher's Next Paper}

\author{
Rohan Pandey$^{1,2}$ \quad Sunjae Kwon$^{1,2}$ \quad Hong Yu$^{1,2,3}$\\
$^{1}$University of Massachusetts, Amherst\\
$^{2}$VA Bedford Health Care\\
$^{3}$University of Massachusetts, Lowell\\
\texttt{\{rohanpandey,sunjaekwon\}@umass.edu, Hong\_Yu@uml.edu}
}

\iclrfinalcopy

\providecommand{\todo}[1]{\textcolor{red}{[TODO: #1]}}

\begin{document}
\newif\ifrevisionmarks
\revisionmarkstrue
\maketitle

\fancyhead{}
\renewcommand{\headrulewidth}{0pt}

\begin{abstract}
Some of the work that shapes a researcher's next paper is not
something they went looking for: it may arrive before the
project has a question anyone could search for.
How much of it can be anticipated from a researcher's past
publications alone?
We introduce MUSES, a benchmark of 1.04 million cases built
from the Semantic Scholar Open Research Corpus (S2ORC),
of which 140,278 form the held-out test split.
Given a researcher's papers before a cutoff, a system ranks
2.33 million earlier candidates for their next paper and is
scored primarily on works the researcher had not cited before.
Because citations alone do not show what shaped a paper,
we also asked authors directly which works shaped or inspired
their papers, yielding 1,633 author-collected evaluable cases,
a distinct construct from the citation-based targets.
A classifier grounded in human coding adds citation contexts
where the next paper describes building on a work.
Across lexical, dense and citation-graph retrievers and
supervised models, we find that a researcher's past provides
a useful guide to their next paper's literature, but retrieval
becomes harder for works they had not previously cited.
In the main comparison, history-based retrievers place a newly
cited work in their top 100 for 30--37\% of cases, against
5\% for popularity.
Among author-collected cases, one retriever finds a work the
authors themselves named in 52\% of cases
(30\% when targets exclude works the researcher previously cited).
The researcher's own history also outperforms histories from
topically similar researchers.
For works the researcher had not previously cited, even
pooling the evaluated rankings leaves 36--58\% of cases
without a target in any top-100 list, across citation-based
and author-reported targets.
Information about the project helps narrow this gap:
in hindsight, adding the next paper's title improves retrieval,
and adding its abstract improves it further.
MUSES provides a benchmark for developing systems that help
researchers discover literature relevant to their next project.
\end{abstract}

\section{Introduction}\label{sec:intro}

Some of the work that shapes a researcher's next paper is work
they did not know to look for. It may come to their attention
through a conversation, a talk, or a paper stumbled on while
looking for something else---before the project has a question
anyone could search for. Tools that recommend literature try
to make such encounters less a matter of luck. Early in a project,
whether they are personalized feeds~\citep{flicke2025scholarinbox}
or ideation agents that retrieve prior papers as
inspiration~\citep{wang2024scimon,si2024}, they can draw on
one input: the researcher's own publication record.
How much can that record alone reveal about the literature
of the researcher's next paper---and what does it miss?

Several lines of work address related parts of this problem.
Researcher-history recommenders profile a scholar's interests
from past papers or ratings~\citep{sugiyama2010recentinterests,
jiang2023taprec,flicke2025scholarinbox}, and are evaluated on
relevance to those interests or on recently cited papers.
Citation recommenders and literature search use a draft,
a citation context, or an explicit
question~\citep{R3,R4,zheng2026citerag,zhang2026cbert4rec,
ajith2024litsearch}. These inputs provide clues about the project
or the literature being sought. Ideation systems also retrieve
prior work as inspiration but are typically evaluated on the
ideas they generate~\citep{wang2024scimon,R28,si2024};
ResearchBench compares retrieved inspirations with real papers,
but is given each paper's research question~\citep{liu2026researchbench}.
MUSES asks how a researcher's publication history alone can
help surface literature that might shape their next project---before
they know what to search for. We study this question by testing
what that history can recover from the literature of the
researcher's subsequent paper, without providing its title,
text, or research question.

Research does not always continue in a straight line, so relevance
to a researcher's existing interests need not identify the literature
that shapes their next paper. We introduce MUSES, a benchmark
for retrieving that literature from publication history alone,
named for the Muses, the sources of inspiration it seeks to
anticipate. Researcher histories and candidate papers must precede a fixed
cutoff. We focus primarily on works the researcher had not cited
before, and supplement citation lists with evidence from the
paper's text and its authors.

\begin{figure*}[t]
  \centering
  \includegraphics[width=0.9\textwidth]{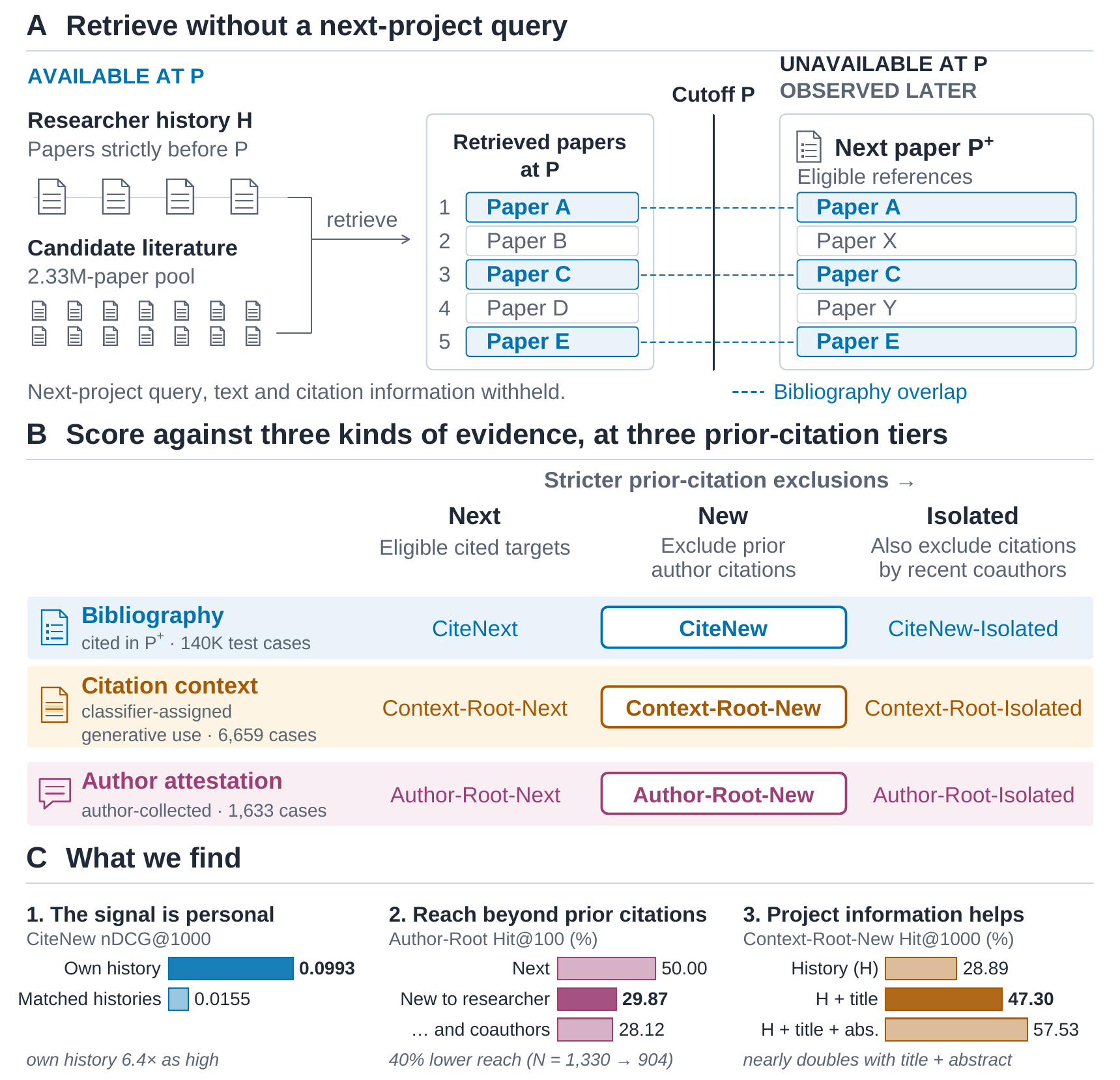}
  \caption{\textbf{MUSES at a glance: task, scoring, and findings.}
(A) At focal paper $P$, a retriever sees only the researcher's
history $H$ and candidates published before $P$.
(B) Rankings are scored against three kinds of evidence from
the next paper $P^+$, at three prior-citation tiers;
boxed cells are the primary targets.
Author-Root uses a separate 2.60M-paper pool.
(C) The three findings (\S\S4.1--4.3), illustrated using
SPECTER2-Proximity with 16 centroids: own versus matched
histories on CiteNew, prior-citation exclusions on Author-Root,
and a hindsight test adding $P^+$'s title and abstract
on Context-Root-New.}
  \label{fig:information-boundary}
\end{figure*}

In our benchmark, each case fixes a researcher, a focal paper $P$,
and their next paper $P^+$. A system sees only the researcher's
publications before $P$ and ranks candidates published before $P$;
the next project's title, text, collaborators, and references stay
hidden (Figure~1A). Built from S2ORC~\citep{lo2020s2orc},
the main benchmark contains about 1.04 million cases across fields,
including 140,278 test cases, with training, validation, and test sets separated by researcher and time and a fixed pool of 2.33 million candidates. Targets are the works
$P^+$ cites, grouped into three nested tiers: all of them; those
the researcher had not cited before, the primary target; and those
neither the researcher nor their recent coauthors had previously cited.

To understand which works shaped or inspired a paper,
we turn to the authors themselves.
We contacted the corresponding authors of papers across
57 journals and received replies for 2,295 surveyed papers,
from which we constructed a distinct author-collected
evaluation set of 1,633 cases.
MUSES pairs this evidence with a scalable complement:
citation contexts in $P^+$ that describe building on a cited
work's idea, framing, or method, labeled by a classifier
grounded in 1,202 human-coded contexts.
The two are complementary rather than interchangeable:
authors report what shaped their paper, whereas citation
contexts show how the paper uses a work (Figure~1B).

We evaluate 24 configurations spanning popularity,
lexical and dense retrieval, citation-graph methods
and supervised models.
A researcher's past provides a useful guide to their
next paper's literature.
In the main comparison (Table~\ref{tab:main-results}),
history-based retrievers place a newly cited work
in their top 100 for 30--37\% of cases,
against 5\% for popularity.
The researcher's own history also outperforms histories
from topically similar researchers, across lexical
and dense retrieval and all three evidence sources.
On the author-collected set, SPECTER2-Proximity with
16 centroids finds a work the authors themselves named
in its top 100 for 52\% of 1,633 cases
(95\% CI 49--54\%).
Substantial gaps remain: for works the researcher
had not previously cited, pooling all 24 rankings
leaves 36--58\% of cases without a target in any
top-100 list, across bibliography, citation-context
and author-reported targets.

Information about the project improves retrieval.
In a hindsight test, adding the next paper's title
to the same history improves recovery, and adding
its abstract provides a further gain (Figure~1C).
Both improvements hold across BM25,
SPECTER2-Proximity and GritLM, for bibliography,
citation-context and author-reported targets,
across all reported tiers.
These results motivate testing whether a short
description of an emerging project, such as a working
title, can provide similar gains beyond history alone.
\section{Related Work}\label{sec:related}

\begin{table*}[t]
\centering
\footnotesize
\setlength{\tabcolsep}{1.3pt}
\renewcommand{\arraystretch}{1.08}

\caption{\textbf{Among the compared benchmarks, MUSES uniquely
combines retrieval without project-specific input,
prior-citation exclusions and evidence beyond bibliography
membership.}
Rows group citation recommendation and search, research
ideation, and researcher-history recommendation.
Parenthesized checks indicate related but non-equivalent
evidence: relevance judgments for RELISH and Sugiyama \& Kan,
and paper-derived inspirations for ResearchBench.
Properties follow the source descriptions:
SciDocs~\citep{R15},
RELISH~\citep{brown2019relish},
LitSearch~\citep{ajith2024litsearch},
CiteRAG~\citep{zheng2026citerag},
ResearchBench~\citep{liu2026researchbench},
\citet{sugiyama2010recentinterests},
and TAPRec~\citep{jiang2023taprec}.}
\label{tab:comparison}

\begin{tabular*}{\textwidth}{
@{\extracolsep{\fill}}
lll
>{\raggedright\arraybackslash}p{0.19\textwidth}
ccc
@{}}
\toprule
Benchmark
& System input
& Candidates
& Evaluated against
& \shortstack{No project-\\specific input}
& \shortstack{Prior-citation\\filter}
& \shortstack{Beyond-bib.\\evidence} \\
\midrule

SciDocs (cite)
& citing paper
& 30/query
& its own references
& \textcolor{badred}{$\times$}
& \textcolor{badred}{$\times$}
& \textcolor{badred}{$\times$} \\

RELISH
& seed article
& ranked lists
& relevance to seed
& \textcolor{badred}{$\times$}
& \textcolor{badred}{$\times$}
& \textcolor{weakamber}{($\checkmark$)} \\

LitSearch
& search question
& 64K
& matching papers
& \textcolor{badred}{$\times$}
& \textcolor{badred}{$\times$}
& \textcolor{badred}{$\times$} \\

CiteRAG
& abstract or draft
& 554K
& its own references
& \textcolor{badred}{$\times$}
& \textcolor{badred}{$\times$}
& \textcolor{badred}{$\times$} \\

\midrule

ResearchBench
& question + survey
& 75/target
& its inspirations
& \textcolor{badred}{$\times$}
& \textcolor{badred}{$\times$}
& \textcolor{weakamber}{($\checkmark$)} \\

\midrule

Sugiyama\,\&\,Kan
& own papers
& 597
& relevance to interests
& \textcolor{goodgreen}{$\checkmark$}
& \textcolor{badred}{$\times$}
& \textcolor{weakamber}{($\checkmark$)} \\

TAPRec
& dated own papers
& sampled
& recently cited papers
& \textcolor{goodgreen}{$\checkmark$}
& \textcolor{badred}{$\times$}
& \textcolor{badred}{$\times$} \\

\midrule

\textbf{MUSES}
& own papers only
& 2.33M
& citations, generative-use labels, author reports
& \textcolor{goodgreen}{$\checkmark$}
& \textcolor{goodgreen}{$\checkmark$}
& \textcolor{goodgreen}{$\checkmark$} \\

\bottomrule
\end{tabular*}
\end{table*}

\paragraph{Citation recommendation and literature search.}
Context-aware recommenders use the words around a citation~\citep{R3,R4}, CiteRAG uses target-paper and manuscript context~\citep{zheng2026citerag}, CBERT4REC an incomplete manuscript~\citep{zhang2026cbert4rec}, and LitSearch an explicit literature question~\citep{ajith2024litsearch}. Each uses project-specific information that is not available in MUSES's history-only setting; Table~\ref{tab:comparison} compares these information contracts.

\paragraph{Recommending from a researcher's past work.}
\citet{sugiyama2010recentinterests} derive interests from a researcher's recent work, TAPRec models changing interests from historical publications, with coauthor signals for cold-start users~\citep{jiang2023taprec}, and Scholar Inbox personalizes a continuing literature feed from user ratings~\citep{flicke2025scholarinbox}; each is judged on relevance to a history or on recently cited papers, not on what a specific later paper uses. Scientific-document encoders learn from citation structure~\citep{R15,R16,R17}, whereas general-purpose embedders are instruction-tuned for broad text similarity~\citep{muennighoff2025gritlm,zhang2025qwen3embedding}; MUSES evaluates both encoder families, and one-vector versus several-interest history summaries, against that outcome (Table~\ref{tab:comparison}).

\paragraph{Literature-based discovery and research ideation.}
Literature-based discovery treats connections latent across literatures as undiscovered knowledge~\citep{swanson1986,smalheiser2012}. Ideation work also builds knowledge bases of recombined ideas~\citep{R29}. ResearchBench evaluates inspiration retrieval as part of hypothesis generation from an explicit research question and background~\citep{liu2026researchbench}, SciPaths forecasts enabling prior-work pathways given a target scientific contribution~\citep{chamoun2026scipaths}, and benchmarks such as DiscoveryBench and AstaBench likewise evaluate discovery and research capabilities once a task or goal has been specified~\citep{discoverybench,astabench}; each of these is given some description of the project it serves. MUSES instead evaluates retrieval before a project-specific query is available.

\paragraph{Evidence beyond bibliography membership.}
Citation-intent, citation-frame, and citation-function research classifies the roles
citations play in published text~\citep{R19,R26,R2,vodicka2026citationfunction}, and \citet{valenzuela2015meaningful} identify citations judged
meaningful to the citing paper from their context. These classify a citation once
written; MUSES instead makes generative use a retrieval target to be recovered without
access to the next paper or its citation context. Author-report research asks a related
question: \citet{wakeling2025citationperception} ask authors how their own work was cited, whereas MUSES asks
authors which prior works shaped or inspired their paper. The two therefore remain
separate, complementary evidence sources.
\section{The MUSES Benchmark}

MUSES asks how well a retriever can anticipate the literature of a researcher’s next paper using only what the researcher has already published. We distinguish what the retriever can observe,
which prior citations are excluded from the targets, and what evidence connects those targets to the next paper.

\subsection{Task: Ranking Literature from a Researcher's Past Work}

The task evaluates retrieval without a project-specific
query: a system knows the researcher's publication
record and nothing about the next paper. Each case pairs a researcher's focal paper $P$ with the next paper they lead, $P^+$ (their immediately subsequent last-author paper), both drawn from S2ORC (Appendix~A). The retriever sees the researcher's papers published before $P$, including papers in any author position. We denote this publication history by $H$ and use the date of $P$ as the retrieval cutoff. 

The retriever uses $H$ to rank papers from a fixed shared pool $G$ that were published before $P$. Formally, for researcher $a$ and cutoff date $t$, the eligible candidate set is
\[
C(t)=\{d\in G:\operatorname{pubdate}(d)<t\},
\]
and the retriever produces the ranking $f(H(a,t),C(t))$. In the prospective task, no information from $P^+$ is used to produce the ranking. Its references and citation contexts define the evaluation targets, while its title and abstract are introduced only in the retrospective hindsight diagnostic. 

\subsection{Targets: Which Works Count as Correct}

A ranker could score well simply by resurfacing works the researcher already cites. We therefore organize the targets into three nested tiers that progressively remove prior citation connections. \textsc{CiteNext} contains every work $P^+$ cites that was published before $P$. \textsc{CiteNew}, the primary tier, removes works the researcher had cited before $P$. \textsc{CiteNew-Isolated} also removes works cited before $P$ by anyone who coauthored with the researcher in the preceding five years, since such works could have reached the researcher through a collaborator. Together, these tiers evaluate how well retrieval extends beyond the researcher’s and recent coauthors’ prior citation histories (Appendix B.1).

\subsection{	Targets grounded in evidence of influence}

The tiers above use only bibliography membership. A citation
shows that a paper used a work, not that the work shaped it;
many references acknowledge background, tools, or contrasts.
Two further sources supplement this evidence: one asks authors which works shaped or inspired the paper, and the other identifies generative use in its citation contexts. For cited works, both use the same three prior-citation tiers (\S3.2), yielding Author-Root-Next, -New and -Isolated,
and Context-Root-Next, -New and -Isolated. 

\paragraph{What authors say shaped their paper.}
After $P^+$ was published, we invited its corresponding
author to name the works that shaped or inspired it,
offering a shortlist of up to eight references from its
bibliography and the option to name any other work.
We received responses for 2,295 papers from 2,243 respondents and matched the reported works to
bibliographic records (Appendix~D.1).
A case is evaluable with a response, a usable history,
and an eligible attested work in the pool:
1,633 cases for Author-Root-All (which also keeps
attested works $P^+$ does not cite),
1,330 for Author-Root-Next, 904 for Author-Root-New,
and 665 for Author-Root-Isolated.
These author-collected cases form a distinct set,
not restricted to last authors, and are ranked against
a 2,603,132-paper pool that adds the attested works
to the benchmark candidates (Appendix~D).
The panel spans life sciences, health and medicine,
physical sciences, social sciences and humanities,
and computer science and mathematics; Appendix~D.1 reports field composition and selection limitations;
Appendix~E.4 reports retrieval by field.

\paragraph{How the next paper uses a work.}
The second source reads the citation contexts of $P^+$,
using a six-category taxonomy of citation function;
MUSES groups these into generative and non-generative use.
Generative contexts describe how the paper builds on
prior ideas, framing or methods.
A cited work enters the Context-Root set when at least
one of its contexts is labelled generative.
The human-coded data contain 1,202 citation contexts.
At benchmark scale, a Qwen3-8B classifier with a LoRA
adapter, distilled from a larger teacher, assigns the
labels (validation: Appendices~B.2--B.3).
The label describes how the paper presents its use
of a work, which may differ from the influence
reported by its authors.
Only 7.6\% of author-endorsed works with available
citing text receive a generative-use label.
This limited overlap supports treating author reports
and citation-context labels as complementary evidence;
Appendix~D.3 reports the comparison and human audit.

\subsection{Eligibility, Splits, and the Candidate Pool}

In the main S2ORC benchmark, a case is eligible when
the researcher has at least three preceding authorship
records (Appendix~A.1), $P^+$ appears within five years
of $P$, cites at least ten references, and is a research
article. We use author-disjoint splits, assigning each researcher by their career midpoint (the midpoint between their earliest and latest dated publications): training before 2018, validation in 2018--2020, and test in 2021--2023. The focal paper $P$ must also fall within its assigned period. The midpoint determines the split and is not a retrieval input (Appendix~A.1).

The shared candidate pool is fixed across the bibliography and Context-Root cases: it contains works cited by benchmark next papers across splits that have a title and abstract, plus 100,000 background papers sampled across publication years, for 2.33 million candidates in total. Each case's targets thus compete with other cases' targets as well as with background papers. For each case, only candidates published before $P$ are eligible for retrieval. Author-Root uses a separate pool of 2,603,132 papers,
extending the shared pool with identified author-reported
works and additional background papers
(Appendices~A.2 and~D.1).

We evaluate retrieval on cases with a usable publication history and at least one target in the eligible candidate pool. Only targets in that pool and published before $P$ are included in metric computation. Within each tier, all methods are evaluated on the same set of cases. The number of evaluated cases is reported alongside the results (Appendix~E).

\subsection{Retrievers}

We evaluate 24 retrieval configurations across several
method families. These methods characterize performance
across retrieval families rather than bound the task.
Table~2 presents eight representative configurations;
Appendix~E reports the complete results.
Popularity provides a history-independent baseline,
ranking candidates by citations received before $P$.

\noindent\textbf{Text-based retrieval.}
BM25, SPLADE and Qwen3-Embedding-8B match candidates
against a compact history query: titles, then abstracts,
from up to 20 recent papers, truncated to 128 words.
We also evaluate BM25 without the word limit.
For dense retrieval, we summarize history-paper embeddings
as either one mean vector or up to 16 centroids from
spherical clustering, allowing separate lines of work
to remain distinct.
Candidates receive their highest cosine similarity
to a centroid.
We test both summaries for GritLM-7B, SciNCL and
SPECTER2-Proximity, and 16 centroids for
Qwen3-Embedding-8B.
SPECTER2-Proximity uses the SPECTER2 encoder with its
proximity adapter.
LateOn uses token-level late interaction with individual
history-paper titles, retaining the highest score.

\noindent\textbf{Citation-graph retrieval.}
Personalized PageRank, co-citation, bibliographic coupling
and Adamic--Adar rank candidates using the researcher's
history and citation links among papers dated before
the case cutoff.

\noindent\textbf{Supervised retrieval.}
Linear-Proximity learns linear maps of history and
candidate SPECTER2-Proximity vectors from training-split
citations and ranks with 16 centroids.
We also test a citation-neighbour variant and its
retrained control without neighbour features.
We fine-tune SPECTER2-Proximity and apply LoRA adaptation
to Qwen3-Embedding-8B.
These encoders and their frozen counterparts score
candidates by maximum similarity to individual history
papers. Appendix~C details training and scoring;
Appendices~H.1--H.2 examine sensitivity to history
representation and query construction.

\subsection{Metrics}

Before a project exists, a researcher needs one good entry point into the relevant literature more than a complete list, so the primary metric asks whether any target appears near the top. Reach, or Hit@$K$, is the share of cases with at least one target in the top $K$. Coverage, or macro Recall@$K$, is the fraction of each case's eligible targets found in the top $K$, averaged equally across cases. We use $K=100$ as the primary reporting depth, with $K=10$ and $K=1000$ showing performance at shallower and deeper ranks. We also report nDCG, which gives greater weight to relevant works retrieved at higher ranks. Appendix~E.1 provides the formal definitions of all metrics, including MRR@1000, and Appendix~E reports the complete results.

\subsection{Three Diagnostics: Matched Histories, Pooled Rankings, and Hindsight}

\textbf{Is the signal specific to the researcher,
rather than to similar researchers?}
We compare retrieval from the researcher's own publication
history with retrieval from three content-matched histories
of other researchers.
Within each case, the candidate pool, temporal cutoff
and targets remain fixed, and performance is averaged
over the three matched histories.
We evaluate SPECTER2-Proximity, GritLM and BM25 across
Bibliography, Context-Root and Author-Root targets.
The primary measure is the paired difference in nDCG@1000
between the researcher's own history and the mean over
matched histories, with Hit@100 as a supporting measure
(Appendix~F.1).

\noindent\textbf{How much do the rankings collectively reach?}
We pool all 24 configurations across the Bibliography,
Context-Root and Author-Root tiers.
Within each tier, all configurations use the same cases
and targets. A case is reached at depth $K$ if any ranking
retrieves a target within its top $K$.
Pooling combines each method's top-$K$ results,
so it can contain more than $K$ distinct papers.
The result measures complementary recovery across
methods; it is neither a combined ranker nor an
upper bound on the task (Appendix~G.1).

\noindent\textbf{How much does project-specific information
add in hindsight?}
We compare publication history alone ($H$), history
plus the next paper's title ($H+T$), and history plus
its title and abstract ($H+T+A$).
Within each retriever, the candidate pool, temporal
cutoff and history representation remain fixed.
We test two dense retrievers, SPECTER2-Proximity
and GritLM, and the lexical retriever BM25, across
Bibliography, Context-Root and Author-Root targets,
using the same cases within each comparison.
This retrospective comparison measures how much retrieval
improves with information about the next paper that is
unavailable at the focal cutoff (Appendix~F.2).
\providecommand{\todo}[1]{\textcolor{red}{[TODO: #1]}}

\begin{table*}[t]
\centering
\footnotesize
\caption{
\textbf{History-based retrieval beats popularity by
5.7--7.1$\times$ on CiteNew.}
Reach (Hit@100, \%) for eight representative configurations.
$N$ denotes evaluation cases, not target works.
The bibliography and Context-Root columns use their respective
evaluation cohorts: cases with at least one eligible target and
at least one researcher paper published strictly before $P$.
Within each column, all systems are evaluated on identical cases.
All systems rank temporally eligible candidates from the shared
2.33M-paper pool.
Citations denotes history-conditioned citation-graph retrieval.
The shaded row is the reference retriever used in later analyses.
Complete results across configurations, metrics, and depths
appear in Appendix~E.
Bold indicates the largest observed value among the displayed
configurations.
}
\label{tab:main-results}
\setlength{\tabcolsep}{2.4pt}
\begin{tabular}{@{}lll ccc c@{}}
\toprule
& & &
\multicolumn{3}{c}{\textbf{Bibliography}} &
\textbf{\shortstack{Citation\\context}}
\\
\cmidrule(lr){4-6}
\cmidrule(l){7-7}

Family
& Retriever
& Representation
& \textcolor{broadblue}{CiteNext}
& \textcolor{broadblue}{CiteNew}
& \textcolor{broadblue}{\shortstack{CiteNew-\\Isolated}}
& \textcolor{rootorange}{\shortstack{Context-Root-\\New}}
\\

& & &
\scriptsize\emph{\shortstack{any\\citation}}
& \scriptsize\emph{\shortstack{new to\\researcher}}
& \scriptsize\emph{\shortstack{new to\\coauthors}}
& \scriptsize\emph{\shortstack{generative\\use}}
\\

& & &
\scriptsize $N{=}135{,}111$
& \scriptsize $134{,}314$
& \scriptsize $133{,}166$
& \scriptsize $6{,}659$
\\
\midrule

Prior only
& Popularity
& none
& 7.01
& 5.21
& 2.50
& 2.30
\\
\midrule

Lexical/sparse
& BM25
& compact query
& 33.96
& 29.71
& 26.79
& 12.43
\\

& SPLADE
& compact query
& 34.21
& 29.85
& 26.85
& 13.31
\\
\midrule

General dense
& Qwen3-Embedding-8B
& compact query
& 36.66
& 32.51
& 29.48
& 16.40
\\

& GritLM-7B
& 16 centroids
& 40.40
& 35.52
& \textbf{32.22}
& 17.03
\\
\midrule

\rowcolor{probeband}
Citation dense
& SPECTER2-Proximity
& 16 centroids
& 37.43
& 32.28
& 28.99
& 15.26
\\
\midrule

Citation graph
& Personalized PageRank
& Citations
& \textbf{46.25}
& \textbf{37.18}
& 31.05
& \textbf{22.53}
\\
\midrule

Supervised
& Linear-Proximity
& learned maps + K16
& 34.51
& 29.57
& 26.30
& 14.19
\\
\bottomrule
\end{tabular}
\end{table*}

\section{Results}\label{sec:experiments}
\suppressfloats[t]

Three results define the benchmark: researcher history carries prospective signal not reproduced by matched other-researcher histories (\S\ref{sec:exp-main-results}); reach declines under stricter prior-citation filters and remains incomplete across the evaluated rankings (\S\ref{sec:exp-novelty}); and project-specific information substantially expands recovery in hindsight (\S\ref{sec:exp-retrospective}).

\subsection{Past work anticipates the next paper's sources, and the signal is personal}\label{sec:exp-main-results}

\textbf{History-based retrieval beats popularity by 5.7--7.1$\times$.}
At depth 100, the seven history-based rankings in
Table~\ref{tab:main-results} find a CiteNew target for
29.57--37.18\% of cases, compared with 5.21\% for
popularity (Hit@100).
All 23 history-based configurations outperform popularity
across every reported tier, metric and depth for Bibliography,
Context-Root and Author-Root (Appendix~E).

\textbf{The signal is specific to the researcher.}
Replacing a researcher's history with histories from
topically similar researchers sharply reduces retrieval
quality.
On the 130,171 CiteNew cases with matched histories,
ranking quality (nDCG@1000) is 6.3--6.9 times as high
with the researcher's own history as with the mean over
three matched histories, across BM25, SPECTER2-Proximity
and GritLM (Appendix~F.1).
Figure~2A shows the SPECTER2-Proximity results.
The own-history advantage holds for all three retrievers
across every reported tier of Bibliography, Context-Root
and Author-Root, for both nDCG@1000 and Hit@100
(Appendix~F.1).

\textbf{Several retrievers find a work authors say shaped
or inspired their paper in over half of cases.}
For works that responding authors named as shaping their
next paper, including works they had previously cited,
SPECTER2-Proximity with 16 centroids achieves Hit@100
of 51.62\% on 1,633 cases (Author-Root-All).
For named works not previously cited by the researcher,
its Hit@100 is 29.87\% on 904 cases (Author-Root-New).
Seven evaluated configurations exceed 50\% on
Author-Root-All (Appendix~E.4).
Author reports also extend beyond the bibliography:
among 313 papers with complete reference lists,
89 of 839 author-named works are not cited
(10.6\%; Appendix~D.3).

\begin{figure*}[t]
  \centering
  \includegraphics[width=\textwidth]{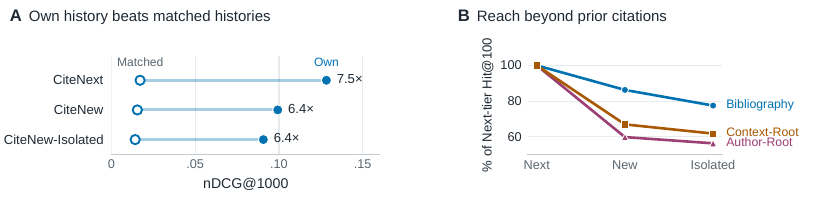}
\caption{\textbf{The signal in a researcher's history is personal,
and reach falls as previously cited works are excluded.}
(A) Own-history versus mean matched-history nDCG@1000
on the bibliography tiers; labels give the ratio.
Each case uses three content-matched histories from other
researchers ($N=130{,}937$, $130{,}171$, and $129{,}063$).
(B) Hit@100 across prior-citation tiers, indexed to each
source's Next-tier value.
Both panels use SPECTER2-Proximity with 16 centroids.}
  \label{fig:personal}
\end{figure*}

\Needspace{4\baselineskip}
\subsection{Retrieval beyond prior citations}\label{sec:exp-novelty}\label{sec:exp-root-gap}

\textbf{In all three kinds of evidence, reach falls for works
the researcher had not cited before.}
For SPECTER2-Proximity with 16 centroids, bibliography
Hit@100 falls from 37.43\% on CiteNext to 32.28\% on
CiteNew and 28.99\% on CiteNew-Isolated (Figure~2B).
The decline from Next to New holds for all 23 history-based
configurations across Bibliography, Context-Root and
Author-Root; further excluding works previously cited by
recent coauthors generally lowers reach again
(Appendix~E).
The Next-to-New decline also persists when comparing the
same cases and accounting for fewer targets, in both
Hit@100 and macro Recall@100 (Appendix~G.3).

\textbf{Retrieval finds entry points, not complete reading lists.}
When retrieving works the researcher had not cited before
(CiteNew; $N=134{,}314$), SPECTER2-Proximity with
16 centroids finds at least one in its top 100 for
32.28\% of cases (Hit@100), while recovering only
9.45\% of each case's targets on average
(macro Recall@100; Figure~3B).
At depth 1,000, it reaches 51.86\% of cases but recovers
only 21.06\% of their targets on average.
Even at this depth, no evaluated configuration recovers
a third of the targets per case on average in any
bibliography tier (Appendix~E).

\textbf{More than a third of cases have no target in any
ranking's top 100.}
Across the New tiers of Bibliography, Context-Root and
Author-Root, pooling all 24 evaluated rankings leaves
approximately 36--58\% of cases unreached at depth 100,
depending on the evidence source (Figure~3A).
At depth 1,000, 16--31\% remain unreached
(Appendix~G.1). Different rankings recover different cases, but their pooled lists still leave substantial gaps.
The pooled result summarizes separate rankings rather
than the performance of a combined retrieval method.

\begin{figure*}[t]
  \centering
  \includegraphics[width=\textwidth]{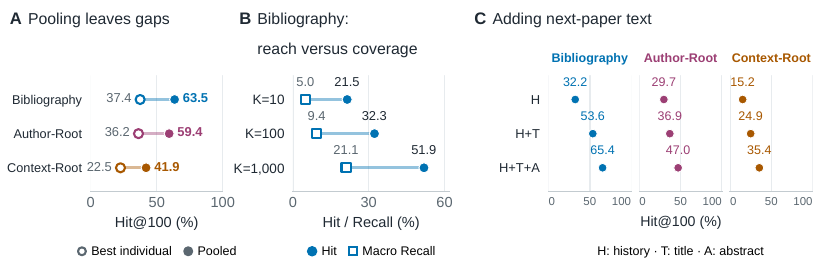}
\caption{\textbf{Retrieval leaves gaps in reach and coverage;
project information improves recovery.}
Results concern works not previously cited by the researcher
(the New tiers).
(A) Best individual (hollow) and pooled (filled) Hit@100
across 24 rankings, shown separately for each evidence source.
Pooling counts cases reached by any ranking's top-100 list;
the best individual is selected separately for each source.
(B) Case reach (Hit) and average target coverage
(macro Recall) on CiteNew at three depths.
(C) Hindsight Hit@100: history ($H$), plus the next paper's
title ($H+T$), plus title and abstract ($H+T+A$).
B and C use SPECTER2-Proximity with 16 centroids.
Bibliography and Context-Root use $N=134{,}314$ and
$6{,}659$; Author-Root uses $N=904$ in A and
$700$ abstract-available cases in C.
All C inputs use the same cases within each source.
Details in Appendices~G.1 and~F.2.}
\label{fig:root-gap}
\end{figure*}

\subsection{Project information substantially improves retrieval}\label{sec:exp-retrospective}

\textbf{The next paper's title improves retrieval, and its
abstract adds further gains.}
For works the researcher had not previously cited
(New tiers), SPECTER2-Proximity with 16 centroids gains
approximately 7--21 percentage points in Hit@100 when
the next paper's title is added, with a further 10--12
points from its abstract.
These ranges span Bibliography, Context-Root and
Author-Root (Figure~3C).
The comparisons use the same cases for all three inputs;
Author-Root includes cases with an available next-paper
abstract.
Both additions improve Hit@100, Hit@1000 and nDCG@1000
for BM25, SPECTER2-Proximity and GritLM across all three
evidence sources and every reported tier (Appendix~F.2).
These gains are measured in hindsight: the next paper's
text is unavailable at the focal cutoff.

\textbf{Do the title and abstract simply name the target works?}
After excluding papers with detected target mentions in
their titles or abstracts, both the title gain and the
additional abstract gain remain positive in Hit@100 and
Hit@1000 for the same three retrievers across all three
evidence sources (Appendix~F.2).
\section{Discussion and Limitations}\label{sec:discussion}

\paragraph{Implications for research agents and retrievers.}
Systems that surface literature early in a project can act
on three findings.
Condition on the individual: the researcher's own history
outperforms histories from topically similar researchers
(\S4.1).
Evaluate works outside prior citation histories: success
at recovering previously cited works can mask weaker
recovery of works the researcher had not cited before
(\S4.2).
Ask about the project early: title and abstract information
improve retrieval in hindsight, motivating prospective
tests of working titles and early project descriptions
(\S4.3). MUSES provides a common evaluation setting for developing
stronger history-based retrievers; prospective user studies
can then test whether these systems help researchers find
useful literature for emerging projects.

\paragraph{Limitations and what would resolve them.}
\textit{Is a classifier label trustworthy?}
Its precision is 0.63 on the contexts used to select it; a
sensitivity analysis with higher-precision labels (0.83) leaves
the conclusions unchanged (Appendix~\ref{app:context-root-precision});
held-out data would better test generalization.
\textit{Are the author reports representative?}
They come from authors who chose to respond, and the
bibliography-derived shortlist may anchor their answers
(Appendix~D.1).
\textit{Could models have seen the answers?}
The pretrained encoders' training data cannot be audited
fully enough to rule out exposure to evaluation-period material.
BM25, which has no pretrained encoder, also shows the tier
decline and incomplete recovery (Appendix~H).
\textit{Would better retrievers close the gap?}
Possibly: the evaluated rankings are not exhaustive, so the
residual describes these systems, inputs and candidate pools,
not the limit of the task.
\textit{Does the last-author convention hold everywhere?}
Less so where authors are ordered alphabetically, as often
in mathematics and economics, and histories inherit S2ORC's
author-disambiguation errors.
\textit{Does the candidate pool limit the conclusions?}
Evaluation covers eligible targets in the candidate pool;
broader pools would test the patterns beyond it.
\textit{What about influence that leaves no citation?}
The tiers remove only recorded citations, and citations
and author reports observe only part of influence;
elicitation not anchored to the bibliography could reach
conversations and uncited reading.
\textit{Which signals are left out?}
Text-based retrievers use titles and abstracts rather than
full text; richer inputs may recover additional works.
\textit{Do entry points help researchers?}
Appendix~D.2 shows examples, but only a user study can
measure the benefit.

\textbf{Conclusion.}
MUSES provides a benchmark for anticipating the literature
of a researcher's next paper from publication history alone.
That history carries a researcher-specific signal, but leaves
substantial gaps for works the researcher had not previously
cited; project information improves recovery in hindsight.
By measuring what history retrieves and misses, MUSES gives
future systems a concrete goal: surface more of that
literature as a project takes shape.

\subsection*{Ethics statement}
MUSES is built from public bibliographic records.
Author responses were collected with consent for
research use. Any public author-derived evaluation
files will contain only identifier-level structured
data needed for evaluation, excluding raw responses,
verbatim response text, free-text rationales,
correspondence, email addresses, and other identifying
content. Retrieval conditioned on an individual
researcher's history could be misused to profile
researchers or rank scientific influence; MUSES
evaluates retrieval of literature for a researcher's
own use.

\subsection*{Reproducibility Statement}
Appendix~A specifies corpus construction, temporal eligibility
and evaluation cases; Appendix~B the citation tiers and
citation-context classifier; Appendix~C the retrieval systems
and training procedures; Appendix~D the author-collected
evaluation; and Appendix~E the metrics and complete results.
Appendix~F describes the matched-history and hindsight
diagnostics, Appendix~G the pooling and target-count analyses,
and Appendix~H the robustness and encoder-exposure checks. 

\subsection*{AI use statement}
We used generative AI tools to edit the paper for readability, and to create and edit code and figures. We have reviewed all AI-assisted work, and we take responsibility for the final content of this work, including text, claims, and artifacts produced with the aid of generative AI.

\bibliographystyle{iclr2027_conference}
\bibliography{bibliography}

@article{R28,
  title={Scideator: Human-llm scientific idea generation grounded in research-paper facet recombination},
  author={Radensky, Marissa and Shahid, Simra and Fok, Raymond and Siangliulue, Pao and Hope, Tom and Weld, Daniel S},
  journal={arXiv preprint arXiv:2409.14634},
  year={2024}
}

@inproceedings{R29,
  author    = {Sternlicht, Noy and Hope, Tom},
  title     = {{CHIMERA}: A Knowledge Base of Scientific Idea Recombinations for Research Analysis and Ideation},
  booktitle = {Proceedings of the 64th Annual Meeting of the Association for Computational Linguistics (Volume 1: Long Papers)},
  pages     = {1871--1905},
  year      = {2026}
}

@inproceedings{R15,
  title={{SPECTER}: Document-level representation learning using citation-informed transformers},
  author={Cohan, Arman and Feldman, Sergey and Beltagy, Iz and Downey, Doug and Weld, Daniel S},
  booktitle={Proceedings of the Annual Meeting of the Association for Computational Linguistics (ACL)},
  pages={2270--2282},
  year={2020}
}

@inproceedings{R16,
  title={Neighborhood contrastive learning for scientific document representations with citation embeddings},
  author={Ostendorff, Malte and Rethmeier, Nils and Augenstein, Isabelle and Gipp, Bela and Rehm, Georg},
  booktitle={Proceedings of the 2022 Conference on Empirical Methods in Natural Language Processing},
  pages={11670--11688},
  year={2022}
}

@inproceedings{R17,
  title={{SciRepEval}: A Multi-Format Benchmark for Scientific Document Representations},
  author={Singh, Amanpreet and D'Arcy, Mike and Cohan, Arman and Downey, Doug and Feldman, Sergey},
  booktitle={Proceedings of the 2023 Conference on Empirical Methods in Natural Language Processing},
  pages={5548--5566},
  year={2023}
}

@inproceedings{R3,
  title={Context-aware citation recommendation},
  author={He, Qi and Pei, Jian and Kifer, Daniel and Mitra, Prasenjit and Giles, Lee},
  booktitle={Proceedings of the 19th international conference on World wide web},
  pages={421--430},
  year={2010}
}

@inproceedings{R4,
  title={Neural citation network for context-aware citation recommendation},
  author={Ebesu, Travis and Fang, Yi},
  booktitle={Proceedings of the 40th international ACM SIGIR conference on research and development in information retrieval},
  pages={1093--1096},
  year={2017}
}

@inproceedings{R19,
  title={Structural scaffolds for citation intent classification in scientific publications},
  author={Cohan, Arman and Ammar, Waleed and Van Zuylen, Madeleine and Cady, Field},
  booktitle={Proceedings of the Conference of the North American Chapter of the Association for Computational Linguistics (NAACL)},
  pages={3586--3596},
  year={2019}
}

@article{R26,
 title={Measuring the evolution of a scientific field through citation frames},
  author={Jurgens, David and Kumar, Srijan and Hoover, Raine and McFarland, Dan and Jurafsky, Dan},
  journal={Transactions of the Association for Computational Linguistics},
  volume={6},
  pages={391--406},
  year={2018}
}

@inproceedings{R2,
  title={In-depth research impact summarization through fine-grained temporal citation analysis},
  author={Arnaout, Hiba and Sternlicht, Noy and Hope, Tom and Gurevych, Iryna},
  booktitle={Proceedings of the 64th Annual Meeting of the Association for Computational Linguistics (Volume 1: Long Papers)},
  pages={6749--6789},
  year={2026}
}

@inproceedings{discoverybench,
  title     = {{DiscoveryBench}: Towards Data-Driven Discovery with Large Language Models},
  author    = {Majumder, Bodhisattwa Prasad and Surana, Harshit and Agarwal, Dhruv and Mishra, Bhavana Dalvi and Meena, Abhijeetsingh and Prakhar, Aryan and Vora, Tirth and Khot, Tushar and Sabharwal, Ashish and Clark, Peter},
  booktitle = {The Thirteenth International Conference on Learning Representations},
  year      = {2025}
}

@inproceedings{astabench,
  title     = {{AstaBench}: Rigorous Benchmarking of {AI} Agents with a Scientific Research Suite},
  author    = {Bragg, Jonathan and D'Arcy, Mike and Balepur, Nishant and Bareket, Dan and Dalvi, Bhavana and Feldman, Sergey and Haddad, Dany and Hwang, Jena D. and Jansen, Peter and Kishore, Varsha and others},
  booktitle = {The Fourteenth International Conference on Learning Representations},
  year      = {2026}
}

@article{swanson1986,
  title={Undiscovered public knowledge},
  author={Swanson, Don R},
  journal={The Library Quarterly},
  volume={56},
  number={2},
  pages={103--118},
  year={1986},
  publisher={University of Chicago Press}
}

@article{smalheiser2012,
  title={Literature-based discovery: Beyond the ABCs},
  author={Smalheiser, Neil R},
  journal={Journal of the American Society for Information Science and Technology},
  volume={63},
  number={2},
  pages={218--224},
  year={2012},
  publisher={Wiley Online Library}
}

@inproceedings{si2024,
  title={Can {LLMs} Generate Novel Research Ideas? {A} Large-Scale Human Study with 100+ {NLP} Researchers},
  author={Si, Chenglei and Yang, Diyi and Hashimoto, Tatsunori},
  booktitle={The Thirteenth International Conference on Learning Representations},
  year={2025}
}

@inproceedings{wang2024scimon,
  author    = {Wang, Qingyun and Downey, Doug and Ji, Heng and Hope, Tom},
  title     = {{SciMON}: Scientific Inspiration Machines Optimized for Novelty},
  booktitle = {ACL},
  year      = {2024}
}

@inproceedings{lo2020s2orc,
  title={S2ORC: The semantic scholar open research corpus},
  author={Lo, Kyle and Wang, Lucy Lu and Neumann, Mark and Kinney, Rodney and Weld, Daniel S},
  booktitle={Proceedings of the 58th annual meeting of the association for computational linguistics},
  pages={4969--4983},
  year={2020}
}

@inproceedings{zheng2026citerag,
  author    = {Zheng, Leqi and Zhang, Jiajun and Chen, Canzhi and Wang, Chaokun and Li, Hongwei and Li, Yuying and Mao, Yaoxin and Yan, Shannan and Song, Zixin and Feng, Zhiyuan and Kang, Zhaolu and Chen, Zirong and Zhang, Hang and Liu, Qiang and Wang, Liang and Liu, Ziyang},
  title     = {What Should I Cite? A {RAG} Benchmark for Academic Citation Prediction},
  booktitle = {Proceedings of the ACM Web Conference 2026},
  year      = {2026},
  doi       = {10.1145/3774904.3792075},
  eprint    = {2601.14949},
  archivePrefix = {arXiv},
  url       = {https://arxiv.org/abs/2601.14949}
}

@article{zhang2026cbert4rec,
  author  = {Zhang, Yang and Ma, Qiang},
  title   = {{On-the-fly} Citation Recommendation Based on Content-dependent Embeddings},
  journal = {Knowledge-Based Systems},
  volume  = {343},
  pages   = {116052},
  year    = {2026},
  doi     = {10.1016/j.knosys.2026.116052},
  url     = {https://doi.org/10.1016/j.knosys.2026.116052}
}

@inproceedings{ajith2024litsearch,
  author    = {Ajith, Anirudh and Xia, Mengzhou and Chevalier, Alexis and Goyal, Tanya and Chen, Danqi and Gao, Tianyu},
  title     = {LitSearch: A Retrieval Benchmark for Scientific Literature Search},
  booktitle = {Proceedings of the 2024 Conference on Empirical Methods in Natural Language Processing},
  address   = {Miami, Florida, USA},
  pages     = {15068--15083},
  publisher = {Association for Computational Linguistics},
  year      = {2024},
  doi       = {10.18653/v1/2024.emnlp-main.840},
  url       = {https://aclanthology.org/2024.emnlp-main.840/}
}

@inproceedings{sugiyama2010recentinterests,
  author    = {Sugiyama, Kazunari and Kan, Min-Yen},
  title     = {Scholarly Paper Recommendation via User's Recent Research Interests},
  booktitle = {Proceedings of the 10th Annual Joint Conference on Digital Libraries},
  pages     = {29--38},
  year      = {2010},
  doi       = {10.1145/1816123.1816129},
  url       = {https://doi.org/10.1145/1816123.1816129}
}

@article{jiang2023taprec,
  author  = {Jiang, Chi and Ma, Xiao and Zeng, Jiangfeng and Zhang, Yin and Yang, Tingting and Deng, Qiumiao},
  title   = {{TAPRec}: Time-aware Paper Recommendation via the Modeling of Researchers' Dynamic Preferences},
  journal = {Scientometrics},
  volume  = {128},
  number  = {6},
  pages   = {3453--3471},
  year    = {2023},
  doi     = {10.1007/s11192-023-04731-4},
  url     = {https://doi.org/10.1007/s11192-023-04731-4}
}

@inproceedings{flicke2025scholarinbox,
  author    = {Flicke, Markus and Angrabeit, Glenn and Iyengar, Madhav and Protsenko, Vitalii and Shakun, Illia and Cicvaric, Jovan and Kargi, Bora and He, Haoyu and Schuler, Lukas and Scholz, Lewin and Agnihotri, Kavyanjali and Cao, Yong and Geiger, Andreas},
  title     = {Scholar Inbox: Personalized Paper Recommendations for Scientists},
  booktitle = {Proceedings of the 63rd Annual Meeting of the Association for Computational Linguistics (Volume 3: System Demonstrations)},
  address   = {Vienna, Austria},
  pages     = {307--317},
  publisher = {Association for Computational Linguistics},
  year      = {2025},
  doi       = {10.18653/v1/2025.acl-demo.30},
  url       = {https://aclanthology.org/2025.acl-demo.30/}
}

@inproceedings{liu2026researchbench,
  author    = {Liu, Yujie and Yang, Zonglin and Xie, Tong and Ni, Jinjie and Gao, Ben and Li, Yuqiang and Tang, Shixiang and Ouyang, Wanli and Cambria, Erik and Zhou, Dongzhan},
  title     = {ResearchBench: Benchmarking {LLMs} in Scientific Discovery via Inspiration-Based Task Decomposition},
  booktitle = {Findings of the Association for Computational Linguistics: ACL 2026},
  pages     = {13187--13207},
  year      = {2026},
  url       = {https://aclanthology.org/2026.findings-acl.644/}
}

@article{chamoun2026scipaths,
  author  = {Chamoun, Eric and Chi, Yizhou and Chen, Yulong and Cao, Rui and Ding, Zifeng and Korakakis, Michalis and Vlachos, Andreas},
  title   = {{SciPaths}: Forecasting Pathways to Scientific Discovery},
  journal = {arXiv preprint arXiv:2605.14600},
  year    = {2026}
}

@inproceedings{vodicka2026citationfunction,
  author    = {Vodi{\v{c}}ka, Daniel and Kral, Pavel and Cerisara, Christophe and {\v{S}}m{\'{i}}d, Jakub},
  title     = {Large Language Models for Citation Function Classification},
  booktitle = {Proceedings of the Fifteenth Language Resources and Evaluation Conference},
  address   = {Palma de Mallorca, Spain},
  pages     = {2430--2439},
  publisher = {ELRA Language Resource Association},
  year      = {2026},
  doi       = {10.63317/4sb25z5kxz3q},
  url       = {https://aclanthology.org/2026.lrec-1.191/}
}

@article{wakeling2025citationperception,
  author  = {Wakeling, Simon and Paramita, Monica Lestari and Pinfield, Stephen},
  title   = {How Do Authors Perceive the Way Their Work Is Cited? Findings from a Large-scale Survey on Quotation Accuracy},
  journal = {Journal of the Association for Information Science and Technology},
  volume  = {76},
  number  = {10},
  pages   = {1396--1410},
  year    = {2025},
  doi     = {10.1002/asi.70000},
  url     = {https://doi.org/10.1002/asi.70000}
}

@inproceedings{valenzuela2015meaningful,
  author    = {Valenzuela, Marco and Ha, Vu and Etzioni, Oren},
  title     = {Identifying Meaningful Citations},
  booktitle = {AAAI Workshop: Scholarly Big Data},
  year      = {2015}
}

@inproceedings{muennighoff2025gritlm,
  author    = {Muennighoff, Niklas and Su, Hongjin and Wang, Liang and Yang, Nan and Wei, Furu and Yu, Tao and Singh, Amanpreet and Kiela, Douwe},
  title     = {Generative Representational Instruction Tuning},
  booktitle = {The Thirteenth International Conference on Learning Representations},
  year      = {2025}
}

@article{zhang2025qwen3embedding,
  author  = {Zhang, Yanzhao and Li, Mingxin and Long, Dingkun and Zhang, Xin and Lin, Huan and Yang, Baosong and Xie, Pengjun and Yang, An and Liu, Dayiheng and Lin, Junyang and Huang, Fei and Zhou, Jingren},
  title   = {{Qwen3 Embedding}: Advancing Text Embedding and Reranking Through Foundation Models},
  journal = {arXiv preprint arXiv:2506.05176},
  year    = {2025}
}

@article{brown2019relish,
  author  = {Brown, Peter and {RELISH Consortium} and Zhou, Yaoqi},
  title   = {Large Expert-Curated Database for Benchmarking Document Similarity Detection in Biomedical Literature Search},
  journal = {Database},
  volume  = {2019},
  pages   = {baz085},
  year    = {2019}
}

@misc{sourty2026denseonlateon,
  title = {{DenseOn} with the {LateOn}: Open State-of-the-Art
           Single and Multi-Vector Models},
  author = {Sourty, Raphael and Chaffin, Antoine and Weller, Orion
            and Moura Junior, Paulo Roberto and Chatelain, Amelie},
  year = {2026},
  howpublished = {\url{https://huggingface.co/blog/lightonai/denseon-lateon}}
}

@inproceedings{chaffin2025pylate,
  author = {Antoine Chaffin and Rapha{\"e}l Sourty},
  title = {{PyLate}: Flexible Training and Retrieval
           for Late Interaction Models},
  booktitle = {Proceedings of the 34th ACM International Conference
               on Information and Knowledge Management},
  pages = {6334--6339},
  year = {2025},
  doi = {10.1145/3746252.3761608}
}

@inproceedings{haveliwala2002topic,
  author = {Haveliwala, Taher H.},
  title = {Topic-Sensitive {PageRank}},
  booktitle = {Proceedings of the 11th International Conference
               on World Wide Web},
  pages = {517--526},
  year = {2002},
  doi = {10.1145/511446.511513}
}

@article{small1973cocitation,
  author = {Small, Henry},
  title = {Co-citation in the Scientific Literature:
           A New Measure of the Relationship Between Two Documents},
  journal = {Journal of the American Society for Information Science},
  volume = {24},
  number = {4},
  pages = {265--269},
  year = {1973},
  doi = {10.1002/asi.4630240406}
}

@article{kessler1963bibliographic,
  author = {Kessler, M. M.},
  title = {Bibliographic Coupling Between Scientific Papers},
  journal = {American Documentation},
  volume = {14},
  number = {1},
  pages = {10--25},
  year = {1963},
  doi = {10.1002/asi.5090140103}
}

@article{adamic2003friends,
  author = {Adamic, Lada A. and Adar, Eytan},
  title = {Friends and Neighbors on the Web},
  journal = {Social Networks},
  volume = {25},
  number = {3},
  pages = {211--230},
  year = {2003},
  doi = {10.1016/S0378-8733(03)00009-1}
}

\clearpage
\appendix
\section*{Appendix}
\label{app:guide}

\etocdepthtag.toc{appendices}

\begingroup
\small
\setcounter{tocdepth}{2}
\etocsettagdepth{main}{none}
\etocsettagdepth{appendices}{subsection}
\etocsettocstyle{}{}
\etocstandardlines

\makeatletter
\renewcommand*\l@section{\@dottedtocline{1}{0em}{2em}}
\makeatother

\tableofcontents
\endgroup

\clearpage

\section{Benchmark construction and temporal eligibility}
\label{app:benchmark-construction}
\label{sec:appendix-s2orc-pipeline}

This section defines a benchmark case, its temporal cutoff,
and the cases included in evaluation. Figure~4 summarizes
benchmark construction, author reports, and citation-context labels.

\begin{figure*}[t]
  \centering
  \includegraphics[width=\textwidth]{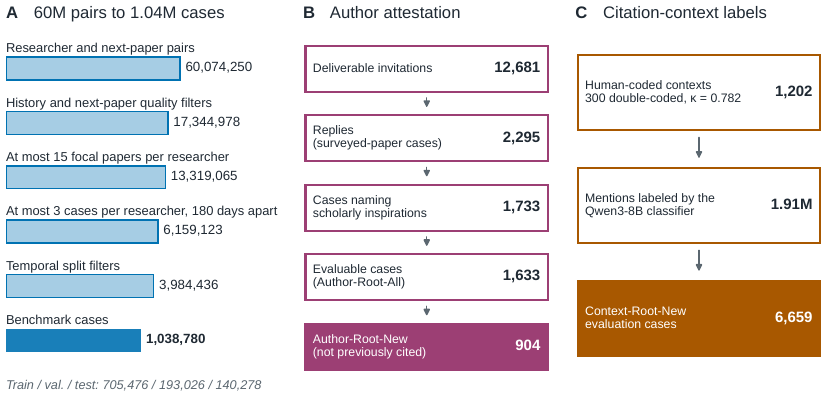}
\caption{\textbf{How MUSES was built.}
(A) Benchmark construction from S2ORC; each stage is nested
in the previous one, and bar length is log-scaled (Table~3).
(B) Author attestation: invitations carried a shortlist of
up to eight references from the next paper's bibliography,
and authors could also name unlisted works (Appendix~D.1).
Counts refer to surveyed-paper cases; one case can yield
several reported works.
(C) Human-coded citation contexts ground the classifier
that labels citation mentions (Appendix~B.3);
these stages count contexts, mentions and evaluation cases.}
  \label{fig:data-construction}
\end{figure*}

\subsection{Case construction and eligibility}
\label{app:corpus-task}

\paragraph{Case definition.}
The benchmark uses paper, authorship, reference,
publication-date, title, and abstract records from a fixed,
processed corpus derived from the Semantic Scholar Open
Research Corpus (S2ORC), release 2026-03-10. An evaluation
case is indexed by a researcher $a$ and a focal paper $P$
on which $a$ is last author. For each researcher, last-author
papers are ordered by recorded publication date, with corpus
identifier used to break ties. The next paper $P^+$ is the
immediately subsequent last-author paper in this ordering.
If $P^+$ fails the eligibility criteria below, no later paper
is substituted. The researcher's retrieval history contains
distinct papers authored in any position and published strictly
before $P$'s recorded publication date; $P$ and papers sharing
its date are excluded.

Formally, we represent each case as $(a, P, P^+)$, where $P^+$
is determined by the selection rule above. The publication
date $t = \operatorname{pubdate}(P)$ defines the information
cutoff.

\paragraph{Eligibility.}
Construction first requires usable publication dates and bibliographic links.
A retained case requires:
\begin{itemize}
    \item at least three preceding authorship records for the researcher;
    \item an immediately subsequent last-author paper $P^+$ within five years
    of $P$;
    \item at least ten references in $P^+$; and
    \item $P^+$ to be a research article rather than a review, meta-analysis,
    news article, editorial, letter/comment, dataset, or book record.
\end{itemize}

\paragraph{Sampling and splits.}
Each author's career midpoint is the calendar midpoint between
their earliest and latest dated publications across all
authorship roles. Authors are deterministically assigned to
author-disjoint temporal splits by this midpoint: training
before 2018, validation in 2018--2020, and test in
2021--2023; retained focal papers must also fall within the
corresponding split's date window. To limit repeated cases
from prolific researchers while retaining career coverage,
eligible focal papers are deterministically spread across at
most 15 positions per author; from these, at most three cases
are retained, with focal-paper dates at least 180 days apart.

After temporal split filtering, a case is retained only if
its next paper $P^+$ has the required metadata---title,
abstract, author list, and reference metadata---and the case
has at least one CiteNext label, that is, a reference in
$P^+$ published before $P$. The resulting benchmark contains
1,038,780 cases: 705,476 train, 193,026 validation, and
140,278 test cases.

The eligibility and sampling thresholds were fixed before retrieval
evaluation and were not tuned to performance. They balance benchmark
size, coverage of researchers' careers, and metadata availability.

\begin{table}[t]
\centering
\small
\caption{\textbf{Cases retained at each construction stage.}
Each stage filters the cases retained at the previous stage.
The final two filters require complete next-paper metadata
and at least one CiteNext target, respectively.}
\label{tab:app-release-attrition}
\begin{tabular}{@{}lrr@{}}
\toprule
Construction stage & Remaining & Removed \\
\midrule
Role-consistent focal / next-paper candidates
  & 60{,}074{,}250 & --- \\
After pre-history and next-paper quality filters
  & 17{,}344{,}978 & 42{,}729{,}272 \\
After career-spread cap of 15 focal papers per author
  & 13{,}319{,}065 & 4{,}025{,}913 \\
After at most 3 cases per author, 180 days apart
  & 6{,}159{,}123 & 7{,}159{,}942 \\
After author-split and focal-date filters
  & 3{,}984{,}436 & 2{,}174{,}687 \\
After next-paper metadata-readiness gate
  & 1{,}057{,}703 & 2{,}926{,}733 \\
After requiring at least one CiteNext label
  & 1{,}038{,}780 & 18{,}923 \\
\midrule
Train / validation / test
  & \multicolumn{2}{r}{705{,}476 / 193{,}026 / 140{,}278} \\
\bottomrule
\end{tabular}
\end{table}

\subsection{Candidate pool and observable information}
\label{app:temporal-eligibility}

\paragraph{Candidate pool.}
Retrieval uses a fixed shared pool $G$ of 2,330,779 papers:
2,230,779 distinct works cited by benchmark next papers
across splits that have title and abstract text, plus
100,000 additional papers with title and abstract sampled
with year stratification (seed 42).

For focal paper $P$ with publication date $t$, the eligible
candidate set is
\[
C(t) = \{d \in G : \operatorname{pubdate}(d) < t\}.
\]
Every ranked candidate therefore belongs to the shared pool
and is dated strictly before $P$. Full-text availability
is not required.

\paragraph{Information available to the retriever.}
Retrievers may use the researcher's papers published
strictly before $P$ and the eligible candidates in
$C(t)$, including their text when the method uses text.
The next paper $P^+$ supplies the evaluation outcome:
its bibliography and citation contexts define targets
but are never retrieval inputs. Its title and abstract
enter only the hindsight diagnostic (Appendix F.2).

\noindent\textbf{Citation records.}
Citation-based methods may additionally use S2ORC citation links
whose citing papers are dated strictly before $P$, excluding
links incident to $P$.
Graph computations may involve papers outside the candidate pool,
but ranked outputs must still belong to $C(t)$.
Appendix~C gives the graph construction and scoring details.

\subsection{Evaluation populations}
\label{app:evaluation-populations}

Retrieval performance is evaluated on cases with at least
one relevant target in the eligible candidate set $C(t)$
and a nonempty researcher publication history before $P$.
Only targets in $C(t)$ are considered relevant for ranking
evaluation. Within each target tier, all retrieval methods
are evaluated on the same cases. The corresponding
evaluation populations are reported alongside the results,
and Appendix~E defines the evaluation metrics.

\section{Target definitions and evidence construction}
\label{app:root-labels}
\suppressfloats[t]


This section defines the prior-citation target tiers and the citation-context
evidence used to construct the Context-Root tiers.

\subsection{Bibliography tiers}

Let $Y_{\mathrm{next}}(P)$ be the set of distinct papers
published before $P$ and cited by its immediately
subsequent last-author paper $P^+$, provided that
$P^+$ passes the case-retention filters. The three tiers are defined as:
\[
\begin{aligned}
Y_{\mathrm{CiteNext}}(P)
    &= Y_{\mathrm{next}}(P),\\
Y_{\mathrm{CiteNew}}(P)
    &= Y_{\mathrm{next}}(P) \setminus B_a(t),\\
Y_{\mathrm{CiteNew\text{-}Isolated}}(P)
    &= Y_{\mathrm{CiteNew}}(P) \setminus S_a(t).
\end{aligned}
\]

Here, $t$ is the publication date of $P$, and $B_a(t)$
contains works cited in the researcher's recorded papers
published before $t$. Similarly, $S_a(t)$ contains works
cited before $t$ by researchers who coauthored with $a$
during the preceding five years. This window identifies
recent coauthors, who need not appear on $P$; their
earlier citation histories are not restricted to that
period. Coauthors who first collaborate with the
researcher on $P^+$ are excluded. These tiers reflect
recorded citation histories rather than establishing
whether a researcher previously encountered or knew
a work.

\subsection{Context-Root labels}

Context-Root records whether a citation context describes the cited
work as serving a generative role in the citing paper. It does not
establish causal influence or reveal the author's private intent.

The six citation roles distinguish three generative uses from
three other uses (Table~4). The classifier predicts whether a
context expresses a generative use, rather than assigning one
of the six individual roles.

A cited work is a Context-Root target of $P^+$ when at
least one of its citation mentions in $P^+$ is labeled
generative. Context-Root-Next, Context-Root-New, and
Context-Root-Isolated retain these targets within
CiteNext, CiteNew, and CiteNew-Isolated, respectively.
A cited work with no recoverable citation context
remains in the bibliography tiers but receives no
Context-Root label; it is not treated as non-generative.

\begin{table}[!htbp]
\centering
\caption{Six citation-context roles. The three
generative-use categories form the Context-Root label;
the other three are not generative.}
\label{tab:context_root_roles}
\small
\begin{tabular}{@{}p{0.30\linewidth}p{0.64\linewidth}@{}}
\toprule
Category & Definition \\
\midrule
\multicolumn{2}{@{}l}{%
\emph{Generative roles (Context-Root)}}\\
\addlinespace[2pt]
Theoretical foundation (TF) &
The citing paper explicitly operates within or
operationalizes the cited theory, framework, or formalism.\\
\addlinespace
Method extension (ME) &
The citing paper explicitly modifies, extends,
generalizes, re-derives, or structurally repurposes
the cited method.\\
\addlinespace
Generative motivation (GM) &
A gap, limitation, finding, suggestion, assumption,
or inspiration in the cited work motivates the
present paper.\\
\midrule
\multicolumn{2}{@{}l}{%
\emph{Other citation roles}}\\
\addlinespace[2pt]
Contrast or comparison (CC) &
The citation supports rhetorical contrast or
benchmarking without generative uptake.\\
\addlinespace
Tool or resource (TR) &
The cited work is used instrumentally as software,
data, a benchmark, or a method applied as-is.\\
\addlinespace
Background context (BC) &
The citation supplies contextual attribution or
evidence without an explicit generative or
instrumental role.\\
\bottomrule
\end{tabular}
\end{table}

\subsection{Human annotation and classifier validation}

\paragraph{Human-labeled development data.}
The development data contain 1,202 citation contexts with human
labels, drawn from a larger stratified sample of 1,900 contexts
that oversampled generative citations. This is not a random
sample of all citation contexts. Annotators used the six roles
in Table~4; TF, ME, and GM are grouped as generative, and
CC, TR, and BC as non-generative.

\paragraph{Annotation agreement.}
Two annotators independently labeled the same 300 citation
contexts without seeing classifier outputs. Cohen's $\kappa$
was 0.782 for the three broad categories (generative,
non-generative, and uncertain) and 0.715 for the individual
citation roles, with uncertain responses retained as a
separate category.

\paragraph{Classifier.}
A Qwen3-8B classifier with a LoRA adapter assigns generative
or non-generative labels using only the citation-context text,
without access to retrieval results. It was trained on 104,976
labels generated by GPT-5.4-mini: 94,479 training examples
and 10,497 validation examples. The 1,202 contexts labeled
by humans were not used to train the classifier.
Inputs are truncated to 512 tokens, and a context is labeled
generative when its predicted probability exceeds 0.5.

\paragraph{Validation and limitations.}
On the human-labeled development data, the Qwen3-8B classifier
achieved precision of 0.6345, recall of 0.9321, and F1 of 0.7550
for generative citations. These data were also used for model
selection. Context-Root targets
therefore reflect classifier predictions rather than individual
human verification.
\FloatBarrier
\section{Retrieval systems and scoring}
\label{app:method-registry}

This section describes the retrieval systems' inputs and scoring,
including training and model selection for the supervised baselines.

\subsection{Retrieval inputs and scoring}
\label{app:method-inputs}

\noindent\textbf{Compact query.}
The query takes up to 20 of the researcher's most recent papers with usable
text, in any author position, published strictly before $P$ (excluding $P$).
It joins their titles, most recent first, followed by their abstracts in the
same order, normalizes whitespace, and truncates the result to 128
whitespace-delimited words. Candidate text is the title followed by the first
1,500 characters of the abstract. BM25, SPLADE, and the compact-query
Qwen3-Embedding-8B baseline use this same source query, followed by
model-specific preprocessing. The word limit applies to the concatenated
history, so the selected papers need not contribute their complete text
to the effective query.

\noindent\textbf{Uncapped BM25.}
This variant retains the same latest-20, title-then-abstract query
without the 128-word limit. Candidate text, index, temporal masking
and BM25 parameters ($k_1=1.2$, $b=0.75$) are unchanged.

\noindent\textbf{One-centroid history.}
The one-centroid variants average the vectors of the researcher's usable
papers published strictly before $P$ into one L2-normalized vector and
rank candidates by cosine similarity. Linear-Proximity applies the same
averaging after its learned history map.

\noindent\textbf{Multi-centroid history.}
The multi-centroid variants summarize the usable pre-$P$ history vectors
with up to 16 spherical centroids, capped at the number of available
vectors. Candidates are scored by their maximum cosine similarity to
a history centroid. The maximum of 16 centroids is fixed rather than
tuned on the reported evaluation. The sensitivity analysis in Appendix~H finds no clear additional aggregate Hit@100
improvement from increasing this maximum to 24 or 64 for
SPECTER2-Proximity across the three bibliography tiers.

\noindent\textbf{Late interaction.}
We evaluate frozen LateOn~\citep{sourty2026denseonlateon}
using PyLate~\citep{chaffin2025pylate} (version 1.3.4).
Each distinct usable pre-$P$ history paper supplies a
whitespace-normalized title query, right-truncated to 29 text
tokens (32 including special tokens). Candidate text is the title
followed by the first 1,500 abstract characters, truncated to
297 text tokens.
For each query--candidate pair, LateOn sums each query token's
maximum similarity to a candidate token. The final candidate score
is the maximum pair score across history titles
(\emph{Title max} in the tables).
We score the full candidate pool exactly and retain the top
1,000 eligible candidates, breaking ties by candidate identifier.
No training on MUSES is performed.

\noindent\textbf{Citation-graph retrieval.}
We evaluate four training-free methods on a background graph of
103.2M papers and 1.31B citation links, retaining only links
admissible under Appendix~A.2 and removing paper self-loops.
Distinct pre-$P$ history works receive equal weight, divided
among their mapped graph records; works without admissible links
are omitted. For imprecise Author-Root dates, the citing paper's
latest possible publication date must precede $P$'s earliest
possible date.

Personalized PageRank~\citep{haveliwala2002topic} ranks papers by
stationary visit probability: it follows a citation link in either
direction with probability $0.9$ and restarts from the weighted
history with probability $0.1$. Reciprocal citations remain
separate links.
Co-citation~\citep{small1973cocitation} counts shared citing papers;
bibliographic coupling~\citep{kessler1963bibliographic} counts
shared references; and Adamic--Adar~\citep{adamic2003friends}
sums $1/\ln\deg(z)$ over shared neighbours, using distinct-neighbour
degree in the case-specific undirected graph.
These three methods aggregate history-weighted pair scores,
excluding self-pairings.
Each graph method returns at most 1,000 eligible papers with
positive scores, ranked by score.

Table~\ref{tab:app-method-registry} summarizes the retrieval systems
and their scoring rules. All systems apply the case-specific
temporal candidate mask.

\begin{table*}[!t]
\centering
\small
\setlength{\tabcolsep}{4pt}
\renewcommand{\arraystretch}{1.08}
\caption{\textbf{Retrieval systems and test-time scoring.}
Rows group configurations that share a retriever and differ only
in history aggregation.
Compact, K1, K16 and Title max follow Appendix~C.1.
Individual-paper MaxSim, matched frozen controls and
supervised training are described in Appendix~C.2.
Configuration coverage by evidence source is reported in Appendix~E.
SPECTER2-Proximity K16 is the reference retriever.}
\label{tab:app-method-registry}

\begin{tabular}{@{}
>{\raggedright\arraybackslash}p{0.30\textwidth}
>{\raggedright\arraybackslash}p{0.28\textwidth}
>{\raggedright\arraybackslash}p{0.36\textwidth}
@{}}
\toprule
Retriever & Input / representation & Scoring \\
\midrule

\multicolumn{3}{@{}l}
{\emph{Popularity, lexical and sparse retrieval}} \\
\addlinespace[2pt]

Popularity &
Pre-cutoff citation counts; no history. &
Rank by citation count. \\

BM25 &
Compact query. &
Lexical BM25 score. \\

BM25 (uncapped) &
Uncapped latest-20 query. &
Lexical BM25 score. \\

SPLADE &
Compact query; sparse vectors. &
Learned sparse matching. \\

\midrule
\multicolumn{3}{@{}l}{\emph{Frozen dense retrievers}} \\
\addlinespace[2pt]

Qwen3-Embedding-8B &
Compact query or K16 history centroids. &
Query--candidate similarity, or maximum centroid cosine. \\

SPECTER2-Proximity &
K1 or K16. &
Cosine to the history mean, or maximum centroid cosine. \\

GritLM-7B &
K1 or K16. &
Cosine to the history mean, or maximum centroid cosine. \\

SciNCL &
K1 or K16. &
Cosine to the history mean, or maximum centroid cosine. \\

SPECTER2-Proximity (frozen control) &
Individual-paper MaxSim; same text inputs and scoring
as the fine-tuned model. &
Maximum cosine over history papers. \\

Qwen3-Embedding-8B (frozen control) &
Individual-paper MaxSim; same text inputs and scoring as the LoRA-adapted model. &
Maximum cosine over history papers. \\

LateOn &
Title max; token vectors. &
Sum of query-token maxima, then maximum over history titles. \\

\midrule
\multicolumn{3}{@{}l}{\emph{Citation-graph retrievers}} \\
\addlinespace[2pt]

Personalized PageRank &
History papers; pre-$P$ citation links. &
Stationary visit probability; restart probability $0.1$. \\

Co-citation &
History papers; pre-$P$ citation links. &
Weighted count of shared citing papers. \\

Bibliographic coupling &
History papers; pre-$P$ citation links. &
Weighted count of shared references. \\

Adamic--Adar &
History papers; pre-$P$ citation links. &
Weighted sum of $1/\ln\deg(z)$ over shared neighbours. \\

\midrule
\multicolumn{3}{@{}l}{\emph{Supervised retrievers}} \\
\addlinespace[2pt]

Linear-Proximity &
Mapped SPECTER2-Proximity vectors; K1 or K16. &
Cosine to the mapped history mean, or maximum
mapped-centroid cosine. \\

Linear-Proximity + neighbours &
Paper and citation-neighbour vectors; learned maps; K16. &
Maximum mapped-centroid cosine. \\

Linear-Proximity (control) &
Same architecture with neighbour inputs zeroed;
separately trained; K16. &
Maximum mapped-centroid cosine. \\

SPECTER2-Proximity (fine-tuned) &
Fine-tuned paper vectors; individual-paper MaxSim. &
Maximum cosine over history papers. \\

Qwen3-Embedding-8B (LoRA) &
LoRA-adapted paper vectors; individual-paper MaxSim. &
Maximum cosine over history papers. \\

\bottomrule
\end{tabular}
\end{table*}

\FloatBarrier

\FloatBarrier

\subsection{Supervised training and scoring}
\label{app:method-linear-set}

We evaluate linear maps over frozen paper vectors, their extension
with citation-neighbour features, and encoder adaptation.
All use the following training objective, with method-specific
representations and training settings.

\noindent\textbf{Training objective.}
For focal case $f$, let $H_f$ contain its usable history papers
published strictly before $P$. The training score of candidate
$c$ is
\[
s(f,c)
=
\max_{h\in H_f}
\cos\!\left(v_Q(h),v_D(c)\right),
\]
where $v_Q$ and $v_D$ are the history and candidate representations.
For positive target $p$ and eligible negative set $N_f$,
\[
\ell(f,p)
=
-\log
\frac{\exp(s(f,p)/\tau)}
{\exp(s(f,p)/\tau)
 + \sum_{n\in N_f}\exp(s(f,n)/\tau)},
\qquad \tau=0.05.
\]
Losses are averaged over the positives used within each case
and then over cases, giving each case equal weight.
Histories are uncapped during training and evaluation.

Each case uses four fixed hard negatives, sampled uniformly
from eligible papers in a frozen SPECTER2-Proximity K16 top-100
ranking, together with eligible in-batch positives and negatives.
Known CiteNext positives, $P$, post-cutoff papers and papers
outside the candidate pool are excluded from negatives.

\noindent\textbf{Linear-Proximity.}
Linear-Proximity learns separate bias-free $768\times768$
history and candidate maps over L2-normalized, frozen
SPECTER2-Proximity vectors. Both maps start at the identity,
and mapped outputs are normalized again.
Training uses all eligible CiteNew positives.
Hard negatives come from rankings built with Semantic Scholar's
released SPECTER2-Proximity vectors; 1,214 cases require
eligible random fallback negatives.

At evaluation, K1 uses the normalized mean of the mapped
history vectors. K16 forms up to 16 spherical centroids from
those vectors and scores candidates by maximum centroid cosine.
Both use the checkpoint selected with K16 validation rankings.

\noindent\textbf{Citation-neighbour features.}
Linear-Proximity + neighbours augments each frozen paper vector
$x$ with the unnormalized mean $\bar x$ of up to 32 citation
neighbours and an empty-neighbour indicator:
\[
u(x)
=
\left[x;\,\bar x;\,\mathbf{1}\{\text{no neighbours}\}\right].
\]
For empty neighbourhoods, $\bar x=0$.
The 1,537-dimensional input is L2-normalized, projected to
768 dimensions by separate bias-free history and candidate
maps initialized as $[I\mid0\mid0]$, and normalized again.

For Bibliography and Context-Root, neighbours come from the
shared 2,330,779-paper candidate pool, using citations in either
direction from the latest year-end snapshot no later than the
case cutoff. Both the neighbour and the link's citing paper
must be dated by that year end. We exclude $P$ and $P^{+}$,
retaining the first 32 qualifying neighbours ordered by a fixed
identifier hash. The same rules apply to histories and candidates.

Linear-Proximity (control) uses the same architecture with
the neighbour mean and indicator set to zero.
It is trained separately using the same data and selection
procedure. Both neighbour models use K16 at evaluation.

\noindent\textbf{Encoder adaptation.}
We fine-tune the SPECTER2-Proximity base encoder and proximity
adapter, and adapt Qwen3-Embedding-8B with LoRA on attention
and feed-forward projections
(rank 16, $\alpha=32$, dropout 0.05).
Paper text is the title followed by up to 1,500 abstract
characters, with model-specific preprocessing and a 512-token
limit. Both encoders use individual-paper MaxSim during
training and evaluation.

Training uses CiteNext positives and hard negatives mined
with our SPECTER2-Proximity K16 encoder.
Each adapted model has a frozen control with matching text
serialization, histories, candidate pool, eligibility rules
and MaxSim scoring.

\noindent\textbf{Optimization and selection.}
All trained models use AdamW, batch size 64, weight decay 0.01,
gradient clipping at 1.0, and 6\% linear warmup followed by
linear decay.
Table~\ref{tab:app-method-linear-set} summarizes the training
recipes. Selection uses CiteNew nDCG@100 on the author-disjoint
validation split, without test or author-reported labels.
Selected checkpoints retain the learning-rate schedule of
the full training horizon.

\begin{table*}[!t]
\centering
\small
\setlength{\tabcolsep}{4pt}
\renewcommand{\arraystretch}{1.08}
\caption{\textbf{Supervised training and model selection.}
The linear families comprise Linear-Proximity, its
citation-neighbour extension and the separately trained
zero-neighbour control.
Case counts describe the available author-disjoint training sets.
Frozen encoder controls receive no training.}
\label{tab:app-method-linear-set}

\begin{tabular}{@{}
>{\raggedright\arraybackslash}p{0.18\textwidth}
>{\raggedright\arraybackslash}p{0.28\textwidth}
>{\raggedright\arraybackslash}p{0.24\textwidth}
>{\raggedright\arraybackslash}p{0.24\textwidth}
@{}}
\toprule
Component &
Linear families &
SPECTER2-Proximity fine-tuning &
Qwen3-Embedding-8B LoRA \\
\midrule

Training data &
CiteNew; 495,041 cases. &
CiteNext; 514,592 cases. &
CiteNext; 514,592 cases. \\

Positives per case &
All eligible positives. &
Up to four, resampled each update. &
Up to four, resampled each update. \\

Training horizon &
Two epochs; 7,736 updates per epoch. &
4,000 updates. &
2,000 updates. \\

Learning-rate trials &
$\{3\times10^{-5},\,10^{-4},\,3\times10^{-4}\}$. &
$\{2.5\times10^{-6},\,5\times10^{-6},\,10^{-5}\}$;
additional trial $2\times10^{-5}$. &
$\{2.5\times10^{-5},\,5\times10^{-5},\,10^{-4}\}$. \\

Checkpoints evaluated &
End of epochs 1 and 2. &
Updates 2,000 and 4,000. &
Updates 1,000 and 2,000. \\

Validation scoring &
K16. &
Individual-paper MaxSim. &
Individual-paper MaxSim. \\

Selected checkpoint &
$3\times10^{-5}$ at epoch 1 for each model. &
$10^{-5}$ at update 2,000. &
$5\times10^{-5}$ at update 2,000. \\

Evaluation scoring &
Linear-Proximity: K1 and K16.
Neighbour models: K16 only. &
Individual-paper MaxSim. &
Individual-paper MaxSim. \\

\bottomrule
\end{tabular}
\end{table*}
\FloatBarrier
\FloatBarrier
\section{Author reports and target construction}
\label{app:author-attestation}
\label{sec:appendix-attested}

This section describes how author reports were collected
and used to construct retrieval targets.

\subsection{Collection and target construction}

We asked corresponding authors which prior works, if any,
most directly inspired their paper's central hypothesis
or methodological approach.
Invitations included up to eight references from the
paper's bibliography, while allowing authors to name
other works.
Authors could select listed references, indicate that
none were primary inspirations, decline, or express
uncertainty. Outreach included up to two reminders
to nonrespondents.
Participation was voluntary, and the shortlist may
have favored listed works.
Table~\ref{tab:author-collection-summary} summarizes
the collection and evaluation counts.

\paragraph{Panel composition and representativeness.}
Author-Root-All contains 1,633 cases from 1,600 responding
researchers across 57 journals.
Table~\ref{tab:author-root-field-composition} compares
its field composition with a uniform sample of
20,000 papers from the MUSES corpus.
Both populations use the same mapping from the
OpenAlex primary topic of the surveyed paper $P^+$.
Life sciences account for 55.3\% of classified
Author-Root cases, compared with 19.0\% in the corpus
sample. All five field groups are represented, but
the other groups have smaller shares than in MUSES.
This comparison describes field coverage;
voluntary participation and nonresponse limit
generalization to the broader MUSES population.

\begin{table}[!htbp]
\centering
\small
\setlength{\tabcolsep}{5pt}
\renewcommand{\arraystretch}{1.08}
\caption{\textbf{Field composition of Author-Root and MUSES.}
MUSES percentages describe a uniform 20,000-paper corpus
sample. Author-Root counts describe the All panel.
Percentages exclude papers without a field assignment:
3 Author-Root cases; classification coverage in the
MUSES sample is 98.8\%.
Each paper receives one field assignment.}
\label{tab:author-root-field-composition}

\begin{tabular}{@{}lrrr@{}}
\toprule
& \multicolumn{2}{c}{Author-Root}
& MUSES \\
\cmidrule(lr){2-3}
Field & Cases & Share (\%) & Share (\%) \\
\midrule
Life sciences
& 901 & 55.3 & 19.0 \\
Health and medicine
& 172 & 10.6 & 24.4 \\
Physical sciences
& 337 & 20.7 & 29.5 \\
Social sciences and humanities
& 155 & 9.5 & 16.2 \\
Computer science and mathematics
& 65 & 4.0 & 10.9 \\
\midrule
Unclassified & 3 & --- & --- \\
Total & 1,633 & --- & --- \\
\bottomrule
\end{tabular}
\end{table}

\paragraph{Researcher history.}
The surveyed paper is $P^+$.
For each responding researcher, $P$ is their latest
identified research work published strictly before
$P^+$.
Works tied at the latest date share the cutoff.
Cases without a resolvable predecessor are excluded.
The query uses only the researcher's papers published
strictly before $P$, across any authorship position.
Neither $P$, $P^+$, their bibliographies, nor the author's
response is a query input.
Unlike the bibliography benchmark, which follows
last authors, these cases follow the responding
researcher.

\paragraph{Response resolution.}
Reported works were identified from authors' responses
using human review and LLM-assisted interpretation,
then matched to bibliographic records.
Shortlist selections were linked to the surveyed
paper's references; named works were resolved through
DOI and title matching.
Unresolved reports and non-scholarly inspirations were
excluded from retrieval targets.
A missing response or an unselected shortlist item
was treated as unknown, not as evidence that a work
had no influence.

\paragraph{Candidate pool and evaluation eligibility.}
Retrieval uses a fixed pool of 2,603,132 works,
including the bibliography benchmark's pool,
identified reported works, and a publication-year-matched
background sample constructed for the original collection.
A reported work is eligible only if it can be identified
in the pool, has a matching title and an abstract of
at least 100 characters, and was publicly available
strictly before $P$.
For imprecise dates, its latest possible publication
date must precede the earliest possible date of $P$;
same-day or overlapping dates are excluded.
Any externally supplied text must come from a version
available before $P$.

Evaluation includes cases with at least one eligible
target and a pre-$P$ publication history containing
usable title or abstract text.
Unsupported targets and cases without usable history
are excluded rather than counted as retrieval misses.
The author and bibliography evaluations use different
candidate pools, which limits direct comparisons
between their results.

\paragraph{Target sets.}
The four Author-Root target sets are nested.
Author-Root-All contains eligible works reported by
authors as shaping or inspiring $P^+$.
Author-Root-Next retains those cited in $P^+$.
Author-Root-New further excludes works previously
cited by the responding researcher.
Author-Root-Isolated also excludes works previously
cited by recent coauthors, following the prior-citation
rules in Appendix~B.1.
Reports with unknown citation status for a required
filter are excluded from that tier.
\FloatBarrier
\begin{table}[!htbp]
\centering
\small
\setlength{\tabcolsep}{4pt}
\renewcommand{\arraystretch}{1.08}
\caption{\textbf{The author-report collection at a glance.}
Cases refer to surveyed papers.
A respondent may answer for more than one paper,
and a response may identify several works.
Endorsements are counted separately for each surveyed
paper.}
\label{tab:author-collection-summary}

\begin{tabular}{@{}p{0.32\linewidth}p{0.62\linewidth}@{}}
\toprule
Stage & Count or description \\
\midrule

Sources
& Corresponding authors of papers in 57 journals \\

Invitation
& Up to eight references from the surveyed paper's
bibliography; unlisted works could be named \\

Reply options
& Listed references, none of these, decline, unsure,
or another work named by the author \\

Follow-up
& Up to two reminders \\

Deliverable invitations
& 12,681 surveyed-paper cases \\

Replies received
& 2,295 cases from 2,243 respondents \\

Cases identifying scholarly inspirations
& 1,733 cases from 1,700 respondents \\

Cases with an identifiable predecessor
& 1,716 \\

Cases with a temporally eligible reported work
& 1,667 \\

Evaluation cases
& 1,633 All, 1,330 Next, 904 New, 665 Isolated \\

Eligible targets
& 3,900 endorsements in Author-Root-All \\

\bottomrule
\end{tabular}
\end{table}

\subsection{Examples of author-reported connections}

The following examples illustrate prior works that authors
reported as shaping their projects, including methodological
and conceptual connections. Descriptions are generalized
to protect participants. These examples are illustrative;
they do not measure benefits to researchers, and any claim
about time saved would require a user study.

\FloatBarrier
\begin{table}[!htbp]
\centering
\caption{\textbf{Three qualitative examples of author-reported
connections between prior work and subsequent projects.}
Ranks use SPECTER2-Proximity with 16 history centroids
on the 2,603,132-work Author-Root candidate pool.}
\begin{tabular}{
    p{0.13\linewidth}
    p{0.36\linewidth}
    p{0.31\linewidth}
    r
}
\toprule
Field & Subsequent project & Author-reported prior work & Rank \\
\midrule
Genomics
& Local-ancestry association study in admixed populations
& Prior local-ancestry method
& 5 \\

Behavioral science
& Complex-systems study of physical activity and behavior change
& Prior behavior-change theory
& 37 \\

Materials science
& Inverse design of hierarchical 3D architectures
& Prior self-assembly pathway design
& 132 \\
\bottomrule
\end{tabular}
\end{table}

\subsection{Author reports and citation-context evidence}
\label{app:author-context}

We examine whether author reports identify the same
works as generative citation-context labels.
Authors identify works they regard as shaping their
paper; the classifier labels citing passages that
describe building on a work's ideas, framing or methods.
The comparison covers 4,181 endorsements from 1,633 cases.

\noindent\textbf{Related signals, limited overlap.}
Most author-endorsed works are not labelled as generative
in the available citation contexts.
Among 1,640 endorsements with usable citing text,
124 receive this label
(7.6\%; 95\% case-cluster bootstrap interval [6.1, 9.1]).
The label is more common among endorsed works than
among other works cited in the same papers
(1.8\%; within-paper ratio 4.2,
95\% interval [3.4, 5.0]).
The signals are therefore related, but overlap is limited
among endorsements with usable citing text.

Unavailable citing text remains unknown
(Table~\ref{tab:author-context-coverage}).
The 2,452 endorsements without usable citing text do
not enter the overlap estimate.
Other cited works are those not endorsed in this
collection, not works judged uninfluential.

\begin{table}[!htbp]
\centering
\small
\setlength{\tabcolsep}{4pt}
\renewcommand{\arraystretch}{1.08}
\caption{\textbf{Author endorsements and citation evidence.}
Each endorsement links a surveyed paper to an endorsed
work. Categories are mutually exclusive for endorsements;
cases may appear in several categories.
A work is certified absent only if it is absent from
a complete Europe PMC reference list and has no
Semantic Scholar citation link.
Unavailable citing text remains unknown.}
\label{tab:author-context-coverage}

\begin{tabular}{@{}p{0.59\linewidth}rr@{}}
\toprule
Citation evidence & Endorsements & Cases \\
\midrule
Certified absent from the bibliography
& 89 & 60 \\
Citing text available; generative-use label
& 124 & 101 \\
Citing text available; no generative-use label
& 1,516 & 792 \\
No usable citing text; status unknown
& 2,452 & 1,169 \\
\midrule
Total
& 4,181 & 1,633 \\
\bottomrule
\end{tabular}
\end{table}

\noindent\textbf{Could classifier errors explain the limited overlap?}
Two annotators independently reviewed sampled citation
passages, blind to endorsement status and classifier
output.
For 40 sampled endorsements with usable citing text
but no generative-use label, the adjudicated review
found no generative use in the available passages
(0/40; Wilson 95\% interval [0, 8.8\%]).
Of 27 decidable passages labelled generative,
17 were confirmed
(63\%; work-cluster bootstrap interval [42, 84]).
This is a passage-level check, not an estimate of
work-level precision.

Author reports thus provide complementary evidence:
most endorsements with usable citing text lack a
generative-use label, and some endorsed works are
absent from the bibliography.  Among 313 cases with complete reference lists, 89 of 839 author-endorsed works are certified uncited (10.6\%; 95\% case-cluster bootstrap interval [8.0, 13.4]).
\FloatBarrier
\FloatBarrier
\section{Evaluation metrics and complete retrieval results}

\makeatletter
\setlength{\@fptop}{0pt}
\setlength{\@fpsep}{16pt plus 2pt minus 2pt}
\setlength{\@fpbot}{0pt plus 1fil}
\makeatother

This section defines the evaluation metrics and reports
complete retrieval results for the bibliography,
Context-Root, and Author-Root tiers.

\subsection{Evaluation metrics}

For each case $i$, let $E_i$ be the set of eligible targets
in the candidate pool and let $L_i(K)$ denote the first
$K$ retrieved candidates. Let $\mathcal{I}$ be the evaluation
cases for the stated evidence source and tier, with
$N = |\mathcal{I}|$. Each case has usable publication history
before $P$ and at least one eligible target in the pool.
All configurations within a source and tier are evaluated
on the same cases and targets. Author-Root uses its separate
candidate pool and the eligibility rules in Appendix~D.

\noindent\textbf{Reach and target coverage.}
Hit@$K$ measures whether at least one target is retrieved:
\begin{equation}
\operatorname{Hit@}K
=
\frac{1}{N}
\sum_{i \in \mathcal{I}}
\mathbf{1}\!\left[|L_i(K) \cap E_i| > 0\right].
\end{equation}
Macro Recall@$K$ averages the fraction of targets retrieved
within each case:
\begin{equation}
\operatorname{Recall@}K
=
\frac{1}{N}
\sum_{i \in \mathcal{I}}
\frac{|L_i(K) \cap E_i|}{|E_i|}.
\end{equation}

\noindent\textbf{Ranking quality.}
For binary relevance, let $d_{i,r}$ be the candidate at
rank $r$. The discounted cumulative gain is
\begin{equation}
\operatorname{DCG}_i@K
=
\sum_{r=1}^{K}
\frac{\mathbf{1}[d_{i,r} \in E_i]}{\log_2(r+1)}.
\end{equation}
Normalizing by the ideal ranking and averaging across
cases gives
\begin{equation}
\operatorname{nDCG@}K
=
\frac{1}{N}
\sum_{i \in \mathcal{I}}
\frac{\operatorname{DCG}_i@K}
{\displaystyle
 \sum_{r=1}^{\min(K,|E_i|)}
 \frac{1}{\log_2(r+1)}}.
\end{equation}

Mean reciprocal rank measures how early the first relevant
work appears. Let $r_i$ be its rank within the stored
top-1,000 list, with $r_i=\infty$ if no target is retrieved.
Using $1/\infty=0$,
\begin{equation}
\operatorname{MRR@1000}
=
\frac{1}{N}
\sum_{i \in \mathcal{I}} \frac{1}{r_i}.
\end{equation}

We report Hit, macro Recall, and nDCG at $K=10,100,1000$,
and MRR over the top-1,000 ranking. All metrics weight
cases equally.

\clearpage
\subsection{Bibliography results}
\suppressfloats[t]
We evaluate 24 retrieval configurations across CiteNext,
CiteNew and CiteNew-Isolated.
Tables~\ref{tab:muses-bib-hit}--\ref{tab:muses-bib-ranking}
report case reach (Hit), average target coverage (macro Recall),
and ranking quality (nDCG and MRR).
Appendix~C describes the methods and scoring rules.

\noindent\textbf{Reach depends on the tier and retrieval depth.}
Personalized PageRank has the highest observed Hit at
all three depths on CiteNext.
On both CiteNew and CiteNew-Isolated, Qwen3-Embedding-8B
LoRA leads at depth 10, while GritLM-7B K1 leads at depths
100 and 1,000 (Table~\ref{tab:muses-bib-hit}).
On CiteNew, for example, Qwen3 LoRA reaches 23.22\% of
cases at depth 10, whereas GritLM K1 reaches 37.41\% at
depth 100 and 59.35\% at depth 1,000.
Across all configurations, scores decrease from CiteNext
to CiteNew and again to CiteNew-Isolated for every
reported metric and depth.
Appendix~G.3 examines how target count and changes in
eligible cases contribute to these differences.

\noindent\textbf{Reaching a case does not mean recovering
most of its targets.}
At depth 1,000 on CiteNew, GritLM K1 reaches 59.35\% of
cases but retrieves 27.46\% of each case's targets on
average. PageRank has the highest macro Recall@1000
on this tier, at 28.10\% (Table~\ref{tab:muses-bib-recall}).
No configuration recovers a third of the targets per case
on average in any bibliography tier, even at depth 1,000.
Ranking quality also distinguishes configurations:
on CiteNew, Qwen3 LoRA has the highest observed nDCG@10,
nDCG@100 and MRR@1000, whereas GritLM K1 has the highest
displayed nDCG@1000 (Table~\ref{tab:muses-bib-ranking}).
The preferred configuration therefore depends on whether
the objective is to find an entry point, cover more targets,
or place relevant works early.

\noindent\textbf{Representation and adaptation have
method-dependent effects.}
Using 16 centroids rather than one improves SPECTER2-Proximity
and SciNCL Hit at every reported tier and depth.
For GritLM, however, one centroid gives higher Hit at depths
100 and 1,000 across all three tiers.
Qwen3 LoRA improves Hit over its matched frozen MaxSim
counterpart throughout these evaluations.
SPECTER2-Proximity fine-tuning instead lowers Hit at depths
10 and 100 while increasing it at depth 1,000.
These comparisons show why both matched controls and
evaluation at multiple depths matter.

\begin{table}[!htbp]
\centering
\scriptsize
\setlength{\tabcolsep}{2pt}
\caption{\textbf{Bibliography-tier Hit@$K$ (\%).} All rows use the same cases within each tier. Bold indicates the highest displayed value in each column,
including ties. Input and scoring definitions follow Appendix~C; Citations denotes history-conditioned citation-graph retrieval. Frozen MaxSim controls are described in Appendix~C.2. PageRank's CiteNew and CiteNew-Isolated values at depth 1000 are shown to one decimal because their numerical bounds are [59.33, 59.34] and [54.09, 54.10], respectively.}
\label{tab:muses-bib-hit}
\resizebox{\linewidth}{!}{%
\begin{tabular}{@{}llrrrrrrrrr@{}}
\toprule
& & \multicolumn{3}{c}{CiteNext} & \multicolumn{3}{c}{CiteNew} & \multicolumn{3}{c}{\shortstack{CiteNew-\\Isolated}} \\
& & \multicolumn{3}{c}{$N=135{,}111$} & \multicolumn{3}{c}{$N=134{,}314$} & \multicolumn{3}{c}{$N=133{,}166$} \\
\cmidrule(lr){3-5} \cmidrule(lr){6-8} \cmidrule(lr){9-11}
Retriever & Input & @10 & @100 & @1000 & @10 & @100 & @1000 & @10 & @100 & @1000 \\
\midrule
\multicolumn{11}{@{}l}{\emph{Popularity, lexical and sparse retrieval}} \\
Popularity & ---
& 2.96 & 7.01 & 14.65
& 1.77 & 5.21 & 12.10
& 0.72 & 2.50 & 7.18 \\
BM25 & Compact
& 22.24 & 33.96 & 51.59
& 17.44 & 29.71 & 48.79
& 15.23 & 26.79 & 45.87 \\
BM25 & Uncapped
& 26.50 & 36.90 & 53.73
& 20.95 & 32.48 & 51.04
& 18.26 & 29.31 & 48.24 \\
SPLADE & Compact
& 22.73 & 34.21 & 51.55
& 18.06 & 29.85 & 48.57
& 15.82 & 26.85 & 45.70 \\
\midrule
\multicolumn{11}{@{}l}{\emph{Frozen dense retrievers}} \\
Qwen3-Embedding-8B & Compact
& 22.54 & 36.66 & 56.64
& 17.74 & 32.51 & 54.39
& 15.56 & 29.48 & 51.75 \\
Qwen3-Embedding-8B & K16
& 28.68 & 39.26 & 56.58
& 22.38 & 34.25 & 54.00
& 19.55 & 30.97 & 51.18 \\
\addlinespace[2pt]
GritLM-7B & K1
& 27.85 & 41.66 & 61.46
& 22.08 & \textbf{37.41} & \textbf{59.35}
& 19.22 & \textbf{33.96} & \textbf{56.75} \\
\addlinespace[2pt]
GritLM-7B & K16
& 29.23 & 40.40 & 58.93
& 22.93 & 35.52 & 56.44
& 20.05 & 32.22 & 53.74 \\
\addlinespace[2pt]
SPECTER2-Proximity & K1
& 19.06 & 32.41 & 51.65
& 14.25 & 27.90 & 48.83
& 12.20 & 24.74 & 45.72 \\
\addlinespace[2pt]
SPECTER2-Proximity & K16
& 27.47 & 37.43 & 54.77
& 21.48 & 32.28 & 51.86
& 18.70 & 28.99 & 48.79 \\
\addlinespace[2pt]
SciNCL & K1
& 17.25 & 30.94 & 51.11
& 12.62 & 26.56 & 48.33
& 10.69 & 23.47 & 45.38 \\
\addlinespace[2pt]
SciNCL & K16
& 26.50 & 35.37 & 51.95
& 20.73 & 30.28 & 48.92
& 17.97 & 27.02 & 46.02 \\
\midrule
\multicolumn{11}{@{}l}{\emph{Frozen token-level late interaction}} \\
LateOn & Title max
& 15.07 & 23.92 & 39.06
& 11.25 & 20.28 & 36.23
& 9.74 & 18.18 & 33.84 \\
\midrule
\multicolumn{11}{@{}l}{\emph{Citation-graph retrievers}} \\
Personalized PageRank & Citations
& \textbf{33.31} & \textbf{46.25} & \textbf{61.95}
& 20.00 & 37.18 & 59.3
& 17.17 & 31.05 & 54.1 \\
Co-citation & Citations
& 14.55 & 23.70 & 27.29
& 4.24 & 12.91 & 18.18
& 1.73 & 6.56 & 11.06 \\
Bibliographic coupling & Citations
& 19.45 & 31.24 & 43.28
& 12.99 & 25.87 & 39.61
& 10.53 & 22.31 & 36.12 \\
Adamic--Adar & Citations
& 24.10 & 37.87 & 49.81
& 14.77 & 31.19 & 46.36
& 11.52 & 25.88 & 41.62 \\
\midrule
\multicolumn{11}{@{}l}{\emph{Frozen counterparts of adapted encoders}} \\
SPECTER2-Proximity & MaxSim
& 27.62 & 37.46 & 54.78
& 21.65 & 32.29 & 51.83
& 18.85 & 28.99 & 48.76 \\
Qwen3-Embedding-8B & MaxSim
& 28.81 & 39.25 & 56.55
& 22.50 & 34.21 & 53.94
& 19.65 & 30.94 & 51.10 \\
\midrule
\multicolumn{11}{@{}l}{\emph{Supervised retrievers}} \\
Linear-Proximity & K16
& 25.14 & 34.51 & 52.08
& 19.63 & 29.57 & 48.99
& 17.04 & 26.30 & 45.92 \\
Linear-Proximity + neighbours & K16
& 24.82 & 34.77 & 52.42
& 19.30 & 29.68 & 49.23
& 16.72 & 26.36 & 46.04 \\
Linear-Proximity (control) & K16
& 25.12 & 34.52 & 52.09
& 19.61 & 29.57 & 49.00
& 17.02 & 26.30 & 45.93 \\
\addlinespace[2pt]
SPECTER2-Proximity (fine-tuned) & MaxSim
& 26.93 & 36.47 & 55.11
& 21.31 & 31.70 & 52.19
& 18.48 & 28.29 & 49.21 \\
Qwen3-Embedding-8B (LoRA) & MaxSim
& 28.92 & 40.95 & 61.20
& \textbf{23.22} & 36.42 & 58.69
& \textbf{20.20} & 32.92 & 55.81 \\
\bottomrule
\end{tabular}%
}
\end{table}

\begin{table}[!htbp]
\centering
\scriptsize
\setlength{\tabcolsep}{2pt}
\caption{\textbf{Bibliography-tier macro Recall@$K$ (\%).} Cases and method groups match
Table~\ref{tab:muses-bib-hit}. Bold indicates the highest displayed value in each column, including ties. PageRank's CiteNext value at depth 1000 is shown to one decimal because its numerical bounds are [31.59, 31.60].}
\label{tab:muses-bib-recall}
\resizebox{\linewidth}{!}{%
\begin{tabular}{@{}llrrrrrrrrr@{}}
\toprule
& & \multicolumn{3}{c}{CiteNext} & \multicolumn{3}{c}{CiteNew} & \multicolumn{3}{c}{\shortstack{CiteNew-\\Isolated}} \\
& & \multicolumn{3}{c}{$N=135{,}111$} & \multicolumn{3}{c}{$N=134{,}314$} & \multicolumn{3}{c}{$N=133{,}166$} \\
\cmidrule(lr){3-5} \cmidrule(lr){6-8} \cmidrule(lr){9-11}
Retriever & Input & @10 & @100 & @1000 & @10 & @100 & @1000 & @10 & @100 & @1000 \\
\midrule
\multicolumn{11}{@{}l}{\emph{Popularity, lexical and sparse retrieval}} \\
Popularity & ---
& 0.33 & 0.92 & 2.22
& 0.23 & 0.74 & 1.89
& 0.13 & 0.42 & 1.20 \\
BM25 & Compact
& 5.04 & 10.28 & 20.90
& 4.18 & 8.93 & 19.35
& 4.08 & 8.64 & 18.76 \\
BM25 & Uncapped
& 6.10 & 11.28 & 22.16
& 4.99 & 9.76 & 20.53
& 4.86 & 9.39 & 19.92 \\
SPLADE & Compact
& 4.99 & 9.88 & 20.02
& 4.22 & 8.57 & 18.42
& 4.14 & 8.30 & 17.91 \\
\midrule
\multicolumn{11}{@{}l}{\emph{Frozen dense retrievers}} \\
Qwen3-Embedding-8B & Compact
& 5.49 & 12.76 & 27.70
& 4.50 & 11.09 & 25.97
& 4.40 & 10.63 & 25.09 \\
Qwen3-Embedding-8B & K16
& 6.95 & 13.53 & 27.50
& 5.51 & 11.38 & 25.40
& 5.37 & 10.91 & 24.48 \\
\addlinespace[2pt]
GritLM-7B & K1
& 6.88 & 14.26 & 29.19
& 5.56 & \textbf{12.43} & 27.46
& 5.35 & \textbf{11.90} & \textbf{26.61} \\
\addlinespace[2pt]
GritLM-7B & K16
& 7.09 & 13.68 & 27.81
& \textbf{5.65} & 11.60 & 25.76
& \textbf{5.49} & 11.16 & 24.97 \\
\addlinespace[2pt]
SPECTER2-Proximity & K1
& 4.20 & 9.33 & 20.83
& 3.38 & 7.98 & 19.24
& 3.25 & 7.61 & 18.49 \\
\addlinespace[2pt]
SPECTER2-Proximity & K16
& 6.17 & 11.22 & 23.07
& 4.98 & 9.45 & 21.06
& 4.86 & 9.13 & 20.35 \\
\addlinespace[2pt]
SciNCL & K1
& 3.72 & 9.11 & 21.60
& 2.95 & 7.80 & 20.04
& 2.83 & 7.42 & 19.30 \\
\addlinespace[2pt]
SciNCL & K16
& 5.83 & 10.45 & 22.22
& 4.76 & 8.82 & 20.37
& 4.65 & 8.48 & 19.67 \\
\midrule
\multicolumn{11}{@{}l}{\emph{Frozen token-level late interaction}} \\
LateOn & Title max
& 3.37 & 7.26 & 15.52
& 2.73 & 6.27 & 14.37
& 2.65 & 6.03 & 13.93 \\
\midrule
\multicolumn{11}{@{}l}{\emph{Citation-graph retrievers}} \\
Personalized PageRank & Citations
& \textbf{8.77} & \textbf{17.69} & \textbf{31.6}
& 4.67 & 12.21 & \textbf{28.10}
& 4.46 & 10.59 & 25.04 \\
Co-citation & Citations
& 2.87 & 6.31 & 8.32
& 0.76 & 2.67 & 4.41
& 0.36 & 1.39 & 2.54 \\
Bibliographic coupling & Citations
& 4.43 & 9.80 & 16.67
& 3.02 & 7.68 & 14.36
& 2.77 & 7.06 & 13.40 \\
Adamic--Adar & Citations
& 5.87 & 13.30 & 21.94
& 3.53 & 9.79 & 18.62
& 3.10 & 8.50 & 16.58 \\
\midrule
\multicolumn{11}{@{}l}{\emph{Frozen counterparts of adapted encoders}} \\
SPECTER2-Proximity & MaxSim
& 6.21 & 11.24 & 23.08
& 5.02 & 9.46 & 21.06
& 4.91 & 9.14 & 20.35 \\
Qwen3-Embedding-8B & MaxSim
& 6.98 & 13.50 & 27.45
& 5.54 & 11.35 & 25.34
& 5.40 & 10.89 & 24.42 \\
\midrule
\multicolumn{11}{@{}l}{\emph{Supervised retrievers}} \\
Linear-Proximity & K16
& 5.38 & 9.68 & 20.97
& 4.45 & 8.25 & 19.21
& 4.35 & 7.95 & 18.59 \\
Linear-Proximity + neighbours & K16
& 5.41 & 10.07 & 21.85
& 4.39 & 8.39 & 19.79
& 4.27 & 8.02 & 18.95 \\
Linear-Proximity (control) & K16
& 5.38 & 9.68 & 20.97
& 4.45 & 8.25 & 19.20
& 4.35 & 7.95 & 18.58 \\
\addlinespace[2pt]
SPECTER2-Proximity (fine-tuned) & MaxSim
& 5.78 & 10.15 & 22.26
& 4.81 & 8.81 & 20.70
& 4.71 & 8.50 & 20.05 \\
Qwen3-Embedding-8B (LoRA) & MaxSim
& 6.76 & 13.19 & 27.84
& 5.62 & 11.59 & 26.13
& 5.45 & 11.06 & 25.18 \\
\bottomrule
\end{tabular}%
}
\end{table}

\begin{table}[!htbp]
\centering
\scriptsize
\setlength{\tabcolsep}{2pt}
\caption{\textbf{Bibliography-tier ranking quality.} Each tier reports nDCG at depths 10, 100 and 1000 (\%) and MRR@1000 (0--1 scale). Cases and method groups match
Table~\ref{tab:muses-bib-hit}. Bold indicates the highest displayed value in each column, including ties.}
\label{tab:muses-bib-ranking}
\resizebox{\linewidth}{!}{%
\begin{tabular}{@{}llrrrrrrrrrrrr@{}}
\toprule
& & \multicolumn{4}{c}{CiteNext} & \multicolumn{4}{c}{CiteNew} & \multicolumn{4}{c}{\shortstack{CiteNew-\\Isolated}} \\
& & \multicolumn{4}{c}{$N=135{,}111$} & \multicolumn{4}{c}{$N=134{,}314$} & \multicolumn{4}{c}{$N=133{,}166$} \\
\cmidrule(lr){3-6} \cmidrule(lr){7-10} \cmidrule(lr){11-14}
& & \multicolumn{3}{c}{nDCG (\%)} & MRR & \multicolumn{3}{c}{nDCG (\%)} & MRR & \multicolumn{3}{c}{nDCG (\%)} & MRR \\
Retriever & Input & @10 & @100 & @1000 & @1000 & @10 & @100 & @1000 & @1000 & @10 & @100 & @1000 & @1000 \\
\midrule
\multicolumn{14}{@{}l}{\emph{Popularity, lexical and sparse retrieval}} \\
Popularity & ---
& 0.54 & 0.67 & 1.00 & 0.019
& 0.30 & 0.45 & 0.73 & 0.010
& 0.12 & 0.21 & 0.39 & 0.004 \\
BM25 & Compact
& 7.04 & 8.28 & 10.84 & 0.165
& 5.04 & 6.42 & 8.77 & 0.122
& 4.62 & 5.93 & 8.08 & 0.107 \\
BM25 & Uncapped
& 8.85 & 9.85 & 12.48 & 0.206
& 6.16 & 7.47 & 9.92 & 0.149
& 5.57 & 6.85 & 9.10 & 0.128 \\
SPLADE & Compact
& 7.10 & 8.19 & 10.64 & 0.171
& 5.27 & 6.48 & 8.71 & 0.131
& 4.84 & 6.02 & 8.06 & 0.115 \\
\midrule
\multicolumn{14}{@{}l}{\emph{Frozen dense retrievers}} \\
Qwen3-Embedding-8B & Compact
& 7.25 & 9.26 & 12.87 & 0.162
& 5.20 & 7.20 & 10.58 & 0.122
& 4.78 & 6.64 & 9.72 & 0.108 \\
Qwen3-Embedding-8B & K16
& 9.72 & 11.19 & 14.58 & 0.217
& 6.64 & 8.29 & 11.48 & 0.156
& 6.02 & 7.59 & 10.49 & 0.135 \\
\addlinespace[2pt]
GritLM-7B & K1
& 9.36 & 11.16 & 14.74 & 0.207
& 6.41 & 8.44 & \textbf{11.84} & 0.147
& 5.76 & 7.68 & \textbf{10.82} & 0.127 \\
\addlinespace[2pt]
GritLM-7B & K16
& 9.79 & 11.24 & 14.63 & 0.215
& 6.53 & 8.22 & 11.41 & 0.149
& 5.89 & 7.51 & 10.44 & 0.128 \\
\addlinespace[2pt]
SPECTER2-Proximity & K1
& 5.63 & 7.01 & 9.79 & 0.131
& 3.78 & 5.19 & 7.73 & 0.089
& 3.39 & 4.70 & 7.00 & 0.076 \\
\addlinespace[2pt]
SPECTER2-Proximity & K16
& 9.03 & 9.96 & 12.84 & 0.212
& 6.27 & 7.45 & 10.09 & 0.153
& 5.68 & 6.85 & 9.25 & 0.132 \\
\addlinespace[2pt]
SciNCL & K1
& 4.88 & 6.42 & 9.43 & 0.115
& 3.20 & 4.70 & 7.46 & 0.075
& 2.85 & 4.24 & 6.74 & 0.064 \\
\addlinespace[2pt]
SciNCL & K16
& 8.64 & 9.43 & 12.27 & 0.206
& 6.03 & 7.07 & 9.68 & 0.149
& 5.47 & 6.50 & 8.86 & 0.128 \\
\midrule
\multicolumn{14}{@{}l}{\emph{Frozen token-level late interaction}} \\
LateOn & Title max
& 4.72 & 5.69 & 7.66 & 0.115
& 3.25 & 4.30 & 6.11 & 0.079
& 2.94 & 3.93 & 5.59 & 0.068 \\
\midrule
\multicolumn{14}{@{}l}{\emph{Citation-graph retrievers}} \\
Personalized PageRank & Citations
& \textbf{11.70} & \textbf{14.03} & \textbf{17.53} & \textbf{0.240}
& 5.77 & 8.03 & 11.82 & 0.146
& 5.19 & 6.94 & 10.11 & 0.125 \\
Co-citation & Citations
& 3.85 & 4.79 & 5.41 & 0.083
& 0.74 & 1.42 & 1.90 & 0.019
& 0.31 & 0.66 & 0.94 & 0.008 \\
Bibliographic coupling & Citations
& 5.74 & 7.24 & 9.07 & 0.119
& 3.01 & 4.55 & 6.18 & 0.065
& 2.52 & 3.89 & 5.32 & 0.053 \\
Adamic--Adar & Citations
& 7.47 & 9.63 & 11.97 & 0.148
& 3.44 & 5.56 & 7.77 & 0.074
& 2.79 & 4.53 & 6.39 & 0.058 \\
\midrule
\multicolumn{14}{@{}l}{\emph{Frozen counterparts of adapted encoders}} \\
SPECTER2-Proximity & MaxSim
& 9.09 & 10.00 & 12.87 & 0.213
& 6.32 & 7.49 & 10.12 & 0.154
& 5.73 & 6.89 & 9.28 & 0.133 \\
Qwen3-Embedding-8B & MaxSim
& 9.79 & 11.23 & 14.61 & 0.220
& 6.74 & 8.37 & 11.55 & 0.160
& 6.12 & 7.67 & 10.56 & 0.138 \\
\midrule
\multicolumn{14}{@{}l}{\emph{Supervised retrievers}} \\
Linear-Proximity & K16
& 8.01 & 8.77 & 11.50 & 0.196
& 5.71 & 6.69 & 9.16 & 0.143
& 5.19 & 6.16 & 8.40 & 0.123 \\
Linear-Proximity + neighbours & K16
& 7.96 & 8.89 & 11.74 & 0.192
& 5.63 & 6.70 & 9.28 & 0.141
& 5.12 & 6.15 & 8.47 & 0.123 \\
Linear-Proximity (control) & K16
& 8.01 & 8.77 & 11.50 & 0.196
& 5.70 & 6.69 & 9.16 & 0.143
& 5.18 & 6.15 & 8.40 & 0.123 \\
\addlinespace[2pt]
SPECTER2-Proximity (fine-tuned) & MaxSim
& 8.62 & 9.29 & 12.16 & 0.208
& 6.11 & 7.13 & 9.79 & 0.151
& 5.55 & 6.55 & 8.98 & 0.130 \\
Qwen3-Embedding-8B (LoRA) & MaxSim
& 9.53 & 10.90 & 14.38 & 0.220
& \textbf{6.82} & \textbf{8.48} & 11.75 & \textbf{0.163}
& \textbf{6.16} & \textbf{7.73} & 10.71 & \textbf{0.141} \\
\bottomrule
\end{tabular}%
}
\end{table}

\clearpage
\subsection{Context-Root results}
\suppressfloats[t]
We evaluate the same 24 retrieval configurations across
Context-Root-Next, Context-Root-New and Context-Root-Isolated.
Tables~\ref{tab:muses-context-hit}--\ref{tab:muses-context-ranking}
report case reach (Hit), average target coverage (macro Recall),
and ranking quality (nDCG and MRR).
These targets describe classifier-labelled generative use;
Appendices~B.2--B.3 define the labels and their validation.

\noindent\textbf{The leading retriever changes with depth.}
Personalized PageRank has the highest observed Hit at
all three depths on Context-Root-Next.
On New and Isolated, Qwen3-Embedding-8B LoRA leads at
depth 10, while PageRank leads at depths 100 and 1,000
(Table~\ref{tab:muses-context-hit}).
On New, Qwen3 LoRA reaches 10.72\% of cases at depth 10;
PageRank reaches 22.53\% at depth 100 and 47.98\% at
depth 1,000.
All 24 configurations have lower Hit on New than Next
and on Isolated than New at every reported depth.
Appendix~G.3 examines these tier differences after
accounting for target count and changes in eligible cases.

\noindent\textbf{Coverage and early ranking favour different
configurations.}
PageRank also has the highest macro Recall at depths 100
and 1,000 across all three tiers
(Table~\ref{tab:muses-context-recall}).
On New at depth 1,000, it reaches 47.98\% of cases and
retrieves 43.52\% of each case's targets on average:
more than half of cases remain unreached, and average
target coverage remains below half.
Qwen3 LoRA instead leads macro Recall@10, nDCG@10 and
MRR@1000 on New and Isolated.
PageRank leads nDCG@1000 across all three tiers, while
nDCG@100 favours PageRank on New and Qwen3 LoRA on
Isolated (Table~\ref{tab:muses-context-ranking}).
Unlike Hit, the other metrics do not decrease uniformly
from New to Isolated for every configuration.

\noindent\textbf{Adaptation and citation-neighbour features
have mixed effects.}
Qwen3 LoRA improves Hit over its matched frozen MaxSim
counterpart at every reported depth on New and Isolated.
On Next, however, it improves Hit only at depth 1,000.
Adding citation-neighbour features to the linear model
increases Hit at depths 100 and 1,000 relative to its
matched no-neighbour control in all three tiers, but
lowers Hit@10.
These comparisons distinguish improvements in deeper
recovery from improvements near the top of the ranking.

\begin{table}[!htbp]
\centering
\scriptsize
\setlength{\tabcolsep}{2pt}
\caption{\textbf{Context-Root Hit@$K$ (\%).} All rows use the same cases within each tier. Input definitions and method groups follow Appendix~C. Bold indicates the highest displayed value in each column, including ties.}
\label{tab:muses-context-hit}
\resizebox{\linewidth}{!}{%
\begin{tabular}{@{}llrrrrrrrrr@{}}
\toprule
& & \multicolumn{3}{c}{Next} & \multicolumn{3}{c}{New} & \multicolumn{3}{c}{Isolated} \\
& & \multicolumn{3}{c}{$N=7{,}848$} & \multicolumn{3}{c}{$N=6{,}659$} & \multicolumn{3}{c}{$N=5{,}666$} \\
\cmidrule(lr){3-5} \cmidrule(lr){6-8} \cmidrule(lr){9-11}
Retriever & Input & @10 & @100 & @1000 & @10 & @100 & @1000 & @10 & @100 & @1000 \\
\midrule
\multicolumn{11}{@{}l}{\emph{Popularity, lexical and sparse retrieval}} \\
Popularity & ---
& 1.36 & 3.03 & 7.76
& 0.93 & 2.30 & 6.64
& 0.51 & 1.11 & 3.83 \\
BM25 & Compact
& 10.92 & 18.30 & 30.54
& 6.91 & 12.43 & 23.47
& 6.72 & 11.72 & 21.80 \\
BM25 & Uncapped
& 13.81 & 21.06 & 32.93
& 9.03 & 14.60 & 25.87
& 8.37 & 13.25 & 23.88 \\
SPLADE & Compact
& 11.38 & 19.20 & 31.03
& 7.61 & 13.31 & 23.64
& 7.54 & 12.60 & 21.81 \\
\midrule
\multicolumn{11}{@{}l}{\emph{Frozen dense retrievers}} \\
Qwen3-Embedding-8B & Compact
& 12.95 & 24.38 & 43.60
& 8.02 & 16.40 & 35.94
& 7.64 & 15.39 & 33.00 \\
Qwen3-Embedding-8B & K16
& 16.02 & 26.36 & 43.35
& 10.18 & 17.50 & 34.55
& 9.58 & 15.95 & 31.66 \\
\addlinespace[2pt]
GritLM-7B & K1
& 14.93 & 25.99 & 43.08
& 9.36 & 18.04 & 35.44
& 8.95 & 16.64 & 32.85 \\
\addlinespace[2pt]
GritLM-7B & K16
& 16.26 & 25.71 & 41.70
& 10.41 & 17.03 & 32.78
& 10.01 & 15.76 & 30.46 \\
\addlinespace[2pt]
SPECTER2-Proximity & K1
& 10.50 & 18.92 & 34.91
& 6.44 & 12.39 & 27.57
& 6.07 & 11.37 & 25.11 \\
\addlinespace[2pt]
SPECTER2-Proximity & K16
& 15.11 & 22.82 & 37.49
& 9.76 & 15.26 & 28.92
& 9.34 & 14.08 & 26.86 \\
\addlinespace[2pt]
SciNCL & K1
& 9.24 & 18.37 & 35.03
& 5.47 & 12.00 & 27.38
& 5.14 & 10.98 & 25.15 \\
\addlinespace[2pt]
SciNCL & K16
& 14.42 & 20.97 & 36.53
& 9.45 & 13.83 & 27.98
& 9.02 & 12.95 & 25.82 \\
\midrule
\multicolumn{11}{@{}l}{\emph{Frozen token-level late interaction}} \\
LateOn & Title max
& 6.84 & 12.27 & 22.53
& 4.04 & 7.99 & 17.42
& 3.81 & 7.27 & 15.67 \\
\midrule
\multicolumn{11}{@{}l}{\emph{Citation-graph retrievers}} \\
Personalized PageRank & Citations
& \textbf{21.48} & \textbf{38.04} & \textbf{57.67}
& 9.82 & \textbf{22.53} & \textbf{47.98}
& 9.30 & \textbf{18.53} & \textbf{39.94} \\
Co-citation & Citations
& 8.08 & 18.32 & 23.42
& 1.32 & 6.04 & 10.53
& 0.62 & 2.91 & 5.38 \\
Bibliographic coupling & Citations
& 12.18 & 23.90 & 37.13
& 5.86 & 14.93 & 27.74
& 5.07 & 12.97 & 24.71 \\
Adamic--Adar & Citations
& 15.65 & 30.38 & 46.00
& 6.55 & 17.93 & 34.98
& 5.56 & 14.84 & 29.33 \\
\midrule
\multicolumn{11}{@{}l}{\emph{Frozen counterparts of adapted encoders}} \\
SPECTER2-Proximity & MaxSim
& 15.12 & 22.73 & 37.39
& 9.76 & 15.17 & 28.86
& 9.32 & 14.00 & 26.79 \\
Qwen3-Embedding-8B & MaxSim
& 16.08 & 26.15 & 43.22
& 10.18 & 17.36 & 34.34
& 9.60 & 15.87 & 31.42 \\
\midrule
\multicolumn{11}{@{}l}{\emph{Supervised retrievers}} \\
Linear-Proximity & K16
& 13.66 & 20.68 & 37.05
& 9.18 & 14.19 & 29.09
& 8.77 & 13.13 & 26.61 \\
Linear-Proximity + neighbours & K16
& 13.47 & 21.85 & 39.19
& 8.86 & 14.79 & 31.00
& 8.47 & 13.63 & 28.31 \\
Linear-Proximity (control) & K16
& 13.69 & 20.69 & 37.03
& 9.16 & 14.21 & 29.06
& 8.74 & 13.17 & 26.53 \\
\addlinespace[2pt]
SPECTER2-Proximity (fine-tuned) & MaxSim
& 13.77 & 20.49 & 36.33
& 9.34 & 14.43 & 29.21
& 8.98 & 13.40 & 27.34 \\
Qwen3-Embedding-8B (LoRA) & MaxSim
& 15.71 & 25.89 & 44.86
& \textbf{10.72} & 19.06 & 38.20
& \textbf{10.25} & 17.81 & 35.88 \\
\bottomrule
\end{tabular}%
}
\end{table}

\begin{table}[!htbp]
\centering
\scriptsize
\setlength{\tabcolsep}{2pt}
\caption{\textbf{Context-Root macro Recall@$K$ (\%).} Cases and method groups match
Table~\ref{tab:muses-context-hit}. Bold indicates the highest displayed value in each column, including ties.}
\label{tab:muses-context-recall}
\resizebox{\linewidth}{!}{%
\begin{tabular}{@{}llrrrrrrrrr@{}}
\toprule
& & \multicolumn{3}{c}{Next} & \multicolumn{3}{c}{New} & \multicolumn{3}{c}{Isolated} \\
& & \multicolumn{3}{c}{$N=7{,}848$} & \multicolumn{3}{c}{$N=6{,}659$} & \multicolumn{3}{c}{$N=5{,}666$} \\
\cmidrule(lr){3-5} \cmidrule(lr){6-8} \cmidrule(lr){9-11}
Retriever & Input & @10 & @100 & @1000 & @10 & @100 & @1000 & @10 & @100 & @1000 \\
\midrule
\multicolumn{11}{@{}l}{\emph{Popularity, lexical and sparse retrieval}} \\
Popularity & ---
& 1.09 & 2.35 & 5.85
& 0.79 & 1.81 & 5.05
& 0.44 & 0.94 & 3.05 \\
BM25 & Compact
& 8.78 & 15.23 & 26.31
& 6.05 & 10.77 & 20.50
& 6.10 & 10.43 & 19.45 \\
BM25 & Uncapped
& 11.33 & 17.56 & 28.56
& 8.03 & 12.62 & 22.76
& 7.71 & 11.86 & 21.47 \\
SPLADE & Compact
& 9.15 & 15.73 & 26.68
& 6.67 & 11.35 & 20.67
& 6.81 & 11.15 & 19.54 \\
\midrule
\multicolumn{11}{@{}l}{\emph{Frozen dense retrievers}} \\
Qwen3-Embedding-8B & Compact
& 10.54 & 20.77 & 39.49
& 7.05 & 14.39 & 32.66
& 6.94 & 13.80 & 30.44 \\
Qwen3-Embedding-8B & K16
& 13.25 & 22.66 & 39.07
& 9.04 & 15.43 & 31.19
& 8.77 & 14.46 & 29.08 \\
\addlinespace[2pt]
GritLM-7B & K1
& 12.23 & 22.18 & 38.50
& 8.31 & 15.82 & 31.77
& 8.16 & 15.05 & 29.99 \\
\addlinespace[2pt]
GritLM-7B & K16
& 13.49 & 21.87 & 37.13
& 9.21 & 14.91 & 29.28
& 9.15 & 14.25 & 27.75 \\
\addlinespace[2pt]
SPECTER2-Proximity & K1
& 8.49 & 15.64 & 30.26
& 5.62 & 10.76 & 24.31
& 5.46 & 10.15 & 22.66 \\
\addlinespace[2pt]
SPECTER2-Proximity & K16
& 12.39 & 19.23 & 32.77
& 8.66 & 13.35 & 25.54
& 8.57 & 12.64 & 24.22 \\
\addlinespace[2pt]
SciNCL & K1
& 7.34 & 15.24 & 30.54
& 4.71 & 10.46 & 24.30
& 4.58 & 9.89 & 22.75 \\
\addlinespace[2pt]
SciNCL & K16
& 11.75 & 17.63 & 32.04
& 8.38 & 12.14 & 24.94
& 8.26 & 11.68 & 23.43 \\
\midrule
\multicolumn{11}{@{}l}{\emph{Frozen token-level late interaction}} \\
LateOn & Title max
& 5.50 & 10.43 & 19.27
& 3.58 & 7.11 & 15.22
& 3.49 & 6.59 & 14.00 \\
\midrule
\multicolumn{11}{@{}l}{\emph{Citation-graph retrievers}} \\
Personalized PageRank & Citations
& \textbf{17.88} & \textbf{33.63} & \textbf{53.12}
& 8.80 & \textbf{19.90} & \textbf{43.52}
& 8.63 & \textbf{16.76} & \textbf{36.64} \\
Co-citation & Citations
& 6.05 & 14.74 & 19.50
& 1.05 & 4.78 & 8.62
& 0.52 & 2.32 & 4.41 \\
Bibliographic coupling & Citations
& 9.67 & 20.10 & 32.41
& 5.05 & 12.75 & 24.24
& 4.50 & 11.37 & 22.03 \\
Adamic--Adar & Citations
& 12.72 & 26.37 & 41.44
& 5.67 & 15.62 & 31.07
& 4.96 & 13.17 & 26.44 \\
\midrule
\multicolumn{11}{@{}l}{\emph{Frozen counterparts of adapted encoders}} \\
SPECTER2-Proximity & MaxSim
& 12.40 & 19.16 & 32.66
& 8.66 & 13.27 & 25.47
& 8.57 & 12.57 & 24.15 \\
Qwen3-Embedding-8B & MaxSim
& 13.29 & 22.44 & 38.97
& 9.04 & 15.29 & 31.04
& 8.80 & 14.35 & 28.86 \\
\midrule
\multicolumn{11}{@{}l}{\emph{Supervised retrievers}} \\
Linear-Proximity & K16
& 10.98 & 16.95 & 31.78
& 8.06 & 12.29 & 25.28
& 7.96 & 11.70 & 23.73 \\
Linear-Proximity + neighbours & K16
& 10.94 & 18.15 & 34.18
& 7.82 & 12.80 & 27.38
& 7.72 & 12.22 & 25.55 \\
Linear-Proximity (control) & K16
& 11.00 & 16.95 & 31.77
& 8.04 & 12.29 & 25.27
& 7.94 & 11.73 & 23.70 \\
\addlinespace[2pt]
SPECTER2-Proximity (fine-tuned) & MaxSim
& 11.16 & 16.85 & 31.35
& 8.29 & 12.55 & 25.63
& 8.21 & 11.98 & 24.53 \\
Qwen3-Embedding-8B (LoRA) & MaxSim
& 12.75 & 22.03 & 40.37
& \textbf{9.42} & 16.77 & 34.68
& \textbf{9.30} & 16.02 & 33.08 \\
\bottomrule
\end{tabular}%
}
\end{table}

\begin{table}[!htbp]
\centering
\scriptsize
\setlength{\tabcolsep}{2pt}
\caption{\textbf{Context-Root ranking quality.} Each tier reports nDCG at depths 10, 100 and 1000 (\%) and MRR@1000 (0--1 scale). Cases and method groups match Table~\ref{tab:muses-context-hit}. Bold indicates the highest displayed value in each column, including ties.}
\label{tab:muses-context-ranking}
\resizebox{\linewidth}{!}{%
\begin{tabular}{@{}llrrrrrrrrrrrr@{}}
\toprule
& & \multicolumn{4}{c}{Next} & \multicolumn{4}{c}{New} & \multicolumn{4}{c}{Isolated} \\
& & \multicolumn{4}{c}{$N=7{,}848$} & \multicolumn{4}{c}{$N=6{,}659$} & \multicolumn{4}{c}{$N=5{,}666$} \\
\cmidrule(lr){3-6} \cmidrule(lr){7-10} \cmidrule(lr){11-14}
& & \multicolumn{3}{c}{nDCG (\%)} & MRR & \multicolumn{3}{c}{nDCG (\%)} & MRR & \multicolumn{3}{c}{nDCG (\%)} & MRR \\
Retriever & Input & @10 & @100 & @1000 & @1000 & @10 & @100 & @1000 & @1000 & @10 & @100 & @1000 & @1000 \\
\midrule
\multicolumn{14}{@{}l}{\emph{Popularity, lexical and sparse retrieval}} \\
Popularity & ---
& 0.70 & 1.00 & 1.45 & 0.008
& 0.42 & 0.66 & 1.07 & 0.004
& 0.18 & 0.29 & 0.56 & 0.002 \\
BM25 & Compact
& 6.50 & 7.92 & 9.39 & 0.069
& 4.79 & 5.79 & 7.03 & 0.050
& 4.89 & 5.80 & 6.94 & 0.050 \\
BM25 & Uncapped
& 8.39 & 9.75 & 11.21 & 0.087
& 6.08 & 7.04 & 8.34 & 0.061
& 5.96 & 6.83 & 8.05 & 0.059 \\
SPLADE & Compact
& 6.86 & 8.31 & 9.75 & 0.073
& 5.31 & 6.30 & 7.49 & 0.055
& 5.49 & 6.39 & 7.45 & 0.056 \\
\midrule
\multicolumn{14}{@{}l}{\emph{Frozen dense retrievers}} \\
Qwen3-Embedding-8B & Compact
& 7.40 & 9.64 & 12.12 & 0.078
& 5.23 & 6.75 & 9.11 & 0.054
& 5.24 & 6.65 & 8.78 & 0.053 \\
Qwen3-Embedding-8B & K16
& 9.59 & 11.66 & 13.85 & 0.099
& 6.78 & 8.10 & 10.13 & 0.068
& 6.70 & 7.87 & 9.72 & 0.066 \\
\addlinespace[2pt]
GritLM-7B & K1
& 9.00 & 11.18 & 13.34 & 0.095
& 6.15 & 7.71 & 9.76 & 0.062
& 6.05 & 7.46 & 9.35 & 0.060 \\
\addlinespace[2pt]
GritLM-7B & K16
& 9.50 & 11.35 & 13.38 & 0.096
& 6.46 & 7.64 & 9.49 & 0.063
& 6.46 & 7.50 & 9.20 & 0.061 \\
\addlinespace[2pt]
SPECTER2-Proximity & K1
& 6.03 & 7.57 & 9.50 & 0.064
& 3.86 & 4.92 & 6.65 & 0.038
& 3.78 & 4.74 & 6.31 & 0.037 \\
\addlinespace[2pt]
SPECTER2-Proximity & K16
& 9.18 & 10.67 & 12.47 & 0.096
& 6.63 & 7.60 & 9.16 & 0.067
& 6.62 & 7.46 & 8.92 & 0.066 \\
\addlinespace[2pt]
SciNCL & K1
& 5.18 & 6.89 & 8.91 & 0.056
& 3.22 & 4.39 & 6.16 & 0.032
& 3.17 & 4.23 & 5.86 & 0.031 \\
\addlinespace[2pt]
SciNCL & K16
& 8.77 & 10.04 & 11.93 & 0.092
& 6.39 & 7.17 & 8.79 & 0.064
& 6.36 & 7.05 & 8.52 & 0.063 \\
\midrule
\multicolumn{14}{@{}l}{\emph{Frozen token-level late interaction}} \\
LateOn & Title max
& 4.05 & 5.13 & 6.29 & 0.044
& 2.70 & 3.43 & 4.47 & 0.028
& 2.66 & 3.30 & 4.24 & 0.027 \\
\midrule
\multicolumn{14}{@{}l}{\emph{Citation-graph retrievers}} \\
Personalized PageRank & Citations
& \textbf{12.00} & \textbf{15.58} & \textbf{18.20} & \textbf{0.120}
& 6.62 & \textbf{8.90} & \textbf{12.00} & 0.069
& 6.54 & 8.21 & \textbf{10.76} & 0.066 \\
Co-citation & Citations
& 3.74 & 5.74 & 6.40 & 0.040
& 0.55 & 1.34 & 1.85 & 0.006
& 0.26 & 0.64 & 0.91 & 0.003 \\
Bibliographic coupling & Citations
& 5.87 & 8.22 & 9.91 & 0.058
& 2.86 & 4.49 & 6.01 & 0.027
& 2.63 & 4.05 & 5.43 & 0.025 \\
Adamic--Adar & Citations
& 7.67 & 10.75 & 12.81 & 0.075
& 3.13 & 5.23 & 7.27 & 0.030
& 2.81 & 4.51 & 6.23 & 0.027 \\
\midrule
\multicolumn{14}{@{}l}{\emph{Frozen counterparts of adapted encoders}} \\
SPECTER2-Proximity & MaxSim
& 9.19 & 10.67 & 12.47 & 0.096
& 6.66 & 7.62 & 9.18 & 0.067
& 6.66 & 7.49 & 8.96 & 0.066 \\
Qwen3-Embedding-8B & MaxSim
& 9.65 & 11.68 & 13.87 & 0.100
& 6.84 & 8.14 & 10.16 & 0.069
& 6.77 & 7.92 & 9.76 & 0.067 \\
\midrule
\multicolumn{14}{@{}l}{\emph{Supervised retrievers}} \\
Linear-Proximity & K16
& 8.44 & 9.72 & 11.68 & 0.090
& 6.37 & 7.23 & 8.89 & 0.065
& 6.33 & 7.08 & 8.61 & 0.063 \\
Linear-Proximity + neighbours & K16
& 8.50 & 10.05 & 12.16 & 0.091
& 6.30 & 7.33 & 9.19 & 0.065
& 6.25 & 7.17 & 8.85 & 0.064 \\
Linear-Proximity (control) & K16
& 8.42 & 9.70 & 11.65 & 0.089
& 6.34 & 7.21 & 8.87 & 0.064
& 6.30 & 7.07 & 8.58 & 0.063 \\
\addlinespace[2pt]
SPECTER2-Proximity (fine-tuned) & MaxSim
& 8.56 & 9.78 & 11.68 & 0.092
& 6.46 & 7.33 & 9.01 & 0.065
& 6.46 & 7.23 & 8.80 & 0.064 \\
Qwen3-Embedding-8B (LoRA) & MaxSim
& 9.38 & 11.42 & 13.84 & 0.099
& \textbf{7.01} & 8.53 & 10.84 & \textbf{0.071}
& \textbf{6.91} & \textbf{8.30} & 10.45 & \textbf{0.068} \\
\bottomrule
\end{tabular}%
}
\end{table}

\clearpage
\subsection{Author-Root results}
\suppressfloats[t]
We evaluate the same 24 retrieval configurations across
Author-Root-All, Author-Root-Next, Author-Root-New and
Author-Root-Isolated, with 1,633, 1,330, 904 and 665 cases,
respectively. We use the candidate pool and eligibility
rules described in Appendix~D and the metrics defined
in Appendix~E.1.
Tables~\ref{tab:ar-hit}--\ref{tab:ar-mrr} report case reach
(Hit), average target coverage (macro Recall), and ranking
quality (nDCG and MRR).

\noindent\textbf{The leading configuration changes with
the target tier.}
The frozen Qwen3-Embedding-8B MaxSim control has the
highest observed Hit@100 on All and Next, at 58.79\%
and 57.89\%, respectively.
Its LoRA-adapted counterpart leads on New and Isolated,
at 36.17\% and 34.59\%, and also leads these two tiers
at depths 10 and 1,000 (Table~\ref{tab:ar-hit}).
All 23 history-based configurations have lower Hit@100
on New than Next, while the additional coauthor exclusion
generally lowers reach further.
Appendix~\ref{app:target-count} examines how target count
and case composition contribute to these differences.

\noindent\textbf{Coverage and ranking quality qualify
the reach comparison.}
Qwen3 LoRA also leads macro Recall and nDCG at all three
reported depths on New and Isolated
(Tables~\ref{tab:ar-recall}--\ref{tab:ar-ndcg}).
Nevertheless, on New at depth 100, its 36.17\% case reach
corresponds to 28.70\% average target coverage.
At depth 1,000, these values rise to 58.41\% and 48.81\%,
respectively, leaving substantial gaps in both measures.
The same ordering does not hold for every metric:
the frozen and adapted Qwen3 models tie at the displayed
MRR precision on New, and the frozen model has the highest
displayed MRR on All and Isolated (Table~\ref{tab:ar-mrr}).

\noindent\textbf{Adaptation does not improve retrieval
uniformly.}
Relative to its matched frozen MaxSim counterpart,
Qwen3 LoRA improves Hit at every reported depth on New
and Isolated, but lowers it on All and improves it only
at depth 1,000 on Next.
SPECTER2-Proximity fine-tuning lowers Hit relative to its
matched frozen counterpart at every reported tier and depth.
These comparisons concern point estimates; Appendix~H.3
reports paired Hit@100 intervals and checks whether the
adaptation findings depend on researcher overlap with
training and validation data.

\begin{table}[!htbp]
\centering
\scriptsize
\setlength{\tabcolsep}{2pt}
\renewcommand{\arraystretch}{1.08}
\caption{\textbf{Author-Root Hit@$K$ (\%).}
All rows use the same cases within each tier.
Bold indicates the highest displayed value in each column,
including ties.
Input definitions and method groups follow Appendix~C.
Popularity ranks only Semantic Scholar records, and
citation-graph retrievers rank only graph nodes;
other targets remain in the evaluation and count as misses.}
\label{tab:ar-hit}

\resizebox{\linewidth}{!}{%
\begin{tabular}{@{}llrrrrrrrrrrrr@{}}
\toprule
& & \multicolumn{3}{c}{All} & \multicolumn{3}{c}{Next} & \multicolumn{3}{c}{New} & \multicolumn{3}{c}{Isolated} \\
& & \multicolumn{3}{c}{$N=1{,}633$} & \multicolumn{3}{c}{$N=1{,}330$} & \multicolumn{3}{c}{$N=904$} & \multicolumn{3}{c}{$N=665$} \\
\cmidrule(lr){3-5}
\cmidrule(lr){6-8}
\cmidrule(lr){9-11}
\cmidrule(lr){12-14}
Retriever & Input & @10 & @100 & @1000 & @10 & @100 & @1000 & @10 & @100 & @1000 & @10 & @100 & @1000 \\

\midrule
\multicolumn{14}{@{}l}{\emph{Popularity, lexical and sparse retrieval}} \\
Popularity & ---
& 0.06 & 0.18 & 0.43
& 0.08 & 0.23 & 0.53
& 0.11 & 0.33 & 0.44
& 0.15 & 0.30 & 0.45 \\
BM25 & Compact
& 23.76 & 40.91 & 61.11
& 23.38 & 39.62 & 59.77
& 12.94 & 25.44 & 43.25
& 12.18 & 22.26 & 38.20 \\
BM25 & Uncapped
& 30.68 & 48.44 & 66.32
& 29.85 & 46.99 & 65.19
& 17.48 & 29.31 & 47.90
& 16.09 & 26.92 & 43.31 \\
SPLADE & Compact
& 21.49 & 36.86 & 55.36
& 20.15 & 35.56 & 53.83
& 11.28 & 23.23 & 38.38
& 10.68 & 21.35 & 34.74 \\

\midrule
\multicolumn{14}{@{}l}{\emph{Frozen dense retrievers}} \\
Qwen3-Embedding-8B & Compact
& 24.74 & 47.40 & 71.10
& 24.06 & 46.62 & 70.53
& 13.72 & 29.87 & 56.42
& 12.78 & 26.77 & 52.78 \\
Qwen3-Embedding-8B & K16
& 37.29 & 56.15 & 73.67
& 36.09 & 55.19 & 73.38
& 18.92 & 33.52 & 56.31
& 16.99 & 30.53 & 53.38 \\
\addlinespace[2pt]
GritLM-7B & K1
& 28.84 & 49.79 & 71.28
& 27.44 & 48.87 & 70.75
& 14.82 & 31.08 & 54.65
& 13.83 & 28.57 & 51.58 \\
GritLM-7B & K16
& 39.99 & 57.56 & 74.53
& 38.27 & 57.14 & 73.83
& 20.80 & 34.62 & 56.42
& 18.80 & 31.28 & 52.93 \\
\addlinespace[2pt]
SPECTER2-Proximity & K1
& 17.02 & 35.33 & 59.46
& 16.84 & 34.06 & 58.80
& 8.85 & 19.91 & 44.14
& 8.27 & 18.65 & 40.30 \\
SPECTER2-Proximity & K16
& 34.23 & 51.62 & 71.10
& 33.31 & 50.00 & 70.00
& 18.14 & 29.87 & 52.43
& 16.54 & 28.12 & 46.92 \\
\addlinespace[2pt]
SciNCL & K1
& 15.43 & 35.33 & 61.79
& 14.44 & 33.91 & 60.68
& 6.31 & 19.69 & 46.13
& 6.77 & 18.95 & 42.86 \\
SciNCL & K16
& 28.97 & 46.05 & 66.99
& 28.27 & 45.11 & 66.47
& 15.60 & 25.77 & 47.79
& 14.59 & 23.76 & 44.51 \\

\midrule
\multicolumn{14}{@{}l}{\emph{Frozen token-level late interaction}} \\
LateOn & Title max
& 8.45 & 16.35 & 30.80
& 8.12 & 16.17 & 30.53
& 4.87 & 10.07 & 23.12
& 4.81 & 10.08 & 22.71 \\

\midrule
\multicolumn{14}{@{}l}{\emph{Citation-graph retrievers}} \\
Personalized PageRank & Citations
& 23.88 & 34.23 & 44.83
& 24.59 & 34.51 & 45.41
& 11.62 & 18.03 & 32.30
& 10.53 & 15.79 & 28.12 \\
Co-citation & Citations
& 6.55 & 18.19 & 30.99
& 6.47 & 18.12 & 31.35
& 1.88 & 6.86 & 15.27
& 1.05 & 4.51 & 10.23 \\
Bibliographic coupling & Citations
& 12.25 & 26.45 & 37.66
& 12.11 & 26.77 & 37.74
& 4.31 & 15.60 & 25.77
& 3.31 & 13.38 & 22.41 \\
Adamic--Adar & Citations
& 12.31 & 28.72 & 41.89
& 12.41 & 28.95 & 42.18
& 3.65 & 15.93 & 29.09
& 3.01 & 13.53 & 25.11 \\

\midrule
\multicolumn{14}{@{}l}{\emph{Frozen counterparts of adapted encoders}} \\
SPECTER2-Proximity & MaxSim
& 40.85 & 55.48 & 73.06
& 40.68 & 54.21 & 72.33
& 21.68 & 31.53 & 52.88
& 19.85 & 28.72 & 48.27 \\
Qwen3-Embedding-8B & MaxSim
& \textbf{43.29} & \textbf{58.79} & \textbf{74.89}
& \textbf{42.48} & \textbf{57.89} & 74.06
& 22.35 & 34.51 & 54.98
& 19.85 & 31.43 & 51.88 \\

\midrule
\multicolumn{14}{@{}l}{\emph{Supervised retrievers}} \\
Linear-Proximity & K16
& 25.66 & 42.68 & 65.95
& 25.41 & 41.95 & 66.09
& 14.82 & 24.23 & 49.00
& 13.83 & 22.71 & 45.56 \\
Linear-Proximity + neighbours & K16
& 24.68 & 41.40 & 64.67
& 24.14 & 40.30 & 65.19
& 13.16 & 24.56 & 48.23
& 12.78 & 23.46 & 43.91 \\
Linear-Proximity (control) & K16
& 25.78 & 42.87 & 65.77
& 25.26 & 41.88 & 65.94
& 14.60 & 24.12 & 48.56
& 13.53 & 22.56 & 45.11 \\
\addlinespace[2pt]
SPECTER2-Proximity (fine-tuned) & MaxSim
& 39.50 & 52.24 & 68.46
& 38.65 & 51.13 & 68.35
& 20.13 & 29.65 & 49.00
& 18.35 & 27.52 & 45.41 \\
Qwen3-Embedding-8B (LoRA) & MaxSim
& 42.38 & 57.81 & 74.71
& 42.11 & 57.59 & \textbf{75.41}
& \textbf{23.01} & \textbf{36.17} & \textbf{58.41}
& \textbf{20.90} & \textbf{34.59} & \textbf{55.04} \\

\bottomrule
\end{tabular}%
}
\end{table}

\begin{table}[!htbp]
\centering
\scriptsize
\setlength{\tabcolsep}{2pt}
\renewcommand{\arraystretch}{1.08}
\caption{\textbf{Author-Root macro Recall@$K$ (\%).}
Values average the fraction of eligible targets retrieved
within each case. Cases, input definitions, method groups
and record-coverage restrictions match Table~\ref{tab:ar-hit}.
Bold indicates the highest displayed value in each column,
including ties.}
\label{tab:ar-recall}

\resizebox{\linewidth}{!}{%
\begin{tabular}{@{}llrrrrrrrrrrrr@{}}
\toprule
& & \multicolumn{3}{c}{All} & \multicolumn{3}{c}{Next} & \multicolumn{3}{c}{New} & \multicolumn{3}{c}{Isolated} \\
& & \multicolumn{3}{c}{$N=1{,}633$} & \multicolumn{3}{c}{$N=1{,}330$} & \multicolumn{3}{c}{$N=904$} & \multicolumn{3}{c}{$N=665$} \\
\cmidrule(lr){3-5}
\cmidrule(lr){6-8}
\cmidrule(lr){9-11}
\cmidrule(lr){12-14}
Retriever & Input & @10 & @100 & @1000 & @10 & @100 & @1000 & @10 & @100 & @1000 & @10 & @100 & @1000 \\

\midrule
\multicolumn{14}{@{}l}{\emph{Popularity, lexical and sparse retrieval}} \\
Popularity & ---
& 0.03 & 0.11 & 0.20
& 0.04 & 0.13 & 0.25
& 0.06 & 0.22 & 0.30
& 0.08 & 0.23 & 0.34 \\
BM25 & Compact
& 14.43 & 27.46 & 45.57
& 15.39 & 27.72 & 45.61
& 9.90 & 19.25 & 33.74
& 9.73 & 17.78 & 30.78 \\
BM25 & Uncapped
& 19.47 & 33.51 & 51.61
& 20.27 & 33.71 & 51.72
& 13.32 & 21.93 & 38.14
& 12.51 & 20.76 & 35.91 \\
SPLADE & Compact
& 12.61 & 24.10 & 40.56
& 12.58 & 24.51 & 40.37
& 8.79 & 17.88 & 29.96
& 8.38 & 17.27 & 28.43 \\

\midrule
\multicolumn{14}{@{}l}{\emph{Frozen dense retrievers}} \\
Qwen3-Embedding-8B & Compact
& 15.88 & 33.98 & 58.36
& 16.43 & 34.62 & 58.97
& 10.37 & 23.04 & 47.29
& 9.96 & 21.62 & 45.16 \\
Qwen3-Embedding-8B & K16
& 24.13 & 41.50 & 61.35
& 24.53 & 42.18 & 62.09
& 14.10 & 26.32 & 47.24
& 13.09 & 24.63 & 46.01 \\
\addlinespace[2pt]
GritLM-7B & K1
& 17.56 & 35.11 & 58.67
& 18.20 & 35.72 & 59.11
& 11.03 & 23.41 & 45.43
& 10.61 & 22.88 & 44.15 \\
GritLM-7B & K16
& 26.06 & 42.35 & 61.44
& 26.18 & 43.16 & 61.69
& 15.97 & 26.69 & 47.12
& 14.82 & 25.12 & 45.62 \\
\addlinespace[2pt]
SPECTER2-Proximity & K1
& 9.64 & 22.87 & 44.72
& 10.28 & 23.08 & 45.26
& 6.54 & 14.57 & 34.52
& 6.50 & 14.45 & 32.55 \\
SPECTER2-Proximity & K16
& 21.68 & 36.35 & 56.35
& 22.39 & 36.79 & 56.78
& 13.76 & 23.13 & 42.19
& 13.09 & 22.54 & 39.52 \\
\addlinespace[2pt]
SciNCL & K1
& 8.53 & 22.71 & 47.95
& 8.85 & 22.81 & 48.12
& 4.23 & 13.71 & 37.21
& 4.73 & 14.00 & 35.82 \\
SciNCL & K16
& 17.82 & 31.85 & 52.78
& 18.68 & 32.42 & 52.88
& 11.81 & 19.34 & 37.86
& 11.23 & 18.43 & 36.49 \\

\midrule
\multicolumn{14}{@{}l}{\emph{Frozen token-level late interaction}} \\
LateOn & Title max
& 4.28 & 9.88 & 21.59
& 4.46 & 10.25 & 22.31
& 3.54 & 7.10 & 17.40
& 3.80 & 7.63 & 17.42 \\

\midrule
\multicolumn{14}{@{}l}{\emph{Citation-graph retrievers}} \\
Personalized PageRank & Citations
& 13.71 & 20.66 & 27.95
& 15.09 & 22.16 & 29.93
& 8.67 & 13.09 & 23.66
& 8.00 & 11.89 & 21.52 \\
Co-citation & Citations
& 3.18 & 10.36 & 18.03
& 3.37 & 11.11 & 19.35
& 1.02 & 4.64 & 10.34
& 0.56 & 3.19 & 7.59 \\
Bibliographic coupling & Citations
& 6.55 & 15.22 & 22.81
& 6.98 & 16.20 & 24.26
& 2.94 & 10.88 & 17.97
& 2.39 & 9.49 & 16.45 \\
Adamic--Adar & Citations
& 6.60 & 16.93 & 25.85
& 7.18 & 18.13 & 27.44
& 2.46 & 11.03 & 20.85
& 2.01 & 9.74 & 18.65 \\

\midrule
\multicolumn{14}{@{}l}{\emph{Frozen counterparts of adapted encoders}} \\
SPECTER2-Proximity & MaxSim
& 26.45 & 39.77 & 57.76
& 27.63 & 40.29 & 58.21
& 16.36 & 24.17 & 42.30
& 15.87 & 22.85 & 40.10 \\
Qwen3-Embedding-8B & MaxSim
& \textbf{28.28} & \textbf{43.30} & \textbf{62.44}
& \textbf{29.20} & \textbf{43.74} & \textbf{62.48}
& 16.93 & 27.12 & 46.21
& 15.73 & 25.47 & 44.28 \\

\midrule
\multicolumn{14}{@{}l}{\emph{Supervised retrievers}} \\
Linear-Proximity & K16
& 15.61 & 29.12 & 51.51
& 16.75 & 29.87 & 52.82
& 10.33 & 18.53 & 38.92
& 10.33 & 18.05 & 38.17 \\
Linear-Proximity + neighbours & K16
& 14.92 & 27.93 & 49.74
& 15.55 & 28.76 & 51.17
& 9.39 & 18.80 & 38.10
& 9.51 & 18.58 & 36.25 \\
Linear-Proximity (control) & K16
& 15.69 & 29.28 & 51.50
& 16.65 & 29.94 & 52.81
& 10.22 & 18.50 & 38.69
& 10.10 & 18.05 & 37.94 \\
\addlinespace[2pt]
SPECTER2-Proximity (fine-tuned) & MaxSim
& 25.43 & 36.64 & 53.54
& 26.49 & 37.23 & 54.50
& 15.12 & 22.27 & 39.16
& 14.32 & 21.61 & 37.72 \\
Qwen3-Embedding-8B (LoRA) & MaxSim
& 27.67 & 41.78 & 60.98
& 28.84 & 43.48 & 62.45
& \textbf{17.37} & \textbf{28.70} & \textbf{48.81}
& \textbf{16.54} & \textbf{27.93} & \textbf{46.74} \\

\bottomrule
\end{tabular}%
}
\end{table}

\begin{table}[!htbp]
\centering
\scriptsize
\setlength{\tabcolsep}{2pt}
\renewcommand{\arraystretch}{1.08}
\caption{\textbf{Author-Root nDCG@$K$ (\%).}
Cases, input definitions, method groups and record-coverage
restrictions match Table~\ref{tab:ar-hit}.
Bold indicates the highest displayed value in each column,
including ties.}
\label{tab:ar-ndcg}

\resizebox{\linewidth}{!}{%
\begin{tabular}{@{}llrrrrrrrrrrrr@{}}
\toprule
& & \multicolumn{3}{c}{All} & \multicolumn{3}{c}{Next} & \multicolumn{3}{c}{New} & \multicolumn{3}{c}{Isolated} \\
& & \multicolumn{3}{c}{$N=1{,}633$} & \multicolumn{3}{c}{$N=1{,}330$} & \multicolumn{3}{c}{$N=904$} & \multicolumn{3}{c}{$N=665$} \\
\cmidrule(lr){3-5}
\cmidrule(lr){6-8}
\cmidrule(lr){9-11}
\cmidrule(lr){12-14}
Retriever & Input & @10 & @100 & @1000 & @10 & @100 & @1000 & @10 & @100 & @1000 & @10 & @100 & @1000 \\

\midrule
\multicolumn{14}{@{}l}{\emph{Popularity, lexical and sparse retrieval}} \\
Popularity & ---
& 0.02 & 0.03 & 0.05
& 0.02 & 0.04 & 0.06
& 0.03 & 0.06 & 0.07
& 0.04 & 0.06 & 0.08 \\
BM25 & Compact
& 11.49 & 14.70 & 17.47
& 11.85 & 14.86 & 17.55
& 7.32 & 9.47 & 11.52
& 6.92 & 8.74 & 10.54 \\
BM25 & Uncapped
& 16.29 & 19.76 & 22.54
& 16.43 & 19.73 & 22.44
& 9.64 & 11.68 & 13.97
& 9.23 & 11.11 & 13.16 \\
SPLADE & Compact
& 10.65 & 13.46 & 15.97
& 10.67 & 13.53 & 15.90
& 7.54 & 9.66 & 11.40
& 6.86 & 8.86 & 10.39 \\

\midrule
\multicolumn{14}{@{}l}{\emph{Frozen dense retrievers}} \\
Qwen3-Embedding-8B & Compact
& 12.34 & 16.86 & 20.57
& 12.50 & 16.96 & 20.58
& 7.63 & 10.50 & 13.81
& 7.29 & 9.84 & 12.93 \\
Qwen3-Embedding-8B & K16
& 17.71 & 22.07 & 25.14
& 17.60 & 21.95 & 24.97
& 10.38 & 13.11 & 16.05
& 9.95 & 12.49 & 15.36 \\
\addlinespace[2pt]
GritLM-7B & K1
& 13.36 & 17.68 & 21.26
& 13.38 & 17.62 & 21.13
& 7.38 & 10.17 & 13.24
& 7.50 & 10.11 & 12.96 \\
GritLM-7B & K16
& 18.79 & 22.83 & 25.80
& 18.63 & 22.77 & 25.61
& 11.27 & 13.70 & 16.57
& 10.70 & 12.96 & 15.69 \\
\addlinespace[2pt]
SPECTER2-Proximity & K1
& 6.90 & 10.17 & 13.52
& 7.18 & 10.27 & 13.57
& 4.42 & 6.23 & 8.99
& 4.46 & 6.19 & 8.61 \\
SPECTER2-Proximity & K16
& 16.37 & 20.01 & 23.06
& 16.44 & 19.97 & 22.96
& 9.89 & 11.97 & 14.59
& 9.69 & 11.74 & 13.99 \\
\addlinespace[2pt]
SciNCL & K1
& 5.79 & 9.26 & 13.06
& 5.77 & 9.15 & 12.86
& 2.70 & 4.94 & 8.11
& 3.01 & 5.11 & 8.00 \\
SciNCL & K16
& 13.86 & 17.28 & 20.50
& 14.13 & 17.45 & 20.53
& 8.78 & 10.49 & 13.10
& 8.57 & 10.15 & 12.58 \\

\midrule
\multicolumn{14}{@{}l}{\emph{Frozen token-level late interaction}} \\
LateOn & Title max
& 3.48 & 4.85 & 6.60
& 3.49 & 4.88 & 6.64
& 2.52 & 3.38 & 4.80
& 2.78 & 3.66 & 4.99 \\

\midrule
\multicolumn{14}{@{}l}{\emph{Citation-graph retrievers}} \\
Personalized PageRank & Citations
& 10.84 & 12.64 & 13.76
& 11.70 & 13.50 & 14.68
& 6.15 & 7.16 & 8.60
& 5.87 & 6.71 & 7.99 \\
Co-citation & Citations
& 2.24 & 3.98 & 5.17
& 2.33 & 4.17 & 5.43
& 0.75 & 1.51 & 2.31
& 0.40 & 0.98 & 1.58 \\
Bibliographic coupling & Citations
& 4.39 & 6.52 & 7.70
& 4.64 & 6.83 & 8.06
& 1.77 & 3.44 & 4.46
& 1.31 & 2.83 & 3.80 \\
Adamic--Adar & Citations
& 4.34 & 6.88 & 8.27
& 4.65 & 7.27 & 8.70
& 1.49 & 3.33 & 4.71
& 1.15 & 2.82 & 4.03 \\

\midrule
\multicolumn{14}{@{}l}{\emph{Frozen counterparts of adapted encoders}} \\
SPECTER2-Proximity & MaxSim
& 19.08 & 22.59 & 25.32
& 19.76 & 23.02 & 25.68
& 11.85 & 13.66 & 16.17
& 11.33 & 12.90 & 15.19 \\
Qwen3-Embedding-8B & MaxSim
& \textbf{21.55} & \textbf{25.39} & \textbf{28.37}
& 21.53 & \textbf{25.18} & \textbf{28.04}
& 12.64 & 14.95 & 17.67
& 11.89 & 14.04 & 16.60 \\

\midrule
\multicolumn{14}{@{}l}{\emph{Supervised retrievers}} \\
Linear-Proximity & K16
& 12.08 & 15.47 & 18.81
& 12.59 & 15.83 & 19.18
& 7.72 & 9.64 & 12.43
& 7.81 & 9.57 & 12.25 \\
Linear-Proximity + neighbours & K16
& 12.40 & 15.68 & 18.95
& 12.66 & 15.95 & 19.25
& 7.68 & 9.80 & 12.43
& 7.81 & 9.87 & 12.23 \\
Linear-Proximity (control) & K16
& 12.16 & 15.55 & 18.86
& 12.62 & 15.87 & 19.21
& 7.79 & 9.72 & 12.48
& 7.83 & 9.65 & 12.30 \\
\addlinespace[2pt]
SPECTER2-Proximity (fine-tuned) & MaxSim
& 18.40 & 21.32 & 23.89
& 18.72 & 21.47 & 24.03
& 11.06 & 12.67 & 15.03
& 10.74 & 12.34 & 14.48 \\
Qwen3-Embedding-8B (LoRA) & MaxSim
& 20.84 & 24.43 & 27.42
& \textbf{21.54} & 25.15 & 28.03
& \textbf{12.73} & \textbf{15.27} & \textbf{18.12}
& \textbf{12.02} & \textbf{14.51} & \textbf{17.09} \\

\bottomrule
\end{tabular}%
}
\end{table}

\begin{table}[!htbp]
\centering
\scriptsize
\setlength{\tabcolsep}{2pt}
\renewcommand{\arraystretch}{1.08}
\caption{\textbf{Author-Root MRR@1000 (0--1 scale).}
Cases, input definitions, method groups and record-coverage
restrictions match Table~\ref{tab:ar-hit}.
Bold indicates the highest displayed value in each column,
including ties.}
\label{tab:ar-mrr}

\resizebox{\linewidth}{!}{%
\begin{tabular}{@{}llrrrr@{}}
\toprule
& & \multicolumn{1}{c}{All} & \multicolumn{1}{c}{Next} & \multicolumn{1}{c}{New} & \multicolumn{1}{c}{Isolated} \\
& & \multicolumn{1}{c}{$N=1{,}633$} & \multicolumn{1}{c}{$N=1{,}330$} & \multicolumn{1}{c}{$N=904$} & \multicolumn{1}{c}{$N=665$} \\
\cmidrule(lr){3-3}
\cmidrule(lr){4-4}
\cmidrule(lr){5-5}
\cmidrule(lr){6-6}
Retriever & Input & @1000 & @1000 & @1000 & @1000 \\

\midrule
\multicolumn{6}{@{}l}{\emph{Popularity, lexical and sparse retrieval}} \\
Popularity & ---
& 0.000
& 0.000
& 0.000
& 0.000 \\
BM25 & Compact
& 0.154
& 0.149
& 0.082
& 0.075 \\
BM25 & Uncapped
& 0.212
& 0.202
& 0.103
& 0.098 \\
SPLADE & Compact
& 0.146
& 0.140
& 0.089
& 0.079 \\

\midrule
\multicolumn{6}{@{}l}{\emph{Frozen dense retrievers}} \\
Qwen3-Embedding-8B & Compact
& 0.162
& 0.156
& 0.090
& 0.083 \\
Qwen3-Embedding-8B & K16
& 0.213
& 0.204
& 0.117
& 0.113 \\
\addlinespace[2pt]
GritLM-7B & K1
& 0.171
& 0.161
& 0.080
& 0.083 \\
GritLM-7B & K16
& 0.219
& 0.210
& 0.122
& 0.116 \\
\addlinespace[2pt]
SPECTER2-Proximity & K1
& 0.092
& 0.090
& 0.050
& 0.048 \\
SPECTER2-Proximity & K16
& 0.202
& 0.193
& 0.109
& 0.107 \\
\addlinespace[2pt]
SciNCL & K1
& 0.079
& 0.073
& 0.033
& 0.036 \\
SciNCL & K16
& 0.175
& 0.169
& 0.097
& 0.096 \\

\midrule
\multicolumn{6}{@{}l}{\emph{Frozen token-level late interaction}} \\
LateOn & Title max
& 0.051
& 0.048
& 0.028
& 0.030 \\

\midrule
\multicolumn{6}{@{}l}{\emph{Citation-graph retrievers}} \\
Personalized PageRank & Citations
& 0.143
& 0.146
& 0.066
& 0.064 \\
Co-citation & Citations
& 0.033
& 0.032
& 0.010
& 0.007 \\
Bibliographic coupling & Citations
& 0.059
& 0.059
& 0.021
& 0.016 \\
Adamic--Adar & Citations
& 0.057
& 0.058
& 0.019
& 0.015 \\

\midrule
\multicolumn{6}{@{}l}{\emph{Frozen counterparts of adapted encoders}} \\
SPECTER2-Proximity & MaxSim
& 0.224
& 0.222
& 0.128
& 0.119 \\
Qwen3-Embedding-8B & MaxSim
& \textbf{0.257}
& 0.245
& \textbf{0.140}
& \textbf{0.131} \\

\midrule
\multicolumn{6}{@{}l}{\emph{Supervised retrievers}} \\
Linear-Proximity & K16
& 0.154
& 0.151
& 0.089
& 0.088 \\
Linear-Proximity + neighbours & K16
& 0.168
& 0.164
& 0.096
& 0.096 \\
Linear-Proximity (control) & K16
& 0.156
& 0.152
& 0.090
& 0.090 \\
\addlinespace[2pt]
SPECTER2-Proximity (fine-tuned) & MaxSim
& 0.220
& 0.212
& 0.122
& 0.118 \\
Qwen3-Embedding-8B (LoRA) & MaxSim
& 0.251
& \textbf{0.248}
& \textbf{0.140}
& 0.130 \\

\bottomrule
\end{tabular}%
}
\end{table}

\FloatBarrier

\begin{table}[!htbp]
\centering
\small
\setlength{\tabcolsep}{4pt}
\renewcommand{\arraystretch}{1.08}
\caption{\textbf{Author-Root retrieval by field.}
SPECTER2-Proximity K16 Hit@100 (\%), with 95\%
researcher-cluster percentile bootstrap intervals
(1,000 replicates). Fields follow Appendix~D.1.
Three unclassified All cases, including one New case,
are included in the overall results but omitted from
the field rows. All and New have different eligible
case and target sets.}
\label{tab:ar-field}
\resizebox{\linewidth}{!}{%
\begin{tabular}{@{}lrlrl@{}}
\toprule
& \multicolumn{2}{c}{All} & \multicolumn{2}{c}{New} \\
\cmidrule(lr){2-3}\cmidrule(lr){4-5}
Field & $N$ & Hit@100 [95\% CI] & $N$ & Hit@100 [95\% CI] \\
\midrule
All cases & 1,633 & 51.6 [49.4, 54.2] & 904 & 29.9 [26.8, 32.8] \\
\midrule
Life sciences & 901 & 56.4 [52.9, 59.6] & 457 & 35.2 [30.9, 39.4] \\
Health and medicine & 172 & 49.4 [42.1, 56.7] & 100 & 22.0 [14.1, 30.3] \\
Physical sciences & 337 & 46.9 [41.5, 52.2] & 198 & 23.7 [18.2, 29.8] \\
Social sciences and humanities & 155 & 44.5 [36.8, 52.3] & 104 & 30.8 [22.1, 40.4] \\
Computer science and mathematics & 65 & 32.3 [21.5, 44.6] & 44 & 15.9 [6.8, 27.3] \\
\bottomrule
\end{tabular}%
}
\end{table}

\FloatBarrier
\noindent\textbf{Results across fields.}
SPECTER2-Proximity K16 Hit@100 varies across fields
and is lower on New than All in every group
(Table~\ref{tab:ar-field}).
The CS and mathematics estimates have wide intervals,
with only 65 All and 44 New cases.
These comparisons describe the evaluated cases;
they do not establish differences caused by field.

\FloatBarrier
\noindent\textbf{Record coverage.}
Popularity and citation-graph scores reflect both
record coverage and ranking performance.
Eligible targets that these methods cannot rank
remain in the evaluation and count as misses.

\FloatBarrier
\noindent\textbf{Training overlap.}
The Author-Root panel includes researchers present in
the encoder-adaptation training split.
Excluding identified training and validation overlap
preserves the directions and interval conclusions
of the adapted-versus-frozen Hit@100 comparisons
(Appendix~H.3).
\FloatBarrier

\subsection{Patterns across evidence sources}
No single configuration leads throughout these evaluations.
At depth 100 on the New tiers, GritLM-7B K1 has the highest
observed Hit on Bibliography, Personalized PageRank on
Context-Root, and Qwen3 LoRA on Author-Root
(Tables~\ref{tab:muses-bib-hit},
\ref{tab:muses-context-hit} and~\ref{tab:ar-hit}).
Adaptation likewise does not uniformly improve on matched
frozen counterparts.
Despite these changes in method rankings, retrieval beyond
the researcher's prior citations remains incomplete across
all three sources.
These comparisons describe the respective evaluations;
differences in cases, targets and candidate pools prevent
attributing the ranking changes to evidence type alone.
\FloatBarrier
\section{History and project-information diagnostics}

This section examines whether similar researchers'
histories can substitute for a researcher's own history,
and how retrieval changes when information about the
next paper is available.

\subsection{Matched histories of other researchers}

To test whether another researcher's publication history
can substitute for the focal researcher's, we compare
retrieval using the researcher's own history with
retrieval using three content-matched histories from
other researchers. Within each case, the candidate pool,
temporal cutoff and targets remain fixed, and retrieval
performance is averaged over the three matched histories.

\noindent\textbf{Matching procedure.}
Matched researchers share the focal year and
publication-history size band, have distinct researcher
identifiers, and share no earlier works with the focal
researcher.
History size is the number $n$ of distinct authored papers
published before the case cutoff, grouped into bands
$\lfloor\log_2(1+n)\rfloor$.
For Bibliography and Context-Root, matching additionally
uses the recorded primary field of $P$, with missing
fields forming a separate stratum.

Author-Root draws donors from the responding-researcher
panel without an explicit field restriction.
We select the three eligible researchers with the
highest cosine similarity between normalized mean
history embeddings, using papers published before
1 January of the focal year.
Retrieval histories use papers published before the
focal case's exact cutoff.
For Author-Root, a matched researcher's publication
history must be available through that cutoff.

\noindent\textbf{Evaluation.}
We evaluate SPECTER2-Proximity and GritLM with 16 history
centroids, and BM25 with the compact query, across the
Bibliography, Context-Root and Author-Root tiers.
Within each tier, a case must have at least one eligible
target and rankings for the researcher's own history and
all three matched histories; the same cases are used for
all three retrievers.
The primary comparison is the per-case difference in
nDCG@1000 between the researcher's own history and the mean
over the matched histories.
Hit@100 provides a supporting comparison.
We report 95\% confidence intervals from 10,000 bootstrap
replicates resampling both focal and matched researchers.

\begin{table*}[!htbp]
\centering
\footnotesize
\setlength{\tabcolsep}{1.5pt}
\renewcommand{\arraystretch}{1.12}

\caption{\textbf{Own-history versus content-matched retrieval.}
Values are retrieval scores using the researcher's own history
minus the mean score using three matched researchers' histories,
with 95\% confidence intervals below.
Positive values favour the researcher's own history.
Differences in nDCG@1000 use the 0--1 scale;
Hit@100 differences are percentage points.
Within each tier, all retrievers and both metrics use the same
cases. BM25 uses the compact query; the dense retrievers use
16 history centroids. Intervals account for dependence through
both focal and matched researchers.}
\label{tab:matched-history-cross-source}

\begingroup
\newcommand{\mhvalue}[2]{%
  \shortstack{$#1$\\[1pt]{\scriptsize[$#2$]}}}

\begin{tabular*}{\linewidth}
{@{\extracolsep{\fill}}lccc@{\hspace{8pt}}ccc@{}}
\toprule
\multicolumn{7}{@{}l}{\textbf{A. Bibliography}} \\
\multicolumn{7}{@{}l}{%
\scriptsize
$N$: CiteNext $130{,}937$;
CiteNew $130{,}171$;
CiteNew-Isolated $129{,}063$} \\
\addlinespace[5pt]
& \multicolumn{3}{c}{\textbf{nDCG@1000 advantage}}
& \multicolumn{3}{c}{\textbf{Hit@100 advantage (pp)}} \\
\cmidrule(lr){2-4}\cmidrule(lr){5-7}
Retriever
& Next & New & Isolated
& Next & New & Isolated \\
\midrule
BM25
& \mhvalue{0.094}{0.093,\,0.095}
& \mhvalue{0.074}{0.073,\,0.075}
& \mhvalue{0.068}{0.067,\,0.069}
& \mhvalue{28.4}{28.0,\,28.7}
& \mhvalue{24.7}{24.4,\,25.1}
& \mhvalue{22.4}{22.1,\,22.7} \\
\addlinespace[5pt]
GritLM-7B
& \mhvalue{0.129}{0.127,\,0.130}
& \mhvalue{0.098}{0.096,\,0.099}
& \mhvalue{0.089}{0.088,\,0.090}
& \mhvalue{34.6}{34.3,\,35.0}
& \mhvalue{30.4}{30.1,\,30.8}
& \mhvalue{27.7}{27.4,\,28.0} \\
\addlinespace[5pt]
SPECTER2-Proximity
& \mhvalue{0.111}{0.110,\,0.113}
& \mhvalue{0.084}{0.083,\,0.085}
& \mhvalue{0.076}{0.075,\,0.078}
& \mhvalue{31.4}{31.1,\,31.8}
& \mhvalue{27.0}{26.7,\,27.4}
& \mhvalue{24.4}{24.1,\,24.8} \\
\bottomrule
\end{tabular*}

\par\vspace{10pt}

\begin{tabular*}{\linewidth}
{@{\extracolsep{\fill}}lccc@{\hspace{8pt}}ccc@{}}
\toprule
\multicolumn{7}{@{}l}{\textbf{B. Context-Root}} \\
\multicolumn{7}{@{}l}{%
\scriptsize
$N$: Next $7{,}639$;
New $6{,}469$;
Isolated $5{,}505$} \\
\addlinespace[5pt]
& \multicolumn{3}{c}{\textbf{nDCG@1000 advantage}}
& \multicolumn{3}{c}{\textbf{Hit@100 advantage (pp)}} \\
\cmidrule(lr){2-4}\cmidrule(lr){5-7}
Retriever
& Next & New & Isolated
& Next & New & Isolated \\
\midrule
BM25
& \mhvalue{0.086}{0.080,\,0.093}
& \mhvalue{0.064}{0.058,\,0.070}
& \mhvalue{0.063}{0.056,\,0.070}
& \mhvalue{16.9}{15.8,\,18.0}
& \mhvalue{11.3}{10.3,\,12.4}
& \mhvalue{10.7}{9.6,\,11.8} \\
\addlinespace[5pt]
GritLM-7B
& \mhvalue{0.123}{0.116,\,0.130}
& \mhvalue{0.085}{0.079,\,0.092}
& \mhvalue{0.083}{0.076,\,0.090}
& \mhvalue{23.8}{22.5,\,25.1}
& \mhvalue{15.6}{14.5,\,16.9}
& \mhvalue{14.5}{13.2,\,15.7} \\
\addlinespace[5pt]
SPECTER2-Proximity
& \mhvalue{0.110}{0.103,\,0.117}
& \mhvalue{0.078}{0.072,\,0.085}
& \mhvalue{0.076}{0.069,\,0.084}
& \mhvalue{20.9}{19.7,\,22.2}
& \mhvalue{13.7}{12.6,\,14.9}
& \mhvalue{12.6}{11.5,\,13.8} \\
\bottomrule
\end{tabular*}

\par\vspace{10pt}

\setlength{\tabcolsep}{1pt}
\begin{tabular*}{\linewidth}
{@{\extracolsep{\fill}}lcccc@{\hspace{6pt}}cccc@{}}
\toprule
\multicolumn{9}{@{}l}{\textbf{C. Author-Root}} \\
\multicolumn{9}{@{}l}{%
\scriptsize
$N$: All $1{,}518$;
Next $1{,}235$;
New $830$;
Isolated $606$} \\
\addlinespace[5pt]
& \multicolumn{4}{c}{\textbf{nDCG@1000 advantage}}
& \multicolumn{4}{c}{\textbf{Hit@100 advantage (pp)}} \\
\cmidrule(lr){2-5}\cmidrule(lr){6-9}
Retriever
& All & Next & New & Isolated
& All & Next & New & Isolated \\
\midrule
BM25
& \mhvalue{0.173}{0.158,\,0.190}
& \mhvalue{0.173}{0.156,\,0.192}
& \mhvalue{0.111}{0.092,\,0.132}
& \mhvalue{0.101}{0.080,\,0.123}
& \mhvalue{40.9}{37.8,\,44.2}
& \mhvalue{39.6}{36.1,\,43.2}
& \mhvalue{25.1}{21.4,\,29.1}
& \mhvalue{21.7}{17.5,\,26.0} \\
\addlinespace[5pt]
GritLM-7B
& \mhvalue{0.258}{0.240,\,0.276}
& \mhvalue{0.256}{0.236,\,0.276}
& \mhvalue{0.164}{0.143,\,0.186}
& \mhvalue{0.155}{0.130,\,0.182}
& \mhvalue{58.0}{54.8,\,61.2}
& \mhvalue{57.6}{54.1,\,61.1}
& \mhvalue{34.5}{30.3,\,38.8}
& \mhvalue{31.1}{26.5,\,36.0} \\
\addlinespace[5pt]
\shortstack[l]{SPECTER2-\\Proximity}
& \mhvalue{0.231}{0.214,\,0.249}
& \mhvalue{0.230}{0.211,\,0.251}
& \mhvalue{0.144}{0.124,\,0.166}
& \mhvalue{0.138}{0.114,\,0.165}
& \mhvalue{51.9}{48.6,\,55.2}
& \mhvalue{50.3}{46.7,\,54.0}
& \mhvalue{29.9}{25.8,\,34.1}
& \mhvalue{28.1}{23.4,\,32.9} \\
\bottomrule
\end{tabular*}

\endgroup
\end{table*}

\noindent\textbf{Results.}
The researcher's own history outperforms content-matched
histories for all three retrievers across all three
evidence sources and their tiers
(Table~\ref{tab:matched-history-cross-source}).
On CiteNew, the nDCG@1000 advantage ranges from
0.074 to 0.098; on Context-Root-New, from 0.064 to 0.085;
and on Author-Root-New, from 0.111 to 0.164.
All reported 95\% confidence intervals exclude zero
for both nDCG@1000 and Hit@100.
The BM25 results show that this researcher-specific
signal is not confined to dense retrievers.

\subsection{Hindsight diagnostic}

The hindsight test examines how much information about the
next project adds to publication history alone.
$H$ uses the researcher's history; $H+T$ adds the next paper's
title; and $H+T+A$ uses its title and abstract alongside the
same history. Within each retriever, the history representation
and candidate pool remain fixed. We evaluate dense and lexical
retrieval across bibliography, citation-context, and
author-attested targets. The next paper's text is unavailable
at the focal cutoff, so this is a hindsight test rather than
a prospective retrieval method.

\noindent\textbf{Query construction.}
For SPECTER2-Proximity and GritLM, we retain the history
representation described in Appendix~C.
$H+T$ adds one vector encoding the next paper's title;
$H+T+A$ replaces it with a joint title-and-abstract vector.
A candidate's score is its highest similarity to any history
or added vector. For BM25, $H$ uses the compact history query.
$H+T$ appends the next paper's title, and $H+T+A$ appends its
title and abstract, without shortening the history query.
The candidate pool and temporal eligibility rules remain
unchanged within each comparison.

\noindent\textbf{Evaluation.}
All three inputs use the same cases within each source
and tier.
Bibliography and Context-Root use their full evaluation
cohorts; Author-Root uses cases with an available
next-paper abstract.
We also evaluate the Author-Root title comparison on
the full cohort.
We evaluate Hit@100, Hit@1000 and nDCG@1000,
with paired 95\% confidence intervals from
10,000 bootstrap resamples of researchers.

\begin{table}[p]
\centering
\small
\caption{\textbf{Retrieval with information from the next paper.}
Values are Hit@$K$ (\%).
The dense retrievers use 16 history centroids.
$H$ uses publication history alone; $H+T$ adds the next
paper's title; $H+T+A$ adds its title and abstract.
All three inputs use the same cases within each tier.
Author-Root includes only cases with an available
next-paper abstract, including for $H$ and $H+T$,
and uses its own candidate pool.}
\label{tab:hindsight-xretriever-levels}
\setlength{\tabcolsep}{3pt}
\renewcommand{\arraystretch}{1.12}

\begin{tabular*}{\linewidth}
{@{\extracolsep{\fill}}p{94pt}p{44pt}rrrrrr@{}}
\toprule
\multicolumn{8}{@{}l}{\textbf{A. Bibliography}} \\
\addlinespace[2pt]
& & \multicolumn{2}{c}{CiteNext}
  & \multicolumn{2}{c}{CiteNew}
  & \multicolumn{2}{c}{CiteNew-Isolated} \\
& & \multicolumn{2}{c}{$N=135{,}111$}
  & \multicolumn{2}{c}{$N=134{,}314$}
  & \multicolumn{2}{c}{$N=133{,}166$} \\
\cmidrule(lr){3-4}
\cmidrule(lr){5-6}
\cmidrule(lr){7-8}
Retriever & Input
& @100 & @1000 & @100 & @1000 & @100 & @1000 \\
\midrule
SPECTER2-Proximity & $H$
& 37.4 & 54.7 & 32.2 & 51.8 & 28.9 & 48.7 \\
& $H+T$
& 57.7 & 80.1 & 53.6 & 78.4 & 50.4 & 76.3 \\
& $H+T+A$
& 68.8 & 86.2 & 65.4 & 85.0 & 62.6 & 83.4 \\
\addlinespace[6pt]
GritLM-7B & $H$
& 40.4 & 58.9 & 35.5 & 56.4 & 32.2 & 53.7 \\
& $H+T$
& 65.7 & 87.1 & 62.2 & 85.9 & 59.2 & 84.5 \\
& $H+T+A$
& 77.6 & 91.9 & 75.1 & 91.0 & 72.7 & 90.0 \\
\addlinespace[6pt]
BM25 & $H$
& 34.0 & 51.6 & 29.7 & 48.8 & 26.8 & 45.9 \\
& $H+T$
& 44.1 & 66.1 & 40.2 & 64.1 & 37.1 & 61.7 \\
& $H+T+A$
& 79.3 & 89.9 & 77.4 & 89.2 & 75.3 & 88.2 \\
\bottomrule
\end{tabular*}

\par\vspace{12pt}

\begin{tabular*}{\linewidth}
{@{\extracolsep{\fill}}p{94pt}p{44pt}rrrrrr@{}}
\toprule
\multicolumn{8}{@{}l}{\textbf{B. Context-Root}} \\
\addlinespace[2pt]
& & \multicolumn{2}{c}{Next}
  & \multicolumn{2}{c}{New}
  & \multicolumn{2}{c}{Isolated} \\
& & \multicolumn{2}{c}{$N=7{,}848$}
  & \multicolumn{2}{c}{$N=6{,}659$}
  & \multicolumn{2}{c}{$N=5{,}666$} \\
\cmidrule(lr){3-4}
\cmidrule(lr){5-6}
\cmidrule(lr){7-8}
Retriever & Input
& @100 & @1000 & @100 & @1000 & @100 & @1000 \\
\midrule
SPECTER2-Proximity & $H$
& 22.8 & 37.4 & 15.2 & 28.9 & 14.0 & 26.8 \\
& $H+T$
& 31.2 & 53.2 & 24.9 & 47.3 & 24.2 & 46.5 \\
& $H+T+A$
& 40.8 & 62.2 & 35.4 & 57.5 & 35.0 & 57.0 \\
\addlinespace[6pt]
GritLM-7B & $H$
& 25.7 & 41.7 & 17.0 & 32.8 & 15.8 & 30.5 \\
& $H+T$
& 38.2 & 61.7 & 31.3 & 56.4 & 30.9 & 55.3 \\
& $H+T+A$
& 50.2 & 70.8 & 45.5 & 67.4 & 45.3 & 67.1 \\
\addlinespace[6pt]
BM25 & $H$
& 18.3 & 30.5 & 12.4 & 23.5 & 11.7 & 21.8 \\
& $H+T$
& 23.0 & 38.9 & 17.0 & 32.5 & 16.1 & 30.5 \\
& $H+T+A$
& 45.9 & 63.6 & 41.9 & 60.6 & 41.1 & 60.3 \\
\bottomrule
\end{tabular*}

\par\vspace{12pt}

\begin{tabular*}{\linewidth}
{@{\extracolsep{\fill}}p{94pt}p{44pt}rrrrrrrr@{}}
\toprule
\multicolumn{10}{@{}l}{\textbf{C. Author-Root}} \\
\addlinespace[2pt]
& & \multicolumn{2}{c}{All}
  & \multicolumn{2}{c}{Next}
  & \multicolumn{2}{c}{New}
  & \multicolumn{2}{c}{Isolated} \\
& & \multicolumn{2}{c}{$N=1{,}173$}
  & \multicolumn{2}{c}{$N=1{,}033$}
  & \multicolumn{2}{c}{$N=700$}
  & \multicolumn{2}{c}{$N=518$} \\
\cmidrule(lr){3-4}
\cmidrule(lr){5-6}
\cmidrule(lr){7-8}
\cmidrule(lr){9-10}
Retriever & Input
& @100 & @1000 & @100 & @1000
& @100 & @1000 & @100 & @1000 \\
\midrule
SPECTER2-Proximity & $H$
& 52.6 & 72.0 & 50.9 & 70.8
& 29.7 & 52.6 & 28.4 & 47.7 \\
& $H+T$
& 57.8 & 78.9 & 55.8 & 77.3
& 36.9 & 62.3 & 36.9 & 59.5 \\
& $H+T+A$
& 64.8 & 83.8 & 63.7 & 82.6
& 47.0 & 70.4 & 47.1 & 68.5 \\
\addlinespace[6pt]
GritLM-7B & $H$
& 58.7 & 74.6 & 58.0 & 74.2
& 35.7 & 56.3 & 31.9 & 53.5 \\
& $H+T$
& 64.8 & 83.0 & 63.8 & 82.3
& 43.4 & 67.9 & 40.5 & 66.6 \\
& $H+T+A$
& 71.9 & 88.3 & 71.2 & 87.9
& 55.0 & 77.6 & 52.9 & 76.4 \\
\addlinespace[6pt]
BM25 & $H$
& 42.2 & 61.7 & 41.4 & 61.1
& 27.0 & 44.4 & 23.6 & 39.8 \\
& $H+T$
& 52.7 & 70.8 & 52.2 & 70.8
& 37.1 & 56.0 & 32.6 & 52.1 \\
& $H+T+A$
& 79.5 & 89.8 & 79.2 & 89.8
& 68.4 & 83.1 & 63.7 & 80.7 \\
\bottomrule
\end{tabular*}
\end{table}

\noindent\textbf{Results.}
Adding the next paper's title improves Hit@100,
Hit@1000 and nDCG@1000 for all three retrievers across
every reported tier of Bibliography, Context-Root and
Author-Root.
Adding the abstract provides a further improvement
throughout, with all paired 95\% confidence intervals
for both gains excluding zero.
On the abstract-available Author-Root cases, Hit@100
title gains range from 4.84 to 10.75 percentage points
across retrievers and tiers; the additional abstract
gains range from 6.99 to 31.29 points.
Title gains also remain positive on the full Author-Root
cohorts, with all intervals excluding zero.
Across the three evidence sources, retrievers and tiers,
the corresponding nDCG@1000 gains range from
0.016 to 0.082 for the title and from 0.030 to 0.202
for the additional abstract, on the 0--1 scale.

\noindent\textbf{Sensitivity checks.}
Excluding papers with detected target mentions in their
titles or abstracts leaves both Hit@100 and Hit@1000
gains positive across all three retrievers and evidence
sources.
On Author-Root, every confidence interval remains above
zero and the gains change by at most 0.31 percentage
points.
Incomplete bibliographic metadata limits mention detection,
so this check does not exclude every possible target mention.

For SPECTER2-Proximity and GritLM on Bibliography and
Context-Root, we also substitute another researcher's
next-paper title, matched by field and year.
These substituted titles slightly reduce retrieval,
rather than reproducing the gains from the actual title.
\FloatBarrier
\FloatBarrier
\section{Retrieval complementarity and case characteristics}
\suppressfloats[t]

This section examines complementary recovery across
retrievers and how retrieval varies with the evidence
source, publication history and target characteristics.

\subsection{Retriever complementarity and pooled reach}

We examine whether different rankings recover different
cases by pooling the same 24 retrieval configurations
across all three evidence sources.
A case is reached at depth $K$ if any configuration
retrieves an eligible target within its top $K$.
The pooled result summarizes separate rankings;
it is not a combined ranker.

\noindent\textbf{Evaluation.}
Within each evidence source and tier, all configurations
use the same cases and target works, following
Appendix~E.1. A configuration that returns no ranking
for a case contributes no hit.
We compare pooled reach with the highest individual
Hit@$K$ among the included configurations, determined
separately for each tier and depth.
Bibliography and Context-Root use the shared candidate
pool; Author-Root uses its separate pool described in
Appendix~D.
Pooling uses up to 24 top-$K$ lists and therefore a
larger collective retrieval budget than an individual
ranking.

\begin{table}[!htbp]
\centering
\small
\setlength{\tabcolsep}{5pt}
\renewcommand{\arraystretch}{1.08}
\caption{\textbf{Individual and pooled retrieval.}
Values are Hit@$K$ (\%).
Best individual is the highest individual reach among
the same 24 configurations, determined separately for
each tier and depth.
Pooled counts cases reached by any of their top-$K$
lists. All configurations use the same cases within
each row. Comparisons are within each source;
cases and candidate pools differ across sources.
CiteNew pooled Hit@1000 is shown to one decimal because
numerical bounds on PageRank membership do not determine
its second decimal.}
\label{tab:pooled-reach}

\begin{tabular}{@{}lrrrrr@{}}
\toprule
& & \multicolumn{2}{c}{Hit@1000 (\%)}
  & \multicolumn{2}{c}{Hit@100 (\%)} \\
\cmidrule(lr){3-4}
\cmidrule(lr){5-6}
Tier & $N$
& Best individual & Pooled
& Best individual & Pooled \\
\midrule

\multicolumn{6}{@{}l}{\emph{Bibliography}} \\
CiteNext
& 135,111 & 61.95 & 84.18 & 46.25 & 66.29 \\
CiteNew
& 134,314 & 59.35 & 83.5 & 37.41 & 63.53 \\
CiteNew-Isolated
& 133,166 & 56.75 & 81.89 & 33.96 & 59.41 \\

\midrule
\multicolumn{6}{@{}l}{\emph{Context-Root}} \\
Next
& 7,848 & 57.67 & 74.29 & 38.04 & 52.94 \\
New
& 6,659 & 47.98 & 68.88 & 22.53 & 41.93 \\
Isolated
& 5,666 & 39.94 & 63.73 & 18.53 & 36.76 \\

\midrule
\multicolumn{6}{@{}l}{\emph{Author-Root}} \\
All
& 1,633 & 74.89 & 90.69 & 58.79 & 77.40 \\
Next
& 1,330 & 75.41 & 90.90 & 57.89 & 77.29 \\
New
& 904 & 58.41 & 81.86 & 36.17 & 59.40 \\
Isolated
& 665 & 55.04 & 79.25 & 34.59 & 54.74 \\

\bottomrule
\end{tabular}
\end{table}

\noindent\textbf{Results.}
Pooling increases recovery over the best individual
configuration in every tier at both depths
(Table~\ref{tab:pooled-reach}).
On Context-Root-New, Hit@1000 rises from 47.98\%
to 68.88\%, leaving 2,072 of 6,659 cases unreached
(31.12\%).
On Author-Root-New, it rises from 58.41\% to 81.86\%,
leaving 164 of 904 cases unreached (18.14\%).
Thus, the rankings provide complementary recovery,
while some cases remain unreached even by their union.

\FloatBarrier

\subsection{Retrieval across case characteristics}
\suppressfloats[t]

\textbf{Publication history.}
We examine how retrieval varies with the number of distinct
publications by the researcher strictly before the focal paper.
Table~\ref{tab:history-size} reports Hit@100 for
SPECTER2-Proximity with 16 centroids, evaluating CiteNew
and Context-Root-New targets on the same 6,659 Context-Root-New
cases. CiteNew recovery increases across history-size
groups, whereas Context-Root-New recovery varies less.
These are observational associations and do not establish
the effect of adding publications to a researcher's history.

\begin{table}[!htbp]
\centering
\small
\caption{
Retrieval by publication-history size on the same
6,659 Context-Root-New cases.
History size counts distinct publications strictly before
the focal paper. Both columns use SPECTER2-Proximity with
16 centroids and report Hit@100 (\%).
}
\label{tab:history-size}
\begin{tabular}{lrrr}
\toprule
Earlier publications & $N$ & CiteNew & Context-Root-New \\
\midrule
1--2      & 78    & 33.33 & 15.38 \\
3         & 1,805 & 39.06 & 14.24 \\
4--5      & 1,703 & 41.69 & 14.68 \\
6--9      & 1,610 & 46.52 & 16.40 \\
$\geq 10$ & 1,463 & 49.97 & 15.93 \\
\bottomrule
\end{tabular}
\end{table}

\noindent\textbf{Prior citation visibility.}
Targets with no recorded citations before the focal cutoff
are not necessarily missed.
Among the 9,108 eligible Context-Root-New target pairs,
pooling the 24 rankings retrieves 76.22\% of the
1,148 targets with no recorded prior citations within
the top 1,000 of at least one ranking, compared with
63.74\% of the 7,960 previously cited targets.
These percentages measure recovery of individual
targets, not the fraction of cases reached.
The comparison is descriptive: it does not isolate
citation visibility from related characteristics
such as target age.

\FloatBarrier
\Needspace{8\baselineskip}

\subsection{Target count and tier differences}
\label{app:target-count}

Stricter tiers contain fewer eligible targets and may
exclude cases with no remaining targets. We test whether
their lower retrieval scores are explained by target
count, while accounting separately for changes in the
evaluated cases.

\noindent\textbf{Count-matched comparison.}
For each comparison, we evaluate both target sets on
the stricter tier's cases, using identical rankings,
candidate pools and eligibility rules.
Within each case, we calculate the expected score
when retaining targets uniformly at random from the
broader tier until their number matches the stricter tier.
Expectations are computed exactly using the corresponding
target-matching rules.
We report the stricter tier's observed score minus this
expectation. Negative differences indicate lower retrieval
than target-count reduction alone predicts.
Macro recall is unchanged in expectation under random
target removal.
Confidence intervals use a researcher-cluster percentile
bootstrap with 10,000 replicates.

\begin{table}[!htbp]
\centering
\footnotesize
\setlength{\tabcolsep}{3pt}
\renewcommand{\arraystretch}{1.12}

\caption{\textbf{Tier differences after matching target counts.}
Values are observed minus count-matched expected scores,
in percentage points, with 95\% confidence intervals below.
The expectation randomly reduces the broader tier's targets
to the stricter tier's count; both scores use the stricter
tier's cases and identical rankings.
Negative values indicate lower retrieval than target-count
reduction alone predicts.
New versus Next excludes the researcher's prior citations;
Isolated versus New additionally excludes recent coauthors'
prior citations.
BM25 uses the compact query; the dense retrievers use
16 history centroids.
Empty rankings receive zero for both scores.
One near-zero interval endpoint is shown to three decimals.}
\label{tab:tier-count-concise}

\begingroup
\newcommand{\tcvalue}[2]{%
  \shortstack{$#1$\\[1pt]{\scriptsize[$#2$]}}}

\begin{tabular*}{\linewidth}
{@{\extracolsep{\fill}}lcc@{\hspace{12pt}}cc@{}}
\toprule
\multicolumn{5}{@{}l}{\textbf{A. Bibliography}} \\
\multicolumn{5}{@{}l}{%
\scriptsize
$N$: CiteNew $134{,}314$;
CiteNew-Isolated $133{,}166$} \\
\addlinespace[5pt]
& \multicolumn{2}{c}{\textbf{New versus Next}}
& \multicolumn{2}{c}{\textbf{Isolated versus New}} \\
\cmidrule(lr){2-3}
\cmidrule(lr){4-5}
Retriever
& $\Delta$Hit@100 & $\Delta$Recall@100
& $\Delta$Hit@100 & $\Delta$Recall@100 \\
\midrule
BM25
& \tcvalue{-2.14}{-2.23,\,-2.05}
& \tcvalue{-1.16}{-1.20,\,-1.13}
& \tcvalue{-0.65}{-0.72,\,-0.58}
& \tcvalue{-0.19}{-0.22,\,-0.16} \\
\addlinespace[4pt]
SPECTER2-Proximity
& \tcvalue{-2.95}{-3.05,\,-2.85}
& \tcvalue{-1.55}{-1.59,\,-1.51}
& \tcvalue{-0.75}{-0.83,\,-0.68}
& \tcvalue{-0.20}{-0.24,\,-0.17} \\
\addlinespace[4pt]
GritLM-7B
& \tcvalue{-2.99}{-3.09,\,-2.89}
& \tcvalue{-1.81}{-1.86,\,-1.77}
& \tcvalue{-0.86}{-0.94,\,-0.78}
& \tcvalue{-0.29}{-0.33,\,-0.26} \\
\addlinespace[4pt]
Personalized PageRank
& \tcvalue{-7.82}{-7.96,\,-7.68}
& \tcvalue{-5.00}{-5.06,\,-4.94}
& \tcvalue{-3.42}{-3.52,\,-3.31}
& \tcvalue{-1.33}{-1.37,\,-1.30} \\
\bottomrule
\end{tabular*}

\par\vspace{10pt}

\begin{tabular*}{\linewidth}
{@{\extracolsep{\fill}}lcc@{\hspace{12pt}}cc@{}}
\toprule
\multicolumn{5}{@{}l}{\textbf{B. Context-Root}} \\
\multicolumn{5}{@{}l}{%
\scriptsize
$N$: New $6{,}659$; Isolated $5{,}666$} \\
\addlinespace[5pt]
& \multicolumn{2}{c}{\textbf{New versus Next}}
& \multicolumn{2}{c}{\textbf{Isolated versus New}} \\
\cmidrule(lr){2-3}
\cmidrule(lr){4-5}
Retriever
& $\Delta$Hit@100 & $\Delta$Recall@100
& $\Delta$Hit@100 & $\Delta$Recall@100 \\
\midrule
BM25
& \tcvalue{-0.53}{-0.76,\,-0.31}
& \tcvalue{-0.45}{-0.64,\,-0.26}
& \tcvalue{+0.07}{-0.07,\,+0.20}
& \tcvalue{+0.05}{-0.06,\,+0.17} \\
\addlinespace[4pt]
SPECTER2-Proximity
& \tcvalue{-0.59}{-0.83,\,-0.36}
& \tcvalue{-0.48}{-0.69,\,-0.27}
& \tcvalue{+0.05}{-0.10,\,+0.19}
& \tcvalue{+0.05}{-0.08,\,+0.18} \\
\addlinespace[4pt]
GritLM-7B
& \tcvalue{-0.98}{-1.24,\,-0.74}
& \tcvalue{-0.87}{-1.08,\,-0.66}
& \tcvalue{+0.11}{-0.05,\,+0.25}
& \tcvalue{+0.12}{-0.02,\,+0.26} \\
\addlinespace[4pt]
Personalized PageRank
& \tcvalue{-2.85}{-3.17,\,-2.55}
& \tcvalue{-2.46}{-2.72,\,-2.21}
& \tcvalue{-0.52}{-0.72,\,-0.32}
& \tcvalue{-0.41}{-0.57,\,-0.25} \\
\bottomrule
\end{tabular*}

\par\vspace{10pt}

\begin{tabular*}{\linewidth}
{@{\extracolsep{\fill}}lcc@{\hspace{12pt}}cc@{}}
\toprule
\multicolumn{5}{@{}l}{\textbf{C. Author-Root}} \\
\multicolumn{5}{@{}l}{%
\scriptsize
$N$: New $904$; Isolated $665$} \\
\addlinespace[5pt]
& \multicolumn{2}{c}{\textbf{New versus Next}}
& \multicolumn{2}{c}{\textbf{Isolated versus New}} \\
\cmidrule(lr){2-3}
\cmidrule(lr){4-5}
Retriever
& $\Delta$Hit@100 & $\Delta$Recall@100
& $\Delta$Hit@100 & $\Delta$Recall@100 \\
\midrule
BM25
& \tcvalue{-3.08}{-4.35,\,-1.87}
& \tcvalue{-2.17}{-3.14,\,-1.24}
& \tcvalue{-1.02}{-1.86,\,-0.21}
& \tcvalue{-0.51}{-1.18,\,+0.16} \\
\addlinespace[4pt]
SPECTER2-Proximity
& \tcvalue{-4.98}{-6.39,\,-3.60}
& \tcvalue{-3.80}{-4.91,\,-2.71}
& \tcvalue{-0.15}{-1.08,\,+0.78}
& \tcvalue{-0.01}{-0.82,\,+0.79} \\
\addlinespace[4pt]
GritLM-7B
& \tcvalue{-6.41}{-7.88,\,-5.01}
& \tcvalue{-5.46}{-6.62,\,-4.34}
& \tcvalue{-0.96}{-1.92,\,-0.001}
& \tcvalue{-0.70}{-1.52,\,+0.11} \\
\addlinespace[4pt]
Personalized PageRank
& \tcvalue{-3.35}{-4.67,\,-2.06}
& \tcvalue{-2.10}{-3.10,\,-1.11}
& \tcvalue{+0.01}{-0.81,\,+0.81}
& \tcvalue{+0.11}{-0.60,\,+0.82} \\
\bottomrule
\end{tabular*}

\endgroup
\end{table}

\noindent\textbf{Results.}
After excluding works previously cited by the researcher,
both metrics remain below the count-matched expectation
for all four retrievers in all three evidence sources,
with confidence intervals excluding zero
(Table~\ref{tab:tier-count-concise}).
For example, SPECTER2-Proximity K16 has an expected
Author-Root-New Hit@100 of 34.84\%, compared with
29.87\% observed, a difference of $-4.98$ percentage
points using unrounded scores.
The additional exclusion of recent coauthors' citations
has smaller or less consistent effects.
For Context-Root, the Isolated-versus-New intervals
include zero for both metrics for the three text
retrievers, while PageRank retains negative differences. PageRank uses citation links, so its results also reflect
how the prior-citation filters change the targets reachable
through those links.

Changes in the evaluated cases also contribute to
the aggregate declines.
For SPECTER2-Proximity K16, restricting Context-Root-Next
to cases that remain eligible for Context-Root-New
accounts for approximately 71\% of the aggregate
Hit@100 decline, before changing their target sets.

Target count and case composition therefore explain
part of the aggregate tier declines, while the retained
targets also matter.
These associations concern recorded citation history,
not demonstrated researcher awareness.

\FloatBarrier
\section{Robustness and encoder exposure}
\suppressfloats[t]

We examine sensitivity to history representation and query
construction. We then use a training-free baseline and an
exact training-edge overlap audit to assess whether encoder
exposure could explain the reported retrieval patterns.

\subsection{History representation}

\textbf{Number of centroids.}
We vary the maximum number of history centroids
$K \in \{1,8,16,24,64\}$ for SPECTER2-Proximity.
Within each bibliography tier, all settings use the same
publication histories, candidate pool, evaluation cases,
and tie-breaking rule. When fewer than $K$ history papers
are available, we use the available number of centroids.
Table~\ref{tab:centroid-sweep-hit100} reports Hit@100.

\begin{table}[!htbp]
\centering
\small
\caption{
Centroid-count sensitivity for SPECTER2-Proximity.
Values are Hit@100 (\%) on the same cases within each
bibliography tier. $K=16$ is the reference setting.
}
\label{tab:centroid-sweep-hit100}
\begin{tabular}{rrrr}
\toprule
Maximum centroids & CiteNext & CiteNew & CiteNew-Isolated \\
\midrule
1  & 32.37 & 27.85 & 24.67 \\
8  & 37.20 & 32.14 & 28.78 \\
16 & 37.39 & 32.23 & 28.92 \\
24 & 37.39 & 32.23 & 28.91 \\
64 & 37.41 & 32.23 & 28.92 \\
\midrule
$N$ & 135{,}111 & 134{,}314 & 133{,}166 \\
\bottomrule
\end{tabular}
\end{table}

Compared with $K=1$, $K=16$ improves Hit@100 by
4.25--5.02 percentage points across the three tiers.
Results change by approximately 0.02 percentage points
or less with $K=24$ or $K=64$.
We use $K=16$ as a fixed common reference for the
multi-centroid methods we test; we do not claim that
it is optimal for every encoder.
Appendix~E reports the corresponding $K=1$ and $K=16$
results across encoders and evidence sources.

\Needspace{8\baselineskip}
\subsection{Query construction}

We examine sensitivity to the amount of publication history
and the length of compact queries.

\noindent\textbf{Amount of publication history.}
We compare all earlier papers with the latest 20 papers on
the CiteNew cases with more than 20 eligible earlier publications
($N=7{,}838$, or 5.8\% of the tier).
Within this subset, using all history increases Hit@100 by
0.80--2.08 percentage points for SPECTER2-Proximity,
Qwen3, and GritLM with K16, but lowers it for several other
representations. It also lowers nDCG@10 and MRR in every
comparison. Since the history selections coincide for
researchers with at most 20 earlier papers, the estimated
full-population Hit@100 change for these configurations is
approximately 0.12 percentage points or less.
The Hit@100 ordering among the compared configurations
remains unchanged in these checks; the effect of
additional history depends on its representation.

\noindent\textbf{Compact-query length.}
Over the full CiteNew population, increasing the word limit
from 128 to 256 changes Hit@100 by $-1.00$ percentage points
for BM25, $-0.49$ for SPLADE, and $+0.31$ for Qwen3.
A moderately larger compact query therefore does not
consistently improve Hit@100.
Removing the limit improves BM25: against a capped control
built from the same distinct-work histories, CiteNew Hit@100
rises from 29.71\% to 32.48\%
(+2.77 points; 95\% CI [2.59, 2.95]).
Hit@100 gains range from 1.55 to 2.94 points across the six
bibliography and Context-Root tiers.
Intervals for these full-population contrasts use a paired
bootstrap resampling researchers ($B=10{,}000$).
On the 7,838 CiteNew cases with more than 20 earlier
publications, removing the BM25 word limit while retaining
the latest 20 papers increases Hit@100 from 33.90\%
to 38.90\% (+5.00 points; 95\% CI [4.14, 5.86]);
nDCG@10 and MRR also improve in this subset comparison.

\subsection{Encoder exposure}

\textbf{Training-free control.}
BM25 has no learned parameters or pretrained encoder.
Its Hit@100 decreases from 33.96\% on CiteNext to
29.71\% on CiteNew and 26.79\% on CiteNew-Isolated.
On Context-Root-New ($N=6{,}659$), it reaches 12.43\%
at depth 100 and 23.47\% at depth 1000.
The decline across bibliography tiers and the incomplete
recovery of Context-Root targets therefore also occur
without a pretrained encoder.
This control does not rule out effects of training-data
exposure on the learned retrievers.

\textbf{Training-edge overlap.}
We check whether any target citation edge in the final
evaluation also appears as the same directed positive edge
in the documented SciRepEval training data associated with
SPECTER2. In the base citation-prediction data, 4 of
$2{,}435{,}222$ CiteNext edges overlap.
At most 4 of $2{,}195{,}342$ CiteNew edges and at most
4 of $1{,}908{,}576$ CiteNew-Isolated edges overlap.
We find no exact overlap in the SciRepEval configurations
listed for the SPECTER2-Proximity adapter.
Exact training-edge overlap is therefore too rare to explain
the aggregate retrieval results.
This audit does not determine whether the underlying papers
or related text appeared elsewhere in pretraining, and
equivalent training-edge records were not available for
the other pretrained encoders.

\textbf{An encoder trained before the test period.}
The original SPECTER encoder~\citep{R15} was released in 2020 and
trained on earlier citation data, so it cannot have seen citations
from the 2021--2023 test period.
On an earlier version of the benchmark, with 16 history centroids,
it shows the same decline across tiers as SPECTER2
(Table~\ref{tab:earlier-specter}): Hit@100 falls by 24\% from
CiteNext to CiteNew and by 35\% to CiteNew-Isolated, compared with
21\% and 31\% for SPECTER2.
These values come from the earlier benchmark version and are not
comparable with the main tables.

\begin{table}[!htbp]
\centering
\small
\caption{\textbf{A pre-2021 encoder shows the same decline across tiers.}
Hit@100 (\%) with 16 history centroids on an earlier version of the
benchmark; values are not comparable with the main tables.}
\label{tab:earlier-specter}
\begin{tabular}{@{}lrrr@{}}
\toprule
Encoder & CiteNext & CiteNew & CiteNew-Isolated \\
\midrule
SPECTER (2020) & 39.52 & 30.15 & 25.72 \\
SPECTER2 & 53.36 & 42.37 & 36.60 \\
\bottomrule
\end{tabular}
\end{table}

\noindent\textbf{Supervised researcher overlap.}
The Author-Root panel includes researchers present in
the encoder-adaptation training split.
Excluding their 347 cases leaves 1,286 All, 1,051 Next,
753 New and 569 Isolated cases.
Additionally excluding identified validation-researcher
overlap leaves 1,223, 995, 716 and 545 cases, respectively.

Neither exclusion changes the direction of any
adapted-minus-frozen Hit@100 difference or whether its
95\% researcher-cluster interval excludes zero.
For example, Qwen3-Embedding-8B LoRA's advantage on
Author-Root-Isolated is 3.16 percentage points
[0.60, 5.71] on the full panel and 3.85 [0.92, 6.61]
after excluding both training and validation overlap.
These checks address identified overlap;
missing Semantic Scholar identifiers limit matching
coverage.

\subsection{Sensitivity to Context-Root label precision}
\label{app:context-root-precision}

\paragraph{Higher-precision labels.}
We repeat the Context-Root analyses with labels from the
classifier's teacher, GPT-5.4-mini, using the same prompt
(Appendix~B.3).
The teacher reassesses every context with a classifier
score above 0.2, including borderline contexts below the
classifier's 0.5 decision threshold.
A cited work becomes a target when the teacher labels
at least one of its contexts as generative.
On the human-coded development contexts used in
Appendix~B.3, this procedure increases precision from
0.63 to 0.83, with recall decreasing from 0.93 to 0.88.

\paragraph{The main findings persist.}
The higher-precision labels yield 5,545 Next, 4,601 New
and 3,817 Isolated cases.
Table~\ref{tab:context-root-precision} summarizes the
comparisons.
All 23 history-based configurations still outperform
popularity, and their Hit@100 declines from Next to New
to Isolated.
The Next-to-New decline persists after matching target
counts for all four retrievers in
Table~\ref{tab:tier-count-concise}.
Own histories continue to outperform matched histories
for all three tested retrievers.
Pooling all 24 rankings still leaves most New cases
unreached at depth 100.
For all three retrievers, the next paper's title and
abstract continue to improve retrieval, including after
excluding papers with detected target mentions.
Generative-use labels also continue to cover only a
small share of author endorsements
(Appendix~\ref{app:author-context}).

\paragraph{Other sensitivities and absolute scores.}
The conclusions also persist for placebo titles,
pooling in every tier, publication-history size,
prior-citation visibility and uncapped BM25 queries
(Appendices~F--H).
On cases with targets under both label sets, the reference
retriever's Hit@100 is 0.7--2.1 percentage points lower
with the higher-precision labels.

\begin{table}[t]
\centering
\footnotesize
\setlength{\tabcolsep}{4pt}
\renewcommand{\arraystretch}{1.08}

\caption{\textbf{Context-Root findings persist with
higher-precision labels.}
Retrieval results use SPECTER2-Proximity with 16 history
centroids, except pooling, which combines all 24 rankings.
Precision and recall are measured on the human-coded
development contexts in Appendix~B.3.
Values are percentages unless indicated otherwise; precision and recall are proportions;
pp denotes percentage points, and brackets give
95\% confidence intervals.
Cases without an eligible target are excluded separately
under each label set.
The own-history ratio uses the matched-history comparison
described in Table~\ref{tab:matched-history-cross-source}.
The final row concerns author endorsements with usable
citing text.
$H$ denotes history, $T$ the next paper's title and $A$
its abstract.}
\label{tab:context-root-precision}

\begin{tabular*}{\linewidth}{
@{\extracolsep{\fill}}
lcc
@{}}
\toprule
Quantity
& Classifier
& \shortstack{Higher-precision\\labels} \\
\midrule

\multicolumn{3}{@{}l}{
\textit{Label quality on development contexts}} \\
Precision
& 0.63
& 0.83 \\
Recall
& 0.93
& 0.88 \\

\addlinespace[2pt]
\cmidrule(lr){1-3}
\addlinespace[2pt]

\multicolumn{3}{@{}l}{
\textit{Reach falls beyond prior citations (\S4.2)}} \\
Next Hit@100
& 22.82
& 21.59 \\
New Hit@100
& 15.26
& 14.48 \\
Isolated Hit@100
& 14.08
& 13.62 \\
Cases (Next / New / Isolated)
& 7,848 / 6,659 / 5,666
& 5,545 / 4,601 / 3,817 \\

\addlinespace[2pt]
\cmidrule(lr){1-3}
\addlinespace[2pt]

\multicolumn{3}{@{}l}{
\textit{The Next-to-New drop persists after count matching
(\S4.2)}} \\
\shortstack[l]{Observed minus expected\\New Hit@100 (pp)}
& \shortstack{$-0.59$\\
  \scriptsize[$-0.83$, $-0.36$]}
& \shortstack{$-0.36$\\
  \scriptsize[$-0.59$, $-0.12$]} \\

\addlinespace[2pt]
\cmidrule(lr){1-3}
\addlinespace[2pt]

\multicolumn{3}{@{}l}{
\textit{History beats popularity (\S4.1)}} \\
New Hit@100 (history / popularity)
& 15.26 / 2.30
& 14.48 / 2.26 \\

\addlinespace[2pt]
\cmidrule(lr){1-3}
\addlinespace[2pt]

\multicolumn{3}{@{}l}{
\textit{The signal is researcher-specific (\S4.1)}} \\
New nDCG@1000 (own / matched histories)
& $8.4\times$
& $8.6\times$ \\

\addlinespace[2pt]
\cmidrule(lr){1-3}
\addlinespace[2pt]

\multicolumn{3}{@{}l}{
\textit{Pooling leaves gaps (\S4.2)}} \\
New cases unreached at depth 100
& 58.07
& 59.16 \\

\addlinespace[2pt]
\cmidrule(lr){1-3}
\addlinespace[2pt]

\multicolumn{3}{@{}l}{
\textit{Project information helps (\S4.3)}} \\
New Hit@100 gain, $H+T$ over $H$ (pp)
& 9.72
& 9.65 \\
New Hit@100 gain, $H+T+A$ over $H$ (pp)
& 20.24
& 19.52 \\

\addlinespace[2pt]
\cmidrule(lr){1-3}
\addlinespace[2pt]

\multicolumn{3}{@{}l}{
\textit{Labels cover few author endorsements
(Appendix~\ref{app:author-context})}} \\
Endorsements with a generative-use label
& 7.6
& 6.0 \\

\bottomrule
\end{tabular*}
\end{table}
\end{document}